\documentclass[11pt,a4paper]{article}
\usepackage[T1]{fontenc}
\usepackage[utf8]{inputenc}
\usepackage{lmodern}
\usepackage[margin=1in]{geometry}
\usepackage{amsmath,amssymb,amsthm,mathtools}
\usepackage{microtype,enumitem}
\usepackage{xurl}
\usepackage[hidelinks,unicode,pdfdisplaydoctitle]{hyperref}
\hypersetup{
 pdftitle={Approximately Counting Knapsack Solutions in Almost Linear Time},
 pdfauthor={},
 pdfsubject={Randomized approximation algorithms for counting knapsack},
 pdfkeywords={counting knapsack, approximate counting, exponential tilting, randomized algorithms}
}
\newtheorem{theorem}{Theorem}[section]
\newtheorem{lemma}[theorem]{Lemma}
\newtheorem{proposition}[theorem]{Proposition}
\newtheorem{corollary}[theorem]{Corollary}
\theoremstyle{definition}

\theoremstyle{remark}
\newtheorem{remark}[theorem]{Remark}
\newcommand{\E}{\mathbb E}
\newcommand{\Var}{\operatorname{Var}}
\newcommand{\Cov}{\operatorname{Cov}}
\newcommand{\tO}{\widetilde O}
\DeclareMathOperator{\Bern}{Bern}
\DeclareMathOperator{\Bin}{Bin}
\setlist{itemsep=2pt,topsep=4pt}
\allowdisplaybreaks
\newcommand{\one}{\mathbf 1}
\title{Approximately Counting Knapsack Solutions\\in Almost Linear Time}
\author{Lin Chen, Biyou Ma, Yuchen Mao, Guochuan Zhang\footnote{\{chenlin198662,12621057,maoyc,zgc\}@zju.edu.cn}}
\date{\today}
\begin{document}
\maketitle

\begin{abstract}
Given integer weights $w_1,\ldots,w_n\in[1,T]$, we approximate the
number $Z$ of subsets whose total weight is at most the capacity $T$.
For every fixed $\delta>0$ we give a randomized algorithm that
estimates $Z$ to relative error $\varepsilon$ with probability at
least $2/3$ in $O_\delta(n+\varepsilon^{-(1+\delta)})$ time on the
standard word RAM with words of $\Theta(\log(n/\varepsilon)+\log T)$
bits. This improves the time bound
$\tO(n\varepsilon^{-2})$ by Jin, Park, and Saha.

Instead of estimating the core threshold probabilities by sampling,
we compute them under an exponentially tilted product law by Fourier
analysis; this suffices when the total weight is polynomial. For
arbitrary weights we use a discrete inverse Littlewood--Offord theorem
to show that, after an inward rounding that preserves almost all
feasible sets, all but a small power of $n+\varepsilon^{-1}$ of the
items selected under the tilted law lie, in expectation, in a
progression of bounded rank. We find such a progression by sampling
and lattice reduction, count the weights in it with the Fourier
machinery, and add back the remaining weights through a degree-truncated
inclusion--exclusion whose correction terms are estimated by
sampling.
\end{abstract}

\section*{AI statement} The high-level proof route was designed by the authors. The technical details, especially the Fourier analysis and probablistic argument, are attributed to GPT Astra Ultra.  GPT and Claude were used for drafting the paper and also verifying this paper’s main results. The authors take full responsibility for
the entire content, technical claims, and integrity of the paper.

\newpage
\setcounter{tocdepth}{2}
\tableofcontents
\newpage

\section{Introduction}\label{s48:intro}
Let $T$ be a positive integer and $w_1,\ldots,w_n$ integers with
$1\le w_i\le T$, and write
\[
 Z(w,T)=\#\left\{A\subseteq[n]:\sum_{i\in A}w_i\le T\right\}
\]
for the number of subsets of the weights whose sum is at most the
capacity $T$. As in~\cite{JPS}, the bounds on the weights lose no
generality: weights larger than $T$ lie in no feasible subset, and $k$
zero weights multiply the count by $2^k$, so both are removed in one
scan of the input. All time bounds below concern the reduced input. Computing $Z(w,T)$ exactly is \#P-hard~\cite{FJ}, and the
standard dynamic program takes $O(n(T+1))$ arithmetic operations, which
is exponential in the bit length of $T$. The question is therefore how
fast $Z$ can be approximated. A fully polynomial randomized
approximation scheme (FPRAS) estimates $Z(w,T)$ to relative error
$\varepsilon$ with probability at least $2/3$ in time polynomial in the
input size and $\varepsilon^{-1}$; a fully polynomial time
approximation scheme (FPTAS) does so deterministically.

The first randomized approaches used Markov chains: Dyer et
al.~\cite{DFKKPV} gave a mildly exponential algorithm, and Morris and
Sinclair~\cite{MS} obtained the first FPRAS by proving rapid mixing for the random walk on the Boolean cube
truncated by the knapsack constraint. Dyer~\cite{Dyer} gave a different
scheme: round the weights, count the rounded instance by dynamic
programming, and estimate by sampling the fraction of rounded solutions
that satisfy the original constraint; with randomized rounding this
takes $\tO(n^{5/2}+n^2\varepsilon^{-2})$ time. Feng and Jin~\cite{FJ}
refined this approach, building approximate samplers by convolution
and exploiting the structure of subsets near the capacity, and obtained
$\tO(n^{3/2}\varepsilon^{-2})$ time for inputs that fit in
$O(\log(n/\varepsilon))$-bit words. They asked whether
$\tO(n\operatorname{poly}(1/\varepsilon))$ is possible. Jin, Park, and
Saha~\cite{JPS} answered this with an $\tO(n\varepsilon^{-2})$
algorithm on a word RAM with $O(\log(n/\varepsilon)+\log T)$-bit words.
They use exponential tilting to lower the precision needed in FFT
convolutions of rounded counting tables, and sampling to correct for
the rounding.

The first deterministic FPTASes, by Gopalan et al.~\cite{GKMSVV} and
by \v{S}tefankovi\v{c}, Vempala, and Vigoda~\cite{SVV}, run in
$O(n^3\varepsilon^{-1}\log(n/\varepsilon))$ time. Their key idea is to
index the dynamic program by a geometric grid of target counts and to
approximate the least capacity that reaches each target. Rizzi and
Tomescu~\cite{RT} improved the logarithmic factor with floating-point
counts, and Halman~\cite{Halman} gave a simpler scheme that also
handles bounded multiplicities. Gawrychowski, Markin, and
Weimann~\cite{GMW} reached $\tO(n^{5/2}\varepsilon^{-3/2})$ by
combining sparse approximations of the counting functions through
divide-and-conquer convolution, and Jin, Park, and Saha~\cite{JPSdet},
adding a deterministic balancing procedure, reached
$\tO(n^{12/5}\varepsilon^{-7/5})$ unit-cost arithmetic operations,
including arithmetic on intermediate count values.

\subsection{Our result}\label{rw:intro-results}
\begin{theorem}\label{fin:main}
Let $T$ be a positive integer, let $1\le w_1,\ldots,w_n\le T$ be
integers, and let $0<\varepsilon<1$. For every fixed $\delta>0$ there
is a randomized algorithm that returns $\widehat Z$ in
\[
                     O_\delta(n+\varepsilon^{-(1+\delta)})
\]
time and satisfies
$\Pr(|\widehat Z-Z|\le\varepsilon Z)\ge 2/3$.
The bound holds for all capacities and all weight magnitudes.
\end{theorem}

We refer to this bound as the target bound
\begin{equation}\label{eq:target}
 O_\delta(n+\varepsilon^{-(1+\delta)})
 \qquad\text{for each fixed }\delta>0 .
\end{equation}
The randomized algorithm of Jin, Park, and Saha~\cite{JPS} takes
$\tO(n\varepsilon^{-2})$ time. Theorem~\ref{fin:main} gives an additive
bound with accuracy term $\varepsilon^{-(1+\delta)}$ for arbitrarily
small fixed $\delta>0$. The bound is almost linear in
$n+\varepsilon^{-1}$, and in particular it is $O_\delta(n)$, the time
needed to read the input, whenever $\varepsilon^{-1}\le
n^{1/(1+\delta)}$. Two further features deserve emphasis. There is no
lower bound on the feasible density: the bound holds even when
$Z/2^n$ is exponentially small. And there is no restriction on the
weights beyond $w_i\le T$: their magnitudes, the number of distinct
values and their ratios are all arbitrary.

\paragraph{Model of computation.}
We work in the standard word RAM with multiplication and integer
division, as in~\cite{JPS}, and running time counts word operations.
The word length $L_{\mathrm{word}}=\Theta(\log((n+1)/\varepsilon)+\log(T+1))$ is
available as a machine parameter. Every weight, the capacity and every
index fit in one word. A uniformly random word costs one operation.

\subsection{Overview of the proof}\label{rw:intro-overview}
The key to the $\varepsilon^{-(1+\delta)}$ term is that we do not
estimate the threshold probabilities of the core by sampling: we
compute them analytically, by Fourier estimates on a tilted
law, and the main difficulty is to make these estimates relative
rather than additive when the weights are arbitrary. We sketch the
steps; Figure~\ref{rw:fig-deps} in Section~\ref{rw:intro-org} shows
how the sections depend on one another.

\paragraph{Exponential tilting.}
Following Jin, Park, and Saha~\cite{JPS}, for $\lambda\ge0$ let
$Q_\lambda$ include item $i$ independently with probability
$1/(1+e^{\lambda w_i})$ (Section~\ref{s48:tilting}). Then $Z$ equals an
explicitly computable normalizer $C_\lambda$ times an expectation $r$
under $Q_\lambda$. If $\lambda$ is chosen so that the capacity
constraint holds with probability between $0.4$ and $0.6$, then
$r\ge1/(5n)$. So even when $Z/2^n$ is exponentially small, the quantity
to be estimated is only polynomially small. Estimating it by sampling,
however, would take order $\varepsilon^{-2}$ samples.

\paragraph{Hidden structure.}
The second ingredient is structural: every instance hides a
low-rank core. At the balanced tilt, an inward rounding of the weights
at a suitable scale loses only an $O(\varepsilon)$ fraction of feasible
sets and creates a heavy atom in the distribution of the rounded sum.
A weighted form of the discrete inverse Littlewood--Offord theorem of Nguyen and
Vu~\cite{NVinverse} then implies that all but
$O_\eta((n+\varepsilon^{-1}+2)^\eta)$ of the selected sites, in
expectation, lie in a progression of bounded rank
(Proposition~\ref{iv:inner}, Section~\ref{iv:section}). The weights in the progression form a core with short integer
coordinates, so it remains to count instances of polynomial weight
magnitude and to account for the few remaining \emph{exceptional}
sites. Proposition~\ref{iv:inner} is an existence statement; how the
progression is found depends on the size of $n$ relative to
$\varepsilon^{-1}$, and we return to this after the counting tools.

\paragraph{Fourier estimates for thresholds.}
To count the core we do not sample: we evaluate the probability of the threshold event
$\{S\le T\}$, where $S=\sum_iw_iX_i$, by Fourier analysis
(Section~\ref{sp:section}). The threshold indicator is written as a
telescoping sum of smooth windows over dyadic scales, all periodized
with one common random period so that each scale needs few Fourier
coefficients; an entire-function expansion then evaluates the
characteristic function of $S$ at all these frequencies together. This
gives an additive $\varepsilon$ estimate in time nearly
$n+\varepsilon^{-1}$ when $\log(W+2)$ is polylogarithmic in
$n+\varepsilon^{-1}+2$. To make this error relative to $r$ without a
polynomial loss in $\varepsilon$, the Fourier estimate must be applied
to a positive weight whose total mass exceeds $r$ by at most a factor
that the running time can absorb.

\paragraph{All capacities at polynomial weight magnitude.}
At or above the middle capacity $W/2$ the untilted law already has
feasible probability at least one half. Below it,
Section~\ref{dt:section} moves the tilt so that the capacity lies
between one and three standard deviations above the tilted mean. At
this displaced tilt $C_\lambda$ exceeds $Z$ by at most a polynomial
factor (Lemma~\ref{dt:price}), and the tilt is found by a
deterministic linear-time search (Lemma~\ref{dt:search}). A single
tilt is still not enough: the tilted law may put too much mass on a
large infeasible atom just above the threshold.
Theorem~\ref{pw:main} (Section~\ref{pw:section}) attains
\eqref{eq:target} at every capacity when the total weight $W$ is at
most $(n+\varepsilon^{-1}+2)^C$ for a fixed $C$, which is the case for the core. The weights are grouped
by scaled size, a second tilt acts on the deviations inside the groups
separately on regions of group counts, and a conditional local limit
theorem (Lemma~\ref{pw:local}) shows that the mass used by the Fourier
estimate is controlled by $Z$, even at exponentially small density.
This rests on Fourier coefficients over polytopes of multiplicities
(Section~\ref{mc:region}) and on coarse trees that keep a few
exceptional weights at their original values (Lemma~\ref{ex:coarse}),
combined through a degree-truncated signed inclusion--exclusion whose
positive mass before cancellation is at most a constant times the
target (Section~\ref{sec:extra}).

\paragraph{Small inputs.}
When $n\le\varepsilon^{-(1+\delta/2)}$, polylogarithmic factors in $n$
are absorbed into the term $\varepsilon^{-(1+\delta)}$, so a
near-linear algorithm suffices. A sample from the tilted law and
lattice basis reduction construct the rounding scale, a progression
with $O(\log(n/\varepsilon))$ generators, and the exceptional sites
(Section~\ref{cg:section}). Section~\ref{bc:section} then computes the
prefix counts of the core at all residual capacities
(Lemma~\ref{bc:cdf}), returns these quantiles to the original weight
scale with an order-preserving decoder (Lemma~\ref{bc:decode}), and
combines them with the exceptional sites (Proposition~\ref{bc:mixed}).
The result, Proposition~\ref{bc:main}, runs in
$O_\delta(n\log^{C_\delta}(n+2)+\varepsilon^{-(1+\delta)})$ time; on its
own it gives $\tO_\delta(n+\varepsilon^{-(1+\delta)})$ for every input.

\paragraph{Large inputs.}
When $n>\varepsilon^{-(1+\delta/2)}$, the factor $\log^{C_\delta}n$ can
no longer be hidden, and every step must be linear in $n$.
Theorem~\ref{lcg:main} (Section~\ref{lcg:section}) uses a much smaller
sample to identify a progression of \emph{fixed} rank, recognized by
lattice reduction, and then classifies every site in constant time, in
$O_a(n)$ operations overall. The reduction caps the number of selected
exceptional sites at $O_a(n^a)$. Section~\ref{fin:section} accesses the list
only through rank, selection and sampling queries, after linear-time
preprocessing (Lemma~\ref{fin:ordered}). The exceptional trees
now have polynomially many roots, each with its own residual capacity;
rather than scanning the input once per root,
Lemma~\ref{fin:baselines} builds all the core tables from a single scan
with shared Fourier estimates. The proof of Theorem~\ref{fin:main}
closes Section~\ref{fin:section}.

\subsection{Organization}\label{rw:intro-org}
Section~\ref{rw:prelim} fixes notation and conventions and collects
the tools: the balanced tilt, sign
compression and lattice-point sums in fixed dimension, and the
degree-truncated count for a sparse set of exceptional weights,
with its coarse form (Lemma~\ref{ex:coarse}).
Section~\ref{rw:hidden} proves the hidden-structure theorem and
constructs the structure at logarithmic rank.
Section~\ref{rw:fourier} develops Fourier estimates on the tilted law,
Section~\ref{dt:section} the displaced tilt and the conditional local
estimate, and Section~\ref{pw:section} combines them at all capacities
for polynomial weight magnitude (Theorem~\ref{pw:main}).
Section~\ref{bc:section} proves Proposition~\ref{bc:main}, and
Section~\ref{rw:linear} treats large inputs and proves
Theorem~\ref{fin:main}. Section~\ref{sec:discussion} explains why the
proof needs slice estimates and records two limitations of simpler
approaches. Each section opens by stating what it proves
and where the result is used; Figure~\ref{rw:fig-deps} lists the
dependencies.

\begin{figure}[ht]
\centering
\small
\begin{tabular}{@{}lll@{}}
\hline
Section & Main results & Uses Sections \\
\hline
\ref{rw:prelim} Preliminaries & Lemmas~\ref{s48tilt:search}, \ref{s48tilt:price}, \ref{lem:basegeneral}, \ref{ex:coarse} & --- \\
\ref{rw:hidden} Hidden structure & Proposition~\ref{iv:inner}, Theorem~\ref{cg:structure} & \ref{rw:prelim} \\
\ref{rw:fourier} Fourier threshold estimates & Theorem~\ref{sp:additive}, Lemma~\ref{ft:geom} & \ref{rw:prelim} \\
\ref{dt:section} Displaced tilt & Lemmas~\ref{dt:price}, \ref{dt:search}, \ref{pw:local} & \ref{rw:prelim} \\
\ref{pw:section} Polynomial weights & Theorem~\ref{pw:main} & \ref{rw:prelim}, \ref{rw:fourier}, \ref{dt:section} \\
\ref{bc:section} Small-input algorithm & Proposition~\ref{bc:main} & \ref{rw:prelim}--\ref{pw:section} \\
\ref{rw:linear} Large-input algorithm & Theorems~\ref{lcg:main}, \ref{fin:main} & \ref{rw:prelim}--\ref{bc:section} \\
\hline
\ref{sec:discussion} Discussion & Proposition~\ref{whole:slicereduction} & \ref{rw:hidden}--\ref{dt:section}, \ref{rw:linear} \\
\hline
\end{tabular}
\caption{Dependencies between the sections. Section~\ref{sec:discussion}
is not used in the proof of Theorem~\ref{fin:main}.}
\label{rw:fig-deps}
\end{figure}

\paragraph{Further results.}
Proposition~\ref{whole:slicereduction} in Section~\ref{sec:discussion}
shows that the target bound contains additive threshold estimation on
a uniform fixed-cardinality slice as a special case, and
Remarks~\ref{prop:barrier} and~\ref{sp:obstruction} record limitations
of exact weight recovery and of a single positive envelope.
Proposition~\ref{iv:inner} (Section~\ref{rw:hidden}) is purely
structural and may be of independent interest.
Appendix~\ref{app:structured} contains two further results not used
in the proof of Theorem~\ref{fin:main}: a deterministic algorithm for a
fixed number of distinct weights with polylogarithmic dependence on the
accuracy (Theorem~\ref{thm:fixed}), and a deterministic
$O_m(n)+(A+1)\log^{O_m(1)}(n+A+\varepsilon^{-1}+2)$ algorithm when the
weights have coefficient range $0\le a_i\le A$ in a fixed number of
classes (Theorem~\ref{lc:main} and Corollary~\ref{lc:hidden}).

\section{Preliminaries}\label{rw:prelim}
This section fixes notation and conventions
(Subsection~\ref{s48:model}) and collects three tools used throughout:
the balanced tilt (Subsection~\ref{s48:tilting}), sign compression and
lattice-point sums in fixed dimension (Subsection~\ref{sec:fdtools}),
and a degree-truncated count for a sparse set of exceptional weights
(Subsection~\ref{sec:extra}).

\subsection{Notation and conventions}\label{s48:model}
\paragraph{Notation.}
Throughout, $n$ is the number of input weights, $T$ is the capacity,
the weights satisfy $1\le w_i\le T$, $Z$ is the number of feasible
subsets and $\widehat Z$ is the returned estimate; $\varepsilon$ is
the accuracy and $\delta>0$ is the fixed constant of
Theorem~\ref{fin:main}. Items are labeled, so equal weights are
distinct items. We write $[m]=\{1,\ldots,m\}$. For a set $A$ of items,
$|A|$ is its cardinality and $w(A)=\sum_{i\in A}w_i$ is its weight,
and $W=\sum_iw_i$ is the total weight; $\#\{\cdots\}$ is the number of
objects satisfying the stated condition.

We write $\Pr$ and $\E$ for probability and expectation when the
underlying law is clear from the context. When the law has a name,
say $Q$, we write $Q(\mathcal A)$ for the probability of an event
$\mathcal A$, and $\E_Q$ and $\Var_Q$ for expectation and variance
under $Q$. The indicator of $\mathcal A$ is $\one_{\mathcal A}$.
$\Bern(p)$, $\Bin(m,p)$ and $\mathcal N(\mu,\sigma^2)$ are the
Bernoulli, binomial and normal laws. The tilt $\lambda$, the tilted
law $Q_\lambda$ with marginals $p_i$, the normalizers $P(\lambda)$ and
$C_\lambda$, the tilted expectation $r$ and the sum $S$ are defined in
Subsection~\ref{s48:tilting} and are used throughout.

The letter $e$ always denotes the base of the natural logarithm, and
$\mathrm e(x)=\exp(2\pi i x)$; the letter $i$ is the imaginary unit
when it multiplies a phase and is an index otherwise. A wide hat
denotes an estimate of a number and the Fourier transform of a
function, with the normalization stated where it is used.
All logarithms are natural unless a base is indicated, and
$\log^{a}x$ means $(\log x)^{a}$.

Constants written $C$ or $c$, with or without subscripts, are
positive. Numbered constants such as $C_1$ are local: the same number
may denote different constants in different proofs. The size
parameter $R$ and the precision parameter $J$ are likewise chosen anew
in each result that uses them. A subscript on $O$ or $\tO$ lists the
parameters on which the implied constant depends. $\tO$ suppresses
fixed powers of $\log(n+2)+\log(\varepsilon^{-1}+2)$; local sampling
statements also allow $\log(\eta^{-1}+2)$ for their stated coupling
budget $\eta$. In particular, $\tO$ never hides a factor $\log T$.

\paragraph{Conventions.}
We may replace $\varepsilon$ by a smaller dyadic value within a factor
of two, and assume $0<\varepsilon\le1/2$; likewise, any prescribed
error budget may be replaced by a dyadic lower bound within a factor
of two. If $T\ge\sum_i w_i$, then $Z=2^n$ exactly; otherwise the
instance is nontrivial, and then $n\ge2$. A positive count estimate is
returned as a short significand and a binary exponent. We prove
statistical bounds for the ideal constructions and transfer their
success probabilities to the finite algorithm by coupling the entire
execution.

\paragraph{Arithmetic.}
The model is the word RAM of Subsection~\ref{rw:intro-results}.
Numbers that fit in one word, such as the weights, the capacity and
indices, are called original words, to distinguish them from the
short-precision numbers used in probability calculations. These have
polylogarithmically many bits, and operations on them are charged
their multiword cost. Arithmetic on a fixed number of original words
takes $O(1)$ word operations; for division this is long division by
Knuth's Algorithm~D~\cite[Section~4.3.1]{Knuth}. Short-precision evaluations of $\exp$, $\log$,
and bounded analytic functions are performed by range reduction and
convergent power series, at a cost polynomial in the requested
precision; in particular, small probabilities do not require an
iteration proportional to their negative logarithms.

\subsection{A balanced tilt}\label{s48:tilting}
This subsection sets up the change of measure from which all later
estimates start. We replace uniform subsets by a product Bernoulli law
$Q_\lambda$ that is exponentially tilted towards light subsets, and
express the count $Z$ exactly as a known normalizer times an
expectation $r$ under $Q_\lambda$. Lemma~\ref{s48tilt:search} finds,
in $\tO(n)$ work, a tilt under which the capacity constraint holds
with probability between $0.4$ and $0.6$ (or certifies that no tilt is
needed), and Lemma~\ref{s48tilt:price} shows that at such a tilt the
expectation $r$ is at least $1/(5n)$. Informally, after tilting, the
quantity to be estimated is never smaller than an inverse polynomial.
Sections~\ref{iv:section} and~\ref{cg:section} use this choice of
tilt. Section~\ref{dt:section} studies the same
identity~\eqref{s48tilt:identity} at a displaced tilt below the
middle capacity; Sections~\ref{pw:section} and~\ref{lcg:section} use
that tilt there and the fair law otherwise.
The arithmetic conventions are those of
Subsection~\ref{s48:model}.

Let $S=\sum_iw_iX_i$, and for $\lambda\ge0$ define
\[
 Q_\lambda=\bigotimes_i\Bern(p_i(\lambda)),\quad
 p_i(\lambda)=\frac1{1+e^{\lambda w_i}},\quad
 P(\lambda)=\prod_i(1+e^{-\lambda w_i}),\quad
 C_\lambda=e^{\lambda T}P(\lambda).
\]
The exact identity
\begin{equation}\label{s48tilt:identity}
 Z=C_\lambda r,\qquad
 r=\E_{Q_\lambda}\!\left[
   e^{\lambda(S-T)}\one_{\{S\le T\}}\right]
\end{equation}
follows by expanding the product measure over subsets. Indeed, under $Q_\lambda$ a subset $A$ has probability
$e^{-\lambda w(A)}/P(\lambda)$, so each feasible subset contributes
exactly $1/C_\lambda$ to $r$.
Since $e^{\lambda(S-T)}\le1$ on feasible subsets, each has
$Q_\lambda$-probability at least $1/C_\lambda$, so an event contains at
most $C_\lambda$ times its $Q_\lambda$-probability many feasible subsets. For a fixed set $E$ of items and $K\ge0$, Markov's
inequality for $e^{|A\cap E|}$ under $Q_\lambda$ thus gives
\begin{equation}\label{s48tilt:trunc}
 \#\{A\text{ feasible}:|A\cap E|>K\}\le C_\lambda\exp\Bigl((e-1)\sum_{i\in E}p_i-K\Bigr).
\end{equation}

When $\lambda=y/T$ with short dyadic $y$, as in the search below, an
expression $\lambda w_i$ is stored and compared by cross
multiplication with original integers. If it lies in a bounded
interval, binary search for its short approximation uses only
logarithmically many precision steps. If it is beyond a known
exponential cutoff, its contribution is bounded before being
discarded.

\begin{lemma}[Balanced search]\label{s48tilt:search}
With any prescribed constant failure probability, $\tO(n)$ work suffices
either to certify a positive absolute lower bound on $Q_0(S\le T)$ or to
find an actual $\lambda>0$ such that
\begin{equation}\label{s48tilt:balanced}
 Q_\lambda(S\le T)\in[0.4,0.6].
\end{equation}
The search uses $O(\log(n+1))$ queries on every execution.
\end{lemma}
\begin{proof}
The idea is to bisect in the rescaled parameter $y=\lambda T$, on
which the feasibility probability is monotone and $n$-Lipschitz.

Set $q(y)=Q_{y/T}(S\le T)$. Since $Q_{y/T}$ is the exponential family
with density proportional to $e^{-yS/T}$, differentiation gives
$q'(y)=-\Cov_y(S/T,\one_{\{S\le T\}})$. This is nonnegative because the
indicator is a decreasing function of $S/T$, and $|q'(y)|\le n$ because
$0\le S/T\le n$. At $y=2n$, using $xe^{-x}\le1/e$ termwise,
\[
 \E_y(S/T)=\frac1{2n}\sum_i
 \frac{2nw_i/T}{1+e^{2nw_i/T}}\le\frac1{2e},
\]
so $q(2n)>0.8$ by Markov's inequality.

Allow one query at zero and at most $N=1+\lceil\log_2(25n^2)\rceil$
midpoint queries. At each query, estimate $q$ by the empirical mean of
$O(\log(n+1))$ fresh independent samples, each generated by one scan of
the input. For a prescribed failure probability $\eta$, choose the
sample count so that, conditionally on the previous queries, the
probability of error greater than $0.01$ is at most $\eta/[2(N+1)]$
by Hoeffding's inequality.

If the estimate at zero is at least $0.35$, return the fair branch.
Otherwise bisect $[0,2n]$. At a midpoint $m$, return $\lambda=m/T$
if the estimate lies in $[0.45,0.55]$. If it is smaller, replace the
left endpoint by $m$. If it is larger, replace the right endpoint by $m$.
Report failure if all $N$ midpoint queries end without acceptance.

Suppose every query is within $0.01$ of its true value. Returning
the fair branch then certifies $q(0)\ge0.34$. If bisection starts,
$q(0)<0.36<1/2<q(2n)$, and each rejected midpoint preserves a root
of $q=1/2$ in the bracket. Every accepted midpoint has
$q(m)\in[0.44,0.56]$. By the $N$th midpoint the bracket has length
at most $2n/2^{N-1}\le0.08/n$. Its midpoint lies within $0.04/n$
of a root, so the Lipschitz bound gives $|q(m)-1/2|\le0.04$ and
the midpoint is accepted. A union bound over
the $N+1$ possible queries gives failure probability at most $\eta/2$
for the ideal search. Its work is $\tO(n)$ on every execution.

The midpoints $y$ are short dyadic, $\lambda$ is retained as $y/T$,
and each sampled capacity test uses the exact integer weights. Only the
Bernoulli probabilities are approximated; with absolute errors inverse
polynomial in $n$ and the number of samples, a union bound over all
coordinates of all samples couples the finite search to the ideal one
with failure probability at most $\eta/2$ (Section~\ref{s48:model}).
\end{proof}

The next lemma bounds the price of the tilt: at a balanced parameter
the expectation $r$ in \eqref{s48tilt:identity} cannot be small.

\begin{lemma}\label{s48tilt:price}
For every actual parameter satisfying \eqref{s48tilt:balanced},
$1/r\le5n$.
\end{lemma}
\begin{proof}
Write $q=Q_\lambda(S\le T)$ and
$E_i=p_i(1-p_i)Q_{\lambda,-i}(T-w_i<S_{-i}\le T)$.
The Efron--Stein inequality for the feasibility indicator gives
$q(1-q)\le\sum_i E_i$, since $E_i$ is the conditional variance of the
indicator given all coordinates other than $i$, averaged over those
coordinates. We now bound each $E_i$ by $r$.
Call $A$ a crossing subset for $i$ if it omits $i$ and
$T-w_i<w(A)\le T$. For any such $A$, its contribution to $E_i$ is
\[
 p_i(1-p_i)Q_{\lambda,-i}(A)
   =(1-p_i)e^{-\lambda(w(A)+w_i)}/P(\lambda)
   \le1/C_\lambda .
\]
Every crossing subset is feasible. Hence $E_i\le Z/C_\lambda=r$, and
therefore $nr\ge q(1-q)\ge0.24$.
\end{proof}

\paragraph{Notation used in later sections.}
For a group $D$ of items we write
$P_D=\prod_{i\in D}(1+e^{-\lambda w_i})$. We write
$H(s)=e^{\lambda(s-T)}\one_{\{s\le T\}}$ when the argument is a
physical weight. Lemmas~\ref{s48tilt:search} and~\ref{s48tilt:price}
give $r\ge1/(5n)$ on the success event of the search. In the fair branch,
where $\lambda=0$, we have $r=Q_0(S\le T)\ge0.34$ on the same event.

\subsection{Two lattice tools}\label{sec:fdtools}
We record two subroutines that are used repeatedly below. The first (Section~\ref{sec:fdsign}) is a version of the
Frank--Tardos sign compression~\cite{FT}: it replaces a vector of long
integers by a vector with short entries that has the same sign on every
short integer test vector, in any dimension. It is used in
Sections~\ref{cg:section}, \ref{bc:section} and~\ref{lcg:section} and in
Appendix~\ref{app:structured}. The second (Section~\ref{sec:fdsum}) sums a polynomial
over the lattice points of a rational polytope in fixed dimension,
following De Loera, Hemmecke, K\"oppe and Weismantel~\cite{DHKW}. It is
used in Section~\ref{mc:region} and in Appendix~\ref{app:structured}.

\subsubsection{Short coefficients preserving signs}\label{sec:fdsign}
We give a version of the Frank--Tardos sign compression~\cite{FT} that
records the dependence on the dimension and the cost in original words.
Let $z\in\mathbb Z^h$ have entries given by a constant number of
original words each, and let $D\ge10h$ be an integer. We need the signs
of $z\cdot t$ for integer $\|t\|_1\le D$. The idea is to build
independent short integer rows in succession, each of which decides the
sign of $z\cdot t$ for every $t$ on which the earlier rows vanish, and
finally to combine the rows lexicographically.

If $z=0$, return $z'=0$ with an empty row list and radix $M=3$.
Otherwise, at stage $j$, let $\Psi_{<j}$ have the earlier rows as its rows, and
split $z=z_\parallel+z_\perp$ with
$z_\parallel=\Psi_{<j}^\top(\Psi_{<j}\Psi_{<j}^\top)^{-1}\Psi_{<j}z$,
the orthogonal projection onto their span. If $z_\perp=0$, stop.
Otherwise put $v=z_\perp/\|z_\perp\|_\infty$; clearing the positive
determinant of $\Psi_{<j}\Psi_{<j}^\top$ and integer division compute a
short dyadic approximation $v'$ to $v$ within $\tau/4$ in each
coordinate, where $\tau=2^{-\lceil(h+1)(h/4+4+\log_2D)\rceil}$, so that
$2^{h/4}\tau^{1/(h+1)}\le1/(16D)$. The rational lattice in dimension
$h+1$ generated by $(\tau,v')$ and $(0,e_i)$ has determinant
$\tau$. The first reduced-vector bound of Lenstra, Lenstra and
Lov\'asz~\cite{LLL} gives a nonzero vector $(q\tau,qv'-\psi)$ of
Euclidean norm at most $1/(16D)$. Here $q\ne0$, since otherwise the
tail would be a nonzero integer vector, and we choose $q>0$. Since
$q\le1/(16D\tau)$, we get $\|qv-\psi\|_\infty<1/(8D)$ and
$\|\psi\|_\infty\le q+1$. With $\psi_j=\psi$, $q_j=q$ and
$\varrho_j=\|z_\perp\|_\infty/q_j$ this reads
\begin{equation}\label{fd:proj}
 \|z_\perp-\varrho_j\psi_j\|_\infty<\varrho_j/(8D).
\end{equation}
The row $\psi_j$ is independent of the earlier rows: $qv$ is orthogonal
to them and has norm at least one, while
$\|\psi-qv\|_2\le1/(16D)+\sqrt h\,q\tau/4<1$. So there are $s\le h$
rows $\psi_1,\ldots,\psi_s$.

For integer $t$ with $\|t\|_1\le D$ put $k_j=\psi_j\cdot t$. Then
$z_\parallel\cdot t=c_j(k_{<j})$ is a linear function of the earlier
digits, given by the projection formula, $c_{s+1}(k_{\le s})=z\cdot t$,
and \eqref{fd:proj} gives
\begin{equation}\label{fd:digit}
 \bigl|z\cdot t-c_j(k_{<j})-\varrho_jk_j\bigr|<\varrho_j/8
 \qquad(1\le j\le s).
\end{equation}
So $z\cdot t$ has the sign of the first nonzero digit, and vanishes if
all digits do. Take an odd integer $M>8D\max_j\|\psi_j\|_\infty$ and
$z'=\sum_{j=1}^s M^{s-j}\psi_j$. As $|k_j|\le D\|\psi_j\|_\infty<M/8$,
the $k_j$ are the balanced base-$M$ digits of $z'\cdot t$, read off
without carries, and $z'\cdot t$ has the sign of $z\cdot t$.

The rows have $O(h^2+h\log D)$ bits per coordinate. By the determinant
bound, the Gram minors, the cleared projection coefficients, $M$ and
the entries of $z'$ have $\operatorname{poly}(h+\log D)$ bits, and
$O_h(\log D)$ bits for fixed $h$. Every operand involving original
words (projections, normalization, the numerators and denominators of
$c_j$ and $\varrho_j$) is a linear combination of input words with such
short integer coefficients, and the LLL calls only see short rational
lattices. Hence the construction costs $\operatorname{poly}(h+\log D)$
original-word operations.

\subsubsection{Polynomial sums over lattice points}\label{sec:fdsum}
The fixed-dimensional weighted lattice-point theorem of De Loera,
Hemmecke, K\"oppe and Weismantel~\cite[Lemma~3.1 and evaluation at
one]{DHKW} computes the sum of a polynomial over the integer points
of a bounded rational polytope, in time polynomial in the input size
and in the degree of the polynomial, for fixed dimension. We recall
how this works.

Let $\mathcal P\subseteq\mathbb R^d$ be a rational polytope with $d$ fixed, and
let $f$ be a polynomial of total degree $D$, given as a dense list of
monomials with rational coefficients. The lemma produces a short
rational generating function for
$\sum_{\alpha\in \mathcal P\cap\mathbb Z^d}f(\alpha)z^\alpha$, of size
polynomial in the encoding length of $\mathcal P$, in $D$, and in the
coefficient bit sizes. It uses Barvinok's algorithm followed by $D$
applications of the operators $z_j\partial/\partial z_j$.

To evaluate the sum we take the constant term of the Laurent expansion
of this function at $z=(1,\ldots,1)$, after a generic monomial
substitution $z_j=\zeta^{\mu_j}$. Each denominator starts with at most
$d$ binomial factors, and each differentiation increases the number of
factors in a resulting term by at most one, counting multiplicity.
After substitution, each factor vanishes to first order at $\zeta=1$.
Put $t=\zeta-1$ and factor out the vanishing power $t^k$ of each
denominator, with $k\le d+D$. Write the resulting term as
$t^{-k}N(t)/R(t)$, where $R(0)\ne0$. Its constant Laurent coefficient
is the coefficient of $t^k$ in $N(t)/R(t)$. Exact series division uses
the first $k+1$ Taylor coefficients of $N$ and of the factored
denominator $R$. The binomial expansion of $(1+t)^a$ gives the
coefficients of each factored binomial $(1-(1+t)^a)/t$. This again
costs time polynomial in $D$ and in the input size, for fixed $d$.
Rational signed coefficients are allowed, by clearing denominators and
using linearity.

\subsection{Adding a sparse set of unrestricted weights}\label{sec:extra}
This subsection handles the following situation. Most items are
summarized by a positive base measure $F$ on weights, while a sparse
set $E$ of exceptional items keeps its original weights. We want the
number of ways to add at most $K$ exceptional items to the base while
staying below the capacity. Lemma~\ref{lem:basegeneral} estimates this
quantity, $B_K$, to relative error $\varepsilon$ with an explicit
dependence on $K$, accessing $E$ through rank and selection queries;
for polylogarithmic $K$ and polynomial $L$ its cost is $\tO(L+n+\varepsilon^{-1})$.

The main idea is a signed generating function. Its two parity parts
$B_+,B_-$ are nonnegative, their difference is $B_K$, and their sum
is at most $2B_K$, so it suffices to estimate each part to relative
accuracy. Each part is the value of a positive computation tree. We
compress the tables in this tree and correct the bias of the
compression by an unbiased sampled telescoping sum. The lemma is
used in Lemma~\ref{ex:coarse}, Sections~\ref{pw:section}
and~\ref{bc:section}, and Subsection~\ref{fin:section}; the
inequalities~\eqref{eq:cancel} and~\eqref{eq:jump} are also reused
there.

Let $E\subseteq[n]$ be a set of exceptional items, kept with their
original weights, and let the remaining items be summarized by a
positive base measure $F$ on physical weights. We give the
corresponding evaluation explicitly. Here $F(\le x)$ denotes the
cumulative mass of $F$. We sort the base table
once and prepare cumulative mass queries and sampling conditional
on a prefix. Their finite implementation is described in Step~8 of
the proof of Lemma~\ref{lem:basegeneral}.

\begin{lemma}\label{lem:basegeneral}
There is an absolute constant $d_1$ with the following property.
Let $F$ be a positive base table with $L$ entries and short
significands and exponents of polynomial absolute size in
$n+\varepsilon^{-1}+L$, and let $K\ge0$ be known. Suppose the
base table is word-addressable and each mass is in normalized binary
form, with its exponent and the leading working-precision bits of its
significand accessible in $O(1)$ words. Suppose the exceptional list
$E$, with $|E|\le n$, is available through an oracle
that answers an exact rank query at a weight, a selection by rank,
and a uniform draw from a rank interval, each in time at most $q_E$.
Put $J=\lceil\log(n+\varepsilon^{-1}+L+2)\rceil$. Then a relative
$\varepsilon$ approximation with fixed high probability to
\[
 B_K=\sum_{\substack{A\subseteq E\\|A|\le K}} F(\le T-w(A))
\]
costs
\[
   O(L)+(1+\varepsilon^{-1})\,(K+1)^{d_1}J^{d_1}\,(q_E+1)
\]
operations after a single sorting of the base table, and the estimate
is exactly zero when $B_K=0$. In particular, if $E$ is given as a
list, $L\le(n+\varepsilon^{-1})^{O(1)}$ and $K=\log^{O(1)}(n+\varepsilon^{-1}+2)$, the cost is
$\tO(L+n+\varepsilon^{-1})$.\label{lem:base}
\end{lemma}
\begin{proof}
Replace $K$ by $\min(K,|E|)$, since a subset selects at most $|E|$
exceptional sites. For $K=0$ a single cumulative query suffices.

\paragraph{Step 1: a signed generating function.}
Reserve the
$\min(|E|,4(K+1)^2)$ lightest indices of $E$ as a set $I$, and put
$E'=E\setminus I$ (the tail). In formal variables $y,z,s$ with
$s^2=1$ and $y^{K+1}=0$, form
\begin{equation}\label{eq:pol}
 \prod_{i\in I}(1+yz^{w_i})
 \prod_{\ell=1}^{K}
 \sum_{j=0}^{\lfloor K/\ell\rfloor}
 \frac{y^{\ell j}s^{(\ell-1)j}}{\ell^j j!}
       \left(\sum_{i\in E'}z^{\ell w_i}\right)^j .
\end{equation}
Evaluate each coefficient of $z^u$ against $F(\le T-u)$ and sum
over all degrees in $y$ of at most $K$. Separating the terms by the
parity of the exponent of $s$ gives two nonnegative quantities $B_+,B_-$.
Put $\sigma_\ell=\sum_{i\in E'}z^{\ell w_i}$. Since $y^{K+1}=0$, the
$\ell$th tail factor in \eqref{eq:pol} equals
$\exp(s^{\ell-1}y^\ell\sigma_\ell/\ell)$, so the tail product equals
$\prod_{i\in E'}\exp\bigl(\sum_{\ell\ge1}s^{\ell-1}(yz^{w_i})^\ell/\ell\bigr)$.
By the logarithm series this is $\prod_{i\in E'}(1+yz^{w_i})$ at
$s=-1$, and $\prod_{i\in E'}(1-yz^{w_i})^{-1}$ at $s=1$.
Thus $B_+-B_-=B_K$, while
\[
 B_++B_-=\sum_{(R,\mu)}F(\le T-w(R)-w(\mu)),
\]
the sum over pairs of a subset $R\subseteq I$ and a multiset $\mu$ on
$E'$ with $|R|+|\mu|\le K$.

\paragraph{Step 2: the sum of the two parts is at most $2B_K$.}
If $E'=\emptyset$ there are no multisets with repetitions and
$B_++B_-=B_K$. Otherwise $|I|=4(K+1)^2$. Let $V_m$ be the part of the
displayed sum over pairs in which $\mu$ has exactly $m$ copies beyond
the first of its labels; $V_0=B_K$. We claim
\begin{equation}\label{rw:halve}
 V_m\le V_{m-1}/2\qquad(m\ge1).
\end{equation}
Relate a pair $(R,\mu)$ with excess $m$ to every pair obtained by
removing one copy of the smallest repeated label of $\mu$ and adding
to $R$ one index of $I\setminus R$. The size $k=|R|+|\mu|\le K$ is
preserved, the excess drops to $m-1$, and the weight does not
increase, because $I$ consists of the lightest items of $E$; hence the
$F$-value does not decrease. Each pair of excess $m$ has at least
$|I|-k\ge|I|-K$ partners, and each pair of excess $m-1$ is a partner
of at most $k^2\le K^2$ pairs (choose the added index in $R$ and the
label of $\mu$ to restore). Counting related pairs, weighted by the
$F$-value of the excess-$m$ member, gives
$(|I|-K)V_m\le K^2V_{m-1}$, and $K^2\le(4(K+1)^2-K)/2$ proves
\eqref{rw:halve}. Since the relation preserves $k$ and $F$ is an
arbitrary positive measure, the same bound holds separately for each
fixed size $k\le K$ and with $F$ replaced by a point mass, that is, at
every residual capacity and every degree. Summing the geometric series,
\begin{equation}\label{eq:cancel}
 B_++B_-\le 2B_K .
\end{equation}
Hence relative accuracy for each of $B_+$ and $B_-$ yields relative
accuracy for $B_K$.

\paragraph{Step 3: positive computation trees.}
We now show that each parity is the value of a positive computation
tree with $\kappa=O((K+1)^2)$ vertices and $O((K+1)^2)$ states
per vertex. For a factor $\ell$ in \eqref{eq:pol}, create
$m_\ell=\lfloor K/\ell\rfloor$ slots, each either empty or selecting one
tail label at weight $\ell w_i$. Track the number $j$ of nonempty
slots. At the common root of these slots, multiply the mass by
$\ell^{m_\ell-j}(m_\ell-j)!$, set the degree to $\ell j$ and the parity to
$(\ell-1)j\bmod 2$. Summed over the $\binom{m_\ell}j$ slot patterns, this
gives exactly the factor of \eqref{eq:pol} times $D_\ell=\ell^{m_\ell}m_\ell!$.
Join these trees with the binary prefix items of $I$,
adding degrees and parities and discarding degrees above $K$.
The leaves are the at most $4(K+1)^2$ prefix items and the slots;
since $\sum_{\ell\le K}m_\ell\le K(1+\ln K)$, there are
$O(K\log(K+2))$ slots. Arranging the slots of each factor, and then
these subtrees and the prefix items, in balanced binary trees gives
$\kappa=O((K+1)^2)$ vertices and depth $O(\log(K+2))$. A state records
a degree of at most $K$, a parity, and, inside a slot subtree, the
number of nonempty slots. All transitions depend only on the states.
At the end we divide by $D=\prod_\ell D_\ell$. Every integer mass in
the exception tree is at most $D$ times the number of choices of
leaf entries: $2^{|I|}$ for the prefix items, $2^{O(K\log(K+2))}$ for
the set of nonempty slots, and at most $n^K$ for their labels, since
the degree bound allows at most $K$ nonempty slots. Since $\ell
m_\ell\le K$, we have $D\le K^{K(1+\ln K)}$. Hence every mass has
$O((K+1)^2J)$ bits, and so does the logarithm of the ratio of the
largest to the smallest positive cumulative mass in a state.

\paragraph{Step 4: compressing the vertex tables.}
Process the vertices in children-first order, and compress each vertex
table separately in every state, as follows. Start a group at the
first ungrouped entry, of cumulative integer mass $\nu$, and include
consecutive entries through cumulative mass $(1+\alpha)\nu$. Move the
entire exact group mass to the first coordinate of the group. The
resulting CDF $G$ and the preceding CDF $\Phi_0$ obey
\begin{equation}\label{eq:jump}
 \Phi_0\le G\le(1+\alpha)\Phi_0,\qquad 0\le G-\Phi_0\le\alpha G .
\end{equation}
Indeed, at a threshold meeting a partial group, $\Phi_0$ already includes
the first entry of the group, and at a group boundary equality holds.
The initial cumulative mass of the next group is larger by a factor
greater than $1+\alpha$. We take $\alpha$ dyadic of order
$\eta/(10\kappa)$, where $0<\eta\le1/2$. By the mass range of
Step~3, a state then has at most
$1+O((K+1)^2J)/\log(1+\alpha)=O((K+1)^4J\eta^{-1})$ groups. Joining
two child lists forms at most the product of their lengths for each
of the $O((K+1)^4)$ state pairs; with sorting, a join costs
$(K+1)^{O(1)}J^{O(1)}\eta^{-2}$. A slot leaf for the factor $\ell$ is
the list $E\setminus I$ at weights $\ell w_i$, each of unit mass,
and it is never written out. Its group boundaries are integer ranks
in the order of $w_i$, which is also the order of $\ell w_i$;
since $I$ is a prefix of $E$ in this order, they are ranks in $E$ minus $|I|$,
determined recursively by the spacing rule. There are $O(J/\alpha)$
of them, and each representative costs one selection query. The
prefix leaves are singletons obtained by selection.

\paragraph{Step 5: the bias of compression as a telescoping sum.}
Fix one parity. Let $\Psi_0,\ldots,\Psi_\kappa$ be the values of the scalar
(including the factor $D$ and the external $F$) after successive
compressions, so that $\Psi_0$ is the exact value. Positive
convolution, addition over states and integration against $F$ all
preserve CDF inequalities. Thus $\Psi_0\le \Psi_j\le(1+\alpha)^j\Psi_0$.
We read $\Psi_\kappa$, up to the roundoff of Step~8, by weighting each root entry of coordinate $u$
by $F(\le T-u)$. It remains to estimate each change $\Psi_j-\Psi_{j-1}$.
Let $v$ be the $j$th vertex, let $\Phi_{v,a}$ and $G_{v,a}$ be its table
in state $a$ immediately before and after its compression, and call
an \emph{outside choice} $\omega$ a choice of one entry in every input
of the partially compressed tree other than the table at $v$ (these
are disjoint current subtrees or raw leaves), of a base atom, and of
transitions, such that the state at $v$ is some $a(\omega)$. Write $\varrho_\omega$ for the product of its
masses and multiplicities and $Y_\omega$ for its total weight,
including the base atom. Since the ancestors of $v$ are not yet
compressed and the scalar is linear in the table at $v$,
\begin{equation}\label{rw:change}
 \Psi_j-\Psi_{j-1}=\sum_\omega \varrho_\omega\,
 (G_{v,a(\omega)}-\Phi_{v,a(\omega)})(T-Y_\omega).
\end{equation}

\paragraph{Step 6: importance sampling of one change.}
We sample pairs $(\omega,g)$, where $g$ is an entry of
$G_{v,a(\omega)}$, so that every pair with $w(g)\le T-Y_\omega$ has
probability $\varrho_\omega\cdot(\text{mass of }g)/U_j$ for a computable
$U_j\ge \Psi_j$; the total of these weights is $\Psi_j$. Build a fresh coarse
tree on $G_v$ and the other inputs, with the same state transitions,
now using spacing $1/(16\kappa)$ in \eqref{eq:jump}. Each fine input list is first coarsened individually by the same rule,
recording its incoming entries. By the count of Step~4 with
$\alpha=1/(16\kappa)$, the coarse lists have $O((K+1)^4J)$ entries per
state, so a coarse tree costs $(K+1)^{O(1)}J^{O(1)}\eta^{-1}$
operations and $(K+1)^{O(1)}J$ queries. At the root, sample an
entry of representative weight $u$ with probability proportional to its mass
times $F(\le T-u)$; call the sum of these weights $U_j$. If it is zero,
the change is zero. Next draw a base atom conditional on being at most
$T-u$. Independently given the root entry, backtrack through the
coarse groups in proportion to the stored incoming pair masses,
including transition multiplicities. For a candidate pair $(\omega,g)$
the proposal probability multiplied by $U_j$ equals exactly
$\varrho_\omega$ times the mass of $g$, because root prefix factors and
internal group totals cancel. Since coarse compression only moves
representatives down, every pair with $w(g)\le T-Y_\omega$ is
proposed with this probability. Moreover
$U_j\le (1+1/(16\kappa))^\kappa \Psi_j\le e^{1/16}\Psi_j$ by the same CDF
inequality, applied only to the exception tree.

\paragraph{Step 7: the score, its variance, and the cost.}
Recover the actual current input coordinates along the sampled path.
If $w(g)+Y_\omega>T$, return zero. Otherwise, with $a=a(\omega)$ and the
convention that a zero denominator gives zero, return
\[
 U_j\,\frac{(G_{v,a}-\Phi_{v,a})(T-Y_\omega)}{G_{v,a}(T-Y_\omega)} .
\]
For fixed $\omega$, the proposal probabilities of the allowed $g$ sum
to $\varrho_\omega G_{v,a}(T-Y_\omega)/U_j$, so the mean of the score is the
right side of \eqref{rw:change}, that is, $\Psi_j-\Psi_{j-1}$.
Equation~\eqref{eq:jump} bounds every score by
$\alpha U_j=O(\alpha \Psi_0)$.

We use $M=\lceil C_2(1+\varepsilon^{-2}\eta^2)\rceil$
independent draws per vertex, independent across vertices, and
subtract the sum of the averages from $\Psi_\kappa$. The result has mean
$\Psi_0$ and variance $O(\kappa\alpha^2\Psi_0^2/M)$. For a zero scalar all
candidate masses vanish, so the output is zero. Otherwise Chebyshev's
inequality, with $C_2$ large, gives relative error at most
$\varepsilon/20$ with probability at least $0.99$ for this parity. We
choose $\eta$ within a constant factor of $\varepsilon^{1/2}$, so that
$M$ and $\eta^{-2}$ are $O(1+\varepsilon^{-1})$. The fine tree costs
$\kappa$ joins and $(K+1)^{O(1)}J\eta^{-1}$ queries at its slot leaves,
and there are $\kappa$ coarse trees. A path visits $O(\kappa)$
vertices and makes one uniform draw from a rank interval at each slot
leaf. If $v$ is a slot leaf, the score needs one rank query at weight
$\lfloor(T-Y_\omega)/\ell\rfloor$ and a binary search among the group
boundaries. Base atoms and $U_j$ use prefix queries of cost
$O(\log(L+2))$. With $\kappa M$ paths and the arithmetic of Step~8,
the total is the stated bound for an absolute constant $d_1$; the base
preprocessing costs $O(L)$ and is done once. Finally we subtract the
two parity estimates and apply \eqref{eq:cancel}.

\paragraph{Step 8: finite arithmetic.}
We store sums of positive base masses on a
balanced interval tree with short relative roundoff $\gamma$.
The $O(J)$ working bits fit in $O(1)$ words, since every table index
fits in a word. In adding two normalized positive masses, an exponent
gap larger than the working precision plus two allows us to discard
the smaller term within the roundoff tolerance. Otherwise one
shift of $O(J)$ bits aligns the significands. Thus the tree needs
$O(L)$ word operations. By positivity, a node sum has relative loss at most
$O(\gamma\log(L+2))$. A prefix query sums $O(\log(L+2))$ disjoint
node totals. A conditional prefix draw first picks such a node by its
mass and then descends by mass. Uniform integer draws for the
exception list are served by the oracle. Exact masses have
$O((K+1)^2J)$ bits, so schoolbook arithmetic on them costs
$(K+1)^{O(1)}J^{O(1)}$ per operation, and short relative scalars use
$O(J)$ bits. A backtracking step chooses among $h$ incoming pairs with
positive total integer mass $m$ and cumulative masses $m_j$, where
$m_0=0$ and $m_h=m$. Rounding each $m_j/m$ down to a multiple of
$2^{-B}$ by integer division keeps the endpoints and changes the
transition law by at most $h2^{-B}$ in total variation; a uniform
$B$-bit integer $U$ then locates the draw by binary search of $U/2^B$
among the rounded boundaries. The rounded boundaries depend only on
the tree, the state and the incoming list, not on the draw, so they
are computed once per coarse tree, with one exact division per list
entry, and every draw then costs a single binary search. This
preprocessing is within the cost bound of Step~7. If $h_{\max}$ bounds the list lengths
and $N$ the number of such draws, $B=\lceil\log_2(2+Nh_{\max}/\zeta)\rceil$
couples all these transitions to the ideal experiment with failure
probability at most $\zeta$; the union bound applies conditionally at
every step, even when the list is selected by previous draws. By the
counts above, $B=O(J+\log(K+2)+\log(1/\zeta))$. Taking $\zeta$ and the
other local probability errors small enough, with this many bits,
couples all discrete choices to
the ideal experiment within a fixed small failure budget. The $U_j$
and the final root scalar are positive sums; we round them to
relative error $o(\varepsilon/\kappa)$, and individual score ratios to
absolute error $o(\varepsilon/\kappa)$. Since $U_j=O(\Psi_0)$, the resulting
absolute bias is $o(\varepsilon \Psi_0)$ per parity after summation,
with a sufficiently small constant. The final subtraction uses a
common precision on this scale, and \eqref{eq:cancel} controls the
loss. Finally, rounding the positive atoms of a base table whose binary
exponents are polynomially bounded to short relative precision is
sufficient for every count.

For the last claim, sort $E$ in $O(n\log(n+2))$ time. The sorted array
answers ranks by binary search and selections by indexing, and a
uniform draw from a rank interval uses capped rejection, whose
failure probability is charged to the coupling budget of Step~8, so
$q_E=O(J)$. For polylogarithmic $K$ the stated bound is then
$\tO(L+n+\varepsilon^{-1})$.
\end{proof}

Lemma~\ref{lem:basegeneral} corrects the bias of compression by
sampling. In Sections~\ref{pw:section} and~\ref{bc:section} the
exceptional trees are instead compressed once, with spacing
$1/(16\kappa)$, and used without correction. The next lemma records
what these coarse trees provide.

\begin{lemma}[Coarse exceptional trees]\label{ex:coarse}
There is an absolute constant $d_2$ with the following property.
Let $E$ be a list of items with nonnegative integer weights $w_i$,
sorted by weight and accessed by selection queries, let $K\ge0$ be
an integer, and put $J=\lceil\log(|E|+2)\rceil$ and
$M=(K+2)^{d_2}J^{d_2}$. For $0\le j\le K$ let $F_j$ be the counting
measure of the weights $w(A)$ of the $j$-subsets $A\subseteq E$.
A deterministic construction with at most $M$ selection queries and
$M$ further operations (comparisons and additions of weights, and
arithmetic on integers of at most $M$ bits) outputs at most $M$
\emph{root entries} $\omega=(j_\omega,s_\omega,b_\omega,\varphi_\omega)$,
with degree $0\le j_\omega\le K$, sign $s_\omega\in\{1,-1\}$, integer
\emph{representative} $b_\omega\ge0$ and rational mass
$\varphi_\omega>0$, and for each entry a probability law $\pi_\omega$
of a nonnegative integer $V$, such that:
\begin{enumerate}[label=(\alph*)]
\item for each $j\le K$,
  $\sum_{\omega:\,j_\omega=j}s_\omega\varphi_\omega\,\pi_\omega(V\in\cdot\,)=F_j$;
\item $V\ge b_\omega$ holds $\pi_\omega$-almost surely;
\item for each $j\le K$ and every nonincreasing $f:\mathbb R\to[0,\infty)$,
  \[\sum_{\omega:\,j_\omega=j}\varphi_\omega f(b_\omega)
     \le 2e^{1/16}\int f\,dF_j ,\qquad
     \sum_{\omega:\,j_\omega=j}\varphi_\omega\le2\binom{|E|}{j};\]
\item for $\zeta\in\mathbb C$ with $\Re\zeta\ge0$ and $f\in\mathbb R$,
  the values
  $\Psi_\omega=\E_{\pi_\omega}\bigl[e^{-\zeta(V-b_\omega)}\mathrm e(fV)\bigr]$
  for all root entries are obtained in $M$ operations from the averages
  \[a_{I,\ell}=|I|^{-1}\textstyle\sum_{i\in I}e^{-\zeta\ell(w_i-b_I)}\mathrm e(f\ell w_i),\]
  where $(I,\ell)$ runs over at most $M$ pairs, fixed by the
  construction, of a rank interval $I$ of $E$ with minimum weight
  $b_I$ and an integer $1\le\ell\le K$. All intermediate values have
  modulus at most one, and absolute errors at most $\eta$ in the
  $a_{I,\ell}$ change each $\Psi_\omega$ by at most $M\eta$.
\end{enumerate}
\end{lemma}
\begin{proof}
We use the computation tree of Steps~1, 3 and~4 of the proof of
Lemma~\ref{lem:basegeneral}, applied to $E$ with the degree bound $K$
and without a base measure. A root state records a degree $j\le K$
and the parity of the exponent of $s$ in \eqref{eq:pol}. Let
$\kappa=O((K+1)^2)$ be the number of vertices.

\emph{Positive lists.} At the root states of degree $j$ let $F_j^+$
and $F_j^-$ be the exact positive measures of weights of even and odd
parity, including the division by the slot multiplier $D$ of Step~3.
By Step~1, separated by degree, $F_j^+-F_j^-=F_j$. Moreover, at each
threshold $x$,
\[
          (F_j^++F_j^-)(\le x)\le 2 F_j(\le x).
\]
Indeed, the exchange in Step~2, which replaces a repeated tail label
by an unused light prefix label, preserves the degree $j$, so
\eqref{eq:cancel} holds for each degree with $F$ a point mass. This
estimate includes $x=+\infty$, where $F_j(\le x)=\binom{|E|}j$.

\emph{Compression.} Processing the vertices children first, in each
vertex and state we compress the list by the cumulative rule of
\eqref{eq:jump}, with spacing $\alpha=1/(16\kappa)$, retaining the
incoming entries of each group. A group is represented by its lowest
weight. All masses are exact integers before the division by $D$.
The root entries are the groups at the root states: the state gives
$j_\omega$ and, by its parity, $s_\omega=\pm1$; $b_\omega$ is the
representative and $\varphi_\omega$ the group mass divided by $D$.
Compression only moves representatives, so each group is a union of
exact path contributions. Let $\pi_\omega$ be the law of the true
weight $V$ of a path of the group $\omega$, drawn with probability
proportional to its mass. Then (a) is the identity $F_j^+-F_j^-=F_j$.
The representative of an internal group is the least sum of child
representatives among its entries, so by induction every path
terminating at an entry has true weight at least its representative.
This is (b).

\emph{Part (c).} By positive convolution and addition over states, the
coarse CDF per parity and degree lies between the corresponding exact
CDF and $(1+\alpha)^\kappa\le e^{1/16}$ times it, by \eqref{eq:jump}.
Hence
$\sum_{\omega:j_\omega=j}\varphi_\omega\one_{\{b_\omega\le x\}}
\le2e^{1/16}F_j(\le x)$ for every $x$, and the same holds with strict
inequalities by taking limits. A nonincreasing $f\ge0$ is the integral
over $t>0$ of the indicators of the half-lines $\{f>t\}$, which gives
the first bound. The second follows at infinite capacity, since
compression preserves total mass.

\emph{Sizes.} Every mass is an integer of
$O((K+1)^2J)$ bits, by the count of Step~3 with $|E|^K$ choices of
labels. By the count of Step~4 with $\alpha=1/(16\kappa)$, a state has
$O((K+1)^4J)$ groups. There are $\kappa$ vertices with $O((K+1)^2)$
states each, and a vertex stores, for each pair of child states, at
most the product of the two list lengths. A slot leaf has $O(J/\alpha)$
group boundaries, determined by the spacing rule, and each
representative costs one selection query; the prefix leaves are
singletons obtained by selection. Hence the numbers of root entries,
groups and stored pairs, and the cost of the construction, including
the sorting in joins and the exact arithmetic on masses, are at most
$M$ for a suitable absolute $d_2$.

\emph{Part (d).} We compute the values by recursion over the tree. For
a group with representative $b_v$, we compute the conditional value of
$e^{-\zeta(V_v-b_v)}\mathrm e(fV_v)$, where $V_v$ is the true weight of
a path in the group. At an internal group we take the weighted
average, over its stored incoming pairs, of the product of their two
values times $e^{-\zeta(b_{\mathrm{left}}+b_{\mathrm{right}}-b_v)}$.
All differences in this factor are nonnegative, by the choice of
representative, so the factor has modulus at most one. Transition
multiplicities are included in the weights. A slot leaf for the factor
$\ell$ has unit masses at the weights $\ell w_i$, in the order of
$w_i$; its groups are rank intervals $I$ of $E$, the same for every
$\ell$, with representatives $\ell b_I$, and the value of such a group
is $a_{I,\ell}$. Prefix leaves are intervals of length one, and an
empty slot has value one. Induction proves the formula. All
operations average products of numbers of modulus at most one.
Truncating computed values to the closed unit disc, the error at a
group is at most the sum of the two child errors, so errors $\eta$ in
the $a_{I,\ell}$ give errors at most $\kappa\eta$ at the root.
\end{proof}

\section{Hidden structure in arbitrary instances}\label{rw:hidden}
This section shows that every instance has a hidden structure of
bounded rank (Subsection~\ref{iv:section}), and then constructs such
a structure algorithmically at logarithmic rank
(Subsection~\ref{cg:section}). It uses only the balanced tilt and the
lattice tools of Section~\ref{rw:prelim}. The structure reduces the
problem to a core with short integer coordinates and a few exceptional
sites; Sections~\ref{rw:fourier}--\ref{pw:section} develop the
counting tools for such instances, and Sections~\ref{bc:section}
and~\ref{rw:linear} combine the two.

\subsection{A structural reduction for arbitrary capacities}\label{iv:section}
This subsection shows that, for an arbitrary capacity and arbitrary
original-word weights, the relative counting problem has a hidden
low-complexity structure. The tool is the inverse Littlewood--Offord
theorem of Nguyen and Vu: if a weighted sum of independent bits has
an atom of inverse-polynomial probability, then most of the weights lie
in a generalized arithmetic progression of bounded rank. We first prove
a weighted form of this statement (Lemma~\ref{iv:weighted}), in which
the exceptional weights are measured by their total inclusion
probability. We then apply it under the balanced tilt of
Section~\ref{s48:tilting} to a suitable rounding of the instance
(Proposition~\ref{iv:inner}). The result is a single discretization at
which two things hold at once: the threshold is stable, so that
rounding loses few feasible sets, and most of the tilted occupancy
belongs to a bounded-rank progression. Proposition~\ref{iv:inner} is
only an existence statement; Section~\ref{cg:section} turns it into a
construction.

\begin{lemma}\label{iv:weighted}
Fix $a>0$ and $0<\eta<1$. Let $R$ be a positive integer,
$m\le R$, $v_1,\ldots,v_m\in\mathbb Z$, and
$X_i$ independent with $\Pr(X_i=1)=p_i\le1/2$,
$\Pr(X_i=0)=1-p_i$. If some integer value of $\sum v_iX_i$
has probability at least $R^{-a}$, there exist $s=O_{a,\eta}(1)$
input values $b_1,\ldots,b_s$, an integer $1\le d\le R^{C}$,
and an exceptional index set $E$ such that
\[
  \sum_{i\in E}p_i=O_{a,\eta}(R^\eta),\qquad
  d v_i=\sum_{j=1}^s b_j c_{ij},\quad |c_{ij}|\le R^C,\quad
  c_{ij}\in\mathbb Z\quad(i\notin E)
\]
for a constant $C=C(a,\eta)$. An empty list of generators represents
zero.
\end{lemma}
\begin{proof}
The idea is to reduce to fair bits on a random ``active'' subset of
the sites, apply the unweighted inverse theorem there, and then
convert the conclusion ``few active entries lie outside a progression''
into ``small total probability lies outside a container'' by a
Chernoff and union bound over a fixed polynomial-size family of
containers.

Write $X_i=D_iY_i$ with mutually independent
$D_i\sim\Bern(2p_i)$ and fair $Y_i$; this is possible since
$p_i\le1/2$. Sites with $v_i=0$ do not affect the sum and are
represented by zero, so we discard them. Put $\alpha=R^{-a}$.
Fix an atom of probability at least $\alpha$. By averaging, the random
active set $\{i:D_i=1\}$ has probability at least $\alpha/2$
of giving that atom conditional probability at least $\alpha/2$.
Call such an active set successful.
Let $m'$ be the length of the active list and put
$k=\lceil R^\eta\rceil$. If $m'\le k$, the conclusion below holds
trivially for this active set, because every active entry may be
counted as lying outside the container. So assume $m'>k\ge R^\eta$.
Then $R<(m')^{1/\eta}$ and $m'\ge2$, so the conditional maximal atom
probability is at least $\alpha/2=R^{-a}/2\ge(m')^{-C_0}$ with
$C_0=a/\eta+1$. Moreover $(m')^\eta\le R^\eta\le k<m'$, since
$m'\le m\le R$. Thus the parameters lie in the range of the inverse
theorem, and no padding of the list is needed.

The affine transformation from fair bits to signs preserves the
maximal atom probability. We therefore apply the discrete inverse
theorem of Nguyen and Vu \cite[Theorem 2.5]{NVinverse} directly to the
active list of length $m'$, with atom exponent $C_0$ and exceptional
number $k$. It gives a symmetric proper generalized arithmetic
progression of bounded rank that contains all but $k$ entries of the
active list. Since the maximal atom has inverse-polynomial mass, the
size of this progression is bounded by a fixed power of $R$. In
particular, all integer coordinates in its defining box are bounded by
a fixed power of $R$.

The union bound below requires a family of possible containers
that has polynomial cardinality and is fixed before the $D_i$ are
sampled. We obtain it as follows. Consider one progression given by
the inverse theorem. Among its represented active entries, choose
coordinate rows that form a basis of their rational span; the
dimension $s$ of this span is at most the rank. Choose $s$ columns
that form a nonsingular square minor on those rows. Its absolute
determinant $d$ is a positive integer bounded by
$R^{O_{a,\eta}(1)}$. By Cramer's rule, every represented active row is
$1/d$ times an integer linear combination of the basis rows, with
coefficients of magnitude at most $R^{O_{a,\eta}(1)}$. This identity
holds on every column, since these rows lie in the chosen span.
(Inactive entries of the progression need not lie in that span; they
are not used.) Applying the original generator map now gives the
claimed representation in terms of the \emph{input values} on the
basis rows. In the zero-span case we use $d=1$. Thus, for a
sufficiently large fixed $C$, we can use the family
\[
 \left\{d^{-1}\sum_{j=1}^s b_j c_j:
                   c_j\in\mathbb Z,\ |c_j|\le R^C\right\},
\]
taken over bounded $s$, ordered choices of the input values, and
$1\le d\le R^C$.
This family has at most $R^{C'}$ members, for a fixed $C'$ that is
independent of the bit lengths. By the inverse theorem as applied above, each successful active set
has at most $k$ active entries outside one of these containers.

Fix a container $G$ with $\sum_{i:v_i\notin G}p_i>2k$. Then the
expected number of active outside entries exceeds $4k$, and a
lower-tail Chernoff bound shows that at most $k$ of them are active
with probability at most $e^{-c k}$. For sufficiently large $R$, the
union bound over the fixed family gives total probability less than
$\alpha/2$. Since successful active sets occur with probability at
least $\alpha/2$, some container must have outside tilted occupancy
at most $2k$. Bounded $R$ is covered by increasing the constant in
the occupancy bound.
\end{proof}

We now apply Lemma~\ref{iv:weighted} to a rounded version of the
instance. Recall from Section~\ref{s48:tilting} that $Q_\lambda$ is the
tilted product law, $S$ is the random subset sum under it, and
$C_\lambda$ is the price of the tilt.

\begin{proposition}\label{iv:inner}
Consider a nontrivial original-word instance and $\lambda\ge0$. For
$0<\rho<1$ let $h$ be the least positive integer with
$Q_\lambda(|2S-2T-1|\le h)\ge\rho$, and put $\Delta=h/(16(n+1))$.
\begin{enumerate}[leftmargin=*,label=(\alph*)]
\item For every $0<\Delta'\le\Delta$, the rounded weights and capacity
\[
 v_i=\left\lceil\frac{w_i}{\Delta'}\right\rceil,\qquad
 V=\left\lfloor\frac{T+1/2}{\Delta'}\right\rfloor
\]
define a feasible family contained in the original one, which misses
at most $C_\lambda\rho$ original feasible subsets.
\item At $\Delta'=\Delta$, some integer value of $\sum_iv_iX_i$ has
$Q_\lambda$-probability at least $\rho/(17(n+1))$.
\item Let $R=n+\lceil 1/\varepsilon\rceil+2$, $\varepsilon^{-1}\ge2$,
fix $0<\eta<1$, and take $\rho=\varepsilon/(100n)$ at a tilt of
Section~\ref{s48:tilting} with $C_\lambda/Z\le5n$. Then the family of (a) at $\Delta'=\Delta$
has count at least $(1-\varepsilon/20)Z$, and its weights $v_i$
satisfy the conclusion of Lemma~\ref{iv:weighted} with a constant $a$
independent of the instance. Moreover, retaining only configurations
selecting at most $O_\eta(R^\eta+\log R)$ exceptional sites still
leaves at least $(1-\varepsilon/10)Z$ feasible subsets.
\end{enumerate}
\end{proposition}
\begin{proof}
Informally, $h$ is the smallest window around the threshold that
carries non-negligible tilted mass. Rounding at a mesh below $h/n$
then loses only configurations in a smaller window, which is rare,
while the rounded sum is concentrated on few integer values, which
yields the atom needed by Lemma~\ref{iv:weighted}.

(a) If $\sum_{i\in A}v_i\le V$, then $S/\Delta'\le(T+1/2)/\Delta'$,
so $S\le T$ by integrality. Conversely, let $A$ be feasible with
$|2S-2T-1|\ge h$, so that the slack $T+1/2-S$ is at least $h/2$.
Rounding up adds less than $n$ to $S/\Delta'$, and
$n\Delta'\le nh/(16(n+1))<h/2$. Hence $\sum_{i\in A}v_i\le
(T+1/2)/\Delta'$, and $\sum_{i\in A}v_i\le V$ by integrality. The
lost subsets therefore satisfy $|2S-2T-1|<h$, an event of
$Q_\lambda$-probability less than $\rho$ by the minimality of $h$
(empty if $h=1$, since $2S-2T-1$ is odd), and
their number is at most $C_\lambda\rho$ by the remark after \eqref{s48tilt:identity}.

(b) On the event $|2S-2T-1|\le h$ we have
$|S/\Delta-(T+1/2)/\Delta|\le h/(2\Delta)=8(n+1)$, and
$S/\Delta\le\sum_iv_iX_i\le S/\Delta+n$. So $\sum_iv_iX_i$ takes at
most $17(n+1)$ values on this event, and one of them has probability
at least $\rho/(17(n+1))$.

(c) Since $C_\lambda/Z\le5n$, the loss in (a) is at most $\varepsilon Z/20$.
The atom of (b) has probability at least
$\varepsilon/(1700n(n+1))\ge R^{-a}$ for an absolute $a$, and
$p_i\le1/2$ since $\lambda\ge0$, so Lemma~\ref{iv:weighted} applies.
For its exceptional set $E$, \eqref{s48tilt:trunc} shows that a degree
cap larger than $(e-1)\sum_{i\in E}p_i+\log(100n/\varepsilon)
=O_\eta(R^\eta+\log R)$ costs at most a further
$C_\lambda\rho\le\varepsilon Z/20$.
\end{proof}

Proposition~\ref{iv:inner} is an existence statement at constant
rank. The construction in Section~\ref{cg:section} instead uses
logarithmic rank, and in return obtains a polylogarithmic bound on the
exceptional occupancy. Section~\ref{bc:section} then evaluates the
contribution of those exceptional sites by estimating prefix counts of
the core at multiple residual capacities. Recovering exact sums
instead would not suffice: Remark~\ref{prop:barrier} gives instances
where most of the tilted mass is spread over $\exp(\Omega(n))$
distinct sums.

\subsection{Constructing a representation from the tilted weights}\label{cg:section}
This subsection makes the structure of Section~\ref{iv:section}
algorithmic. We now allow logarithmic rank. In return, the
representation and a suitable rounding scale can actually be found,
and the expected occupancy of the exceptional sites is only
polylogarithmic (Theorem~\ref{cg:structure}). The construction uses
the balanced tilt of Section~\ref{s48:tilting}. A pilot sample chooses
the representation and the scale, by greedily selecting pivots with a
short-vector routine (Lemma~\ref{cg:lattice}, built on the sign
compression of Section~\ref{sec:fdsign}). A coupling to a union of
independent pilot samples then proves the required anti-concentration
estimate for the original law. Theorem~\ref{cg:structure} provides the
input to the counting algorithm of Section~\ref{bc:section}, and its
pivot construction is reused in Section~\ref{lcg:section}.

The following lemma finds, in polynomial time in the short
parameters, a bounded integer combination of given generators that
approximates a target almost as well as the best one.

\begin{lemma}\label{cg:lattice}
Let $b_1,\ldots,b_j,w$ be integers occupying a constant number of
original words each, and let $A\ge1$ be an integer. One can compute
an integer vector $a$ with $\|a\|_\infty\le F A$ and
\[
 |w-b\cdot a|\le F\min_{\substack{z\in\mathbb Z^j\\
                                  \|z\|_\infty\le A}} |w-b\cdot z|,
 \qquad F=2^{\,3j+5}.
\]
The cost in original-word operations is polynomial in $j+\log(A+2)$.
\end{lemma}
\begin{proof}
The empty basis is immediate, so assume $j\ge1$. The plan is to replace
the long integers by short ones that preserve the signs of all
relevant integer tests, solve a closest-vector problem approximately
in the short coordinates, and transfer the guarantee back.

First sign-compress the $(j+1)$-vector $(w,b)$ by
Section~\ref{sec:fdsign}, in dimension $h=j+1$, for all integer tests
of one-norm at most $D=10(j+1)F^2(A+1)$. This gives a short vector
$(\bar w,\bar b)$. Since $D\ge10h$ and $\log D=O(j+\log(A+2))$, this
costs $\operatorname{poly}(j+\log(A+2))$ original-word operations, and
the compressed entries have that many bits.

Let $G=2^{j+1}\lceil\sqrt{j+1}\rceil$, so that $2G\le F$. In the
compressed coordinates, try positive scales $s$, starting at a
power of two below $1/(2G)$ and doubling up to
$2(1+|\bar w|+\sum A|\bar b_i|)$. For each scale, apply LLL and Babai's
nearest-plane algorithm~\cite{Babai} in the lattice
\[
       \{(a/A,\bar b\cdot a/s):a\in\mathbb Z^j\}
       \quad\text{with target }(0,\bar w/s).
\]
For a standard LLL-reduced basis, the rank-$j$ nearest-plane
approximation factor is at most $2^{j/2}$. Orthogonal projection of a
target onto the lattice span, if needed, preserves this factor for the
ambient distance. Rational denominators are cleared for the reduction.
Return the first output satisfying $\|a\|_\infty\le GA$ and
$|\bar w-\bar b\cdot a|\le Gs$.

We check the guarantee. If some $z$ of norm at most $A$
has error at most $s$, it witnesses Euclidean distance at most
$\sqrt{j+1}$, so the output at that scale passes. If an exact witness
exists, integrality forces zero output discrepancy already at the
first scale. Otherwise the compressed optimum is a positive integer,
and the first passing scale is at most twice that optimum. Therefore,
for each admissible $z$,
$|\bar w-\bar b\cdot a|\le F|\bar w-\bar b\cdot z|$.
All scales and lattice operations have polynomial short bit length.

It remains to transfer each such inequality to the original
integers. First transfer the signs of both discrepancies. Then
transfer the sign of their signed difference after multiplying the
second discrepancy by $F$. Every such test has one-norm at most $D$,
so the sign compression preserves it. This proves the bound for the
original words.
\end{proof}

We can now state and prove the main result of this subsection.

\begin{theorem}\label{cg:structure}
Let $R=n+\lceil 1/\varepsilon\rceil+2$ and $J=\lceil\log_2 R\rceil$.
In $n\log^{O(1)}R$ original-word operations (work polylogarithmic
in $R$ per site), one can construct integer core weights of the form
$b\cdot a_i$, with $O(J)$ generators and
$\|a_i\|_\infty\le R^{O(1)}$, a list of unrestricted exceptional
weights, a capacity, and a degree bound $K=O(J^3)$.
With probability at least $0.97$ the subsets satisfying the new
capacity and using at most $K$ exceptional sites form a subfamily
of the original feasible sets of count at least $(1-\varepsilon/100)Z$.
The generators, the exceptional weights, and the new capacity
occupy a constant number of original words each. The construction
applies at every capacity and feasible density.
\end{theorem}
\begin{proof}
The idea is as follows. A pilot sample, drawn at roughly the tilted
inclusion rates, is used to greedily pick up to $q$ pivots, each time
taking the pilot weight worst approximated by the current pivots. The
resulting error level $H$ serves as the rounding scale. If the pivots
are separated at scale $H$, then the tilted sum cannot concentrate on
intervals shorter than $H/F$ (inequality \eqref{cg:price}), so rounding
at that scale is cheap. Sites far from the pivot lattice are declared
exceptional; the pilot shows that their total occupancy is small.

\paragraph{Step 1: parameters and the pilot mask.}
Use the tilt of Section~\ref{s48:tilting}, with
$p_i\le1/2$ and $C_\lambda/Z\le5n$, allotting a fixed small
failure probability to its search. If the search reports failure,
return the original weights as the exceptional list, an empty core,
the original capacity and $K=0$. For the success analysis below,
condition on the chosen successful parameter. Set
$q=\lceil\log_2(10^5 n/\varepsilon)\rceil$,
$F=2^{3q+5}$ (which bounds the constant of Lemma~\ref{cg:lattice} for
at most $q$ generators), and $L=\lceil C_1 J\rceil$ for a large absolute
$C_1$. (Below we need $(1-0.005)^L\le\varepsilon/(1100n)$, so that together
with $2^{-q}\le\varepsilon/(10^5n)$ the sum is at most $\varepsilon/(1000n)$. Since
$n/\varepsilon\le R^2\le4^J$, $J\ge3$ and $-\log_2(0.995)>1/140$, the choice
$C_1=1000$ suffices: $1000J/140>2J+\log_2 1100$. This choice affects
$K=O(J^3)$ only through absolute constants.) Form a pilot mask by including each site $i$
independently with probability $2p_i/L$.

\paragraph{Step 2: greedy choice of pivots on the pilot.}
On the pilot sites we choose successive pivots. Initially, each weight
is represented by zero. Given $j$ pivots $b_1,\ldots,b_j$, apply
Lemma~\ref{cg:lattice} at $A=1$ to each pilot weight, and keep
the better of the new and the previous representations (the one with
smaller absolute error), padding old coordinate vectors by zero. The
chosen pivots always have their own exact unit representations. Let
$H_j$ be the maximum remaining error, with $H_j=0$ for an empty pilot.
If this maximum is zero, stop. Otherwise choose a site attaining $H_j$
as the next pivot. Stop as soon as there are $q$ pivots; in this case
put $H=H_{q-1}$, the error just before the last pivot. On an early
stop put $H=0$. Throughout the procedure all coefficients have
magnitude at most $F$. In either case, every pilot weight has an
approximation with error at most $H$ by the final pivots. Write
$\ell\le q$ for the number of final pivots, $b=(b_1,\ldots,b_\ell)$.

\paragraph{Step 3: separation of the pivots.}
The $H_j$ are weakly decreasing. At a step with $H_j>0$, the distance
of the new pivot to every $\{-1,0,1\}$ combination of earlier pivots
is at least $H_j/F$, by Lemma~\ref{cg:lattice}. This also holds
when an earlier representation was the better one. Consequently,
when $H>0$, all binary subset sums of the $q$ pivots differ
pairwise by at least $H/F$. To see this, consider the difference of
two such sums and use their last differing pivot, whose coefficient is
$1$ or $-1$.

\paragraph{Step 4: anti-concentration of the original law.}
Fix tie rules. If $H'$ is the score of an independent pilot mask,
then with probability at least $0.995$ over the first mask we have
$\Pr(H'\ge H)\ge0.005$. This is the elementary upper-quantile
bound, including ties. Suppose $H>0$ and take the union of $L$
independent new pilot masks. Except with probability $(1-0.005)^L$,
this union contains $q$ distinct sites whose binary subset sums have
separation at least $H/F$. The inclusion probabilities of the union
are $1-(1-2p_i/L)^L\le2p_i$, independently across sites. We may
therefore couple the union as a subset of the independent mask with
probabilities $2p_i$. Conditional on this larger mask, independent
fair bits on its sites generate the original product law $Q_\lambda$.
The existence of a separated list is an increasing property of the
mask, and the list may be chosen before revealing the fair bits.
Conditional on all other bits, at most one of its $2^q$ assignments
can put the sum in a given interval of length strictly less than $H/F$.
Thus, for the \emph{original} law and on the pilot quantile event,
\begin{equation}\label{cg:price}
    \sup_{|I|<H/F} Q_\lambda(S\in I)
       \le 2^{-q}+(1-0.005)^L\le \varepsilon/(1000n).
\end{equation}
The new masks are used only in this proof of the inequality and need
not be generated.

\paragraph{Step 5: the container and its outside occupancy.}
For the realized pivots, consider all input sites outside the
container
\[
       \{x:\ |x-b\cdot a|\le H\text{ for some }
                       a\in\mathbb Z^\ell,\ \|a\|_\infty\le F\}.
\]
By Step~2, no pilot site is outside. All possible containers, across
all pilots, belong to a fixed family of size $\exp(O(J^2))$. Indeed,
there are at most $(n+1)^q$ choices of pivots at each possible
length. A nonzero $H$ is the absolute difference of an input
weight and a combination of at most $q$ input pivots with integer
coefficients in $[-F,F]$, and zero is one further choice.
For a fixed outside set of tilted occupancy $\sigma$, the probability
that the pilot misses it is at most $e^{-2\sigma/L}$. A union bound over
the family shows, with failure probability below $0.005$, that the
outside occupancy of the realized container is at most $C_2LJ^2$.

\paragraph{Step 6: the full scan and rounding of the core.}
Now run Lemma~\ref{cg:lattice} with $A=F$ on all input sites.
Keep as core those sites whose returned residual $d_i=w_i-b\cdot a_i$
satisfies $|d_i|\le FH$. Every site in the container passes, including
when $H=0$. Hence, by Step~5, the exceptional list $E$ has
$\sum_{i\in E}p_i\le C_2LJ^2$. For $H=0$ the core representation
is exact. Otherwise let $M=32(n+1)F$ and replace each core weight
by
\[
            b\cdot a_i+\frac H M
                  \left\lceil\frac{M d_i}{H}\right\rceil .
\]
The extra coefficient has magnitude at most $MF+1$. Each weight
increases by less than $H/M$, so a lost feasible subset must have
its original weight in $(T-nH/M,T]$. By \eqref{cg:price}
and the remark after \eqref{s48tilt:identity}, the mass of such subsets in the feasible
count is at most $\varepsilon C_\lambda/(1000n)$. Multiply all weights and the
capacity by $M$ to clear the denominator. The core generators
are then $M b_1,\ldots,M b_\ell,H$. In the exact case we use the original
units. All operations on the original numbers involve only
polynomial-in-$R$ factors or the short arithmetic of
Lemma~\ref{cg:lattice}.

\paragraph{Step 7: degree truncation, cost and success probability.}
By \eqref{s48tilt:trunc} and $\sum_{i\in E}p_i\le C_2LJ^2$, for an
integer $K$ larger than $(e-1)C_2LJ^2+\log(1000n/\varepsilon)$,
truncation at $K$ exceptional sites loses at most
$\varepsilon C_\lambda/(1000n)$ from the original count. Enlarging the
constant budgets covers small inputs. Each pivot round makes at most
$n$ short lattice calls, each with cost polynomial in $J$. Absolute
inverse-polynomial approximations of the pilot inclusion
probabilities couple the entire pilot mask to the ideal one, except
with a further fixed small probability. The failure events are the
tilt search of Step~1 (allot $0.005$), the upper-quantile event of
Step~4 ($0.005$), the container event of Step~5 ($0.005$), and the
pilot coupling (allot $0.005$). The $L$ masks of Step~4 are not
generated, and their miss probability is already part of
\eqref{cg:price}. The total is below $0.03$, so the stated time and
success bounds follow.
\end{proof}

\section{Fourier estimates for thresholds}\label{rw:fourier}\label{sp:section}
This section proves Theorem~\ref{sp:additive}: the probability that a
weighted sum of independent Bernoulli variables, or of a uniformly
random $k$-subset, lies below a threshold can be estimated to
additive error $\varepsilon$ in time $\tO_C(n+\varepsilon^{-1})$, and in time
$O_{\delta,C}(n+\varepsilon^{-(1+\delta)})$ when the weights are polynomially
bounded. The idea has three parts. First, the threshold indicator is
written as a telescoping sum of smooth functions over dyadic scales,
all periodized with one common random period, so that only finitely
many Fourier coefficients of the sum law are needed. Second, all these
coefficients, which are products of one-site characteristic functions,
are evaluated together through an entire-function expansion and fast
nonuniform transforms. Third, the uniform slice is reduced to the
Bernoulli case by Fourier inversion over the cardinality on a short
cyclic group. The section
uses only the conventions of Section~\ref{s48:model}. Its estimates,
the alias bound~\eqref{sp:alias}, and the phase buckets of
Section~\ref{sp:strict} are reused in Sections~\ref{pw:section}
and~\ref{bc:section} and in Subsection~\ref{fin:section}.

Write $\mathrm e(x)=\exp(2\pi i x)$ and $W=\sum_i w_i$.
All bounds in this section allow arbitrary nonnegative integer
weights satisfying the stated bit-length condition. For additive
estimation we keep the original population of $n$ sites, including
zero weights. Likewise, a query on a uniform fixed-cardinality layer
uses its own population, including zero and oversized weights unless
its conditioning and normalization are adjusted explicitly. For
example, the one-item layer of $(0,1,3)$ at cutoff $1$ has probability
$2/3$. We take $T$ to be an integer (replace it by $\lfloor T\rfloor$); trivial cutoffs outside $[0,W)$ are settled directly.

\begin{theorem}\label{sp:additive}
Suppose $\log(W+2)\le\log^C(n+\varepsilon^{-1}+2)$ for fixed $C$. The probability
of $\sum w_iX_i\le T$ can be estimated to additive error $\varepsilon$, with
success at least $0.95$, in $\tO_C(n+\varepsilon^{-1})$ operations in either of
the following laws: independent Bernoulli variables with supplied
short-computable biases, or a uniformly random $k$-subset for
any $0\le k\le n$. If $W\le(n+\varepsilon^{-1}+2)^C$, a common Bernoulli bias
and uniform fixed cardinality both admit the literal bound
$O_{\delta,C}(n+\varepsilon^{-(1+\delta)})$ for every fixed $\delta>0$.
\end{theorem}
Consequently, a constant lower bound on the fair feasible density
gives the relative target at polynomial weight magnitude for
arbitrary capacities.

The proof occupies the rest of this section. The soft bound
$\tO_C(n+\varepsilon^{-1})$ is completed at the end of the subsection on conditioning
on the cardinality, and the literal bound in Section~\ref{sp:strict}.

\subsection{A common random period at all physical scales}
This subsection reduces the threshold probability to finitely many
Fourier coefficients of the sum law at the dyadic scales $\rho_j$ defined
below, with one random period shared by all scales.

Let $g(x)=\pi^{-1/2}\int_x^\infty e^{-y^2}\,dy$ and
$b(x)=g(x)-g(x/2)$. Then $b$ is odd and integrable, with Gaussian
tails. Use a dyadic $H\asymp\sqrt{\log(\varepsilon^{-1}+2)}$ with a sufficiently
large multiplier, and put $\rho_j=2^j/H$. Choose the first $J$ for which
$\rho_J\ge C_1(W+1)/\varepsilon$. For $x=s-T-1/2$ and integral $0\le s\le W$,
\[
 \left|\one_{\{s\le T\}}-\tfrac12-\sum_{j=0}^{J-1} b(x/\rho_j)\right|
                  \le \varepsilon/100.
\]
This holds because the sum telescopes: the finest argument has
magnitude at least $H/2$, and $|g(x/\rho_J)-1/2|\le |x|/\rho_J$.

Choose $\vartheta$ uniformly on $[\Pi,2\Pi]$, where $\Pi=C_2 J/\varepsilon$, and use this same
period at every scale. Put $b_\vartheta(y)=\sum_{z\in\mathbb Z} b(y+z\vartheta)$.
Uniformly in real $y$,
\begin{equation}\label{sp:alias}
 \E_\vartheta|b_\vartheta(y)-b(y)|\le C_3\|b\|_1/\Pi.
\end{equation}
Indeed, substituting $v=z\vartheta$ shows that the contributions with $z>0$
are bounded by
$\Pi^{-1}\sum_{z\ge1}z^{-1}\int_{z\Pi}^{2z\Pi}|b(y+v)|\,dv$.
At every $v$, the integers $z\ge1$ in $[v/(2\Pi),v/\Pi]$ number at most
$1+v/(2\Pi)$ and each has $1/z\le\min(1,2\Pi/v)$, so their reciprocals
sum to at most $2$; the terms with negative $z$ obey the same bound
with $v<0$, so $C_3=2$ suffices.
Integrating \eqref{sp:alias} against any positive finite measure
$\nu$ on $[0,W]$ and summing over scales costs at most
$C_3 J\|b\|_1\|\nu\|/\Pi$ in expected absolute error. This applies
to each fixed threshold. It also applies after integrating those
absolute errors against any fixed positive distribution of thresholds.

With the Fourier convention $\widehat b(y)=\int b(x)\mathrm e(-yx)\,dx$,
integration by parts and the Gaussian transform give
\[
 b_\vartheta(x)=\sum_{\ell\ne0}c_\ell(\vartheta)\mathrm e(\ell x/\vartheta),\qquad
 c_\ell(\vartheta)=
 \frac{e^{-4\pi^2(\ell/\vartheta)^2}-e^{-\pi^2(\ell/\vartheta)^2}}{2\pi i\ell}.
\]
The coefficients follow by integrating the absolutely convergent
periodization over one period. The sum of their magnitudes is $O(1)$,
using the bound $O(|\ell|/\vartheta^2)$ for $|\ell|$ up to $\vartheta$ and a Gaussian
tail beyond $\vartheta$. Truncating to
$|\ell|\le M=\lceil C_4\Pi\sqrt{\log(J/\varepsilon+2)}\rceil$
has uniform error at most $\varepsilon/(100J)$. Thus, at each scale, only the
consecutive characteristic values
$\int\mathrm e(\ell s/(\vartheta \rho_j))\,d\nu(s)$ are required.
The cutoff and the phase $T+1/2$ are applied after computing them.
For a probability law, we evaluate each of them to error
$\varepsilon/(C_5J)$ and take $C_2$ sufficiently large; Markov's inequality
then gives the claimed additive accuracy with any fixed high
probability.

These bounds hold in finite arithmetic. Round the period to a common
dyadic grid of spacing inverse polynomial in $(W+2)(n+\varepsilon^{-1}+2)$, with a
sufficiently large fixed exponent, and couple this rounding to the
continuous uniform law. On $[\Pi,2\Pi]$, the derivative in $\vartheta$ of the
finite Fourier sum at $x/\rho_j$ has magnitude bounded by a fixed
polynomial in $M+\Pi+H(W+1)$, by the displayed coefficients and
termwise differentiation. Hence the discretized sums change uniformly
by at most $\varepsilon/(100J)$ per scale, also after integration against a
measure of unit mass. The precision requires
$O(\log(W+2)+\log(n+\varepsilon^{-1}+2))$ bits. Phases can be reduced modulo one
by integer division before numerical evaluation.

\subsection{Products of characteristic functions on a progression}
The previous subsection requires the characteristic function of the
sum law at many consecutive frequencies. This subsection evaluates all
of them together in near-linear time, by an argument that also
tolerates exact zeros of the individual factors.

Let $\alpha_i$ be phases, let $0\le p_i\le1/2$, and let $y_\ell$ be
prescribed real offsets for $|\ell|\le M$. To inverse-polynomial
accuracy $\xi$, all the products
\[
 \Phi_\ell=\prod_i\{1+p_i[\exp(i y_\ell)\mathrm e(\ell\alpha_i)-1]\}
\]
can be evaluated in $(n+M)\log^{O(1)}((n+M+2)/\xi)$ work.
The offsets can be zero. For arbitrary independent biases, we
complement the sites of bias above $1/2$, reversing their phase and
multiplying by the known deterministic phase. This preserves the
original law of the sum. Complementation also reverses the sign of
the offset, so the complemented sites share the offset $-y_\ell$ and
the others share $y_\ell$. We run the algorithm below separately on
the two groups, each with its own common offset, and multiply the two
products and the deterministic phase.

For $z_i=p_i(\exp(i y_\ell)\mathrm e(\ell\alpha_i)-1)$ put
$s_h=\sum_i z_i^h$ and
$D_\ell=\sum_i p_i(1-p_i)(1-\Re[\exp(i y_\ell)\mathrm e(\ell\alpha_i)])$.
The squared modulus of the factor of site $i$ is
$1-2p_i(1-p_i)(1-\cos\theta_i)$, where $\theta_i$ is its total phase.
Hence $|\Phi_\ell|\le e^{-D_\ell}$, and $\sum_i|z_i|^2\le 2D_\ell$.
Set $L=C_6\log((n+M+2)/\xi)$. If an additive-$1/4$ approximation
to $D_\ell$ exceeds $L$, return zero. On the other rows consider
the entire function
\[
           F(v)=\prod_i(1+v z_i)e^{-v z_i}.
\]
The scalar inequality $|(1+\eta)e^{-\eta}|\le e^{|\eta|^2/2}$ follows by
applying $1+a\le e^a$ to $a=2\Re\eta+|\eta|^2$. Hence on $|v|=2$,
$|F(v)|\le e^{4(L+1)}$. By Cauchy's bounds we may truncate its
Taylor series about $0$ at degree $K=\lceil C_7L\rceil$ and evaluate at $v=1$, with error
$\xi/8$.

The coefficients through this degree are those of the formal series
\[
           \exp\!\left(\sum_{h=2}^K(-1)^{h-1}s_hv^h/h\right).
\]
They are obtained by multiplying the series by its logarithmic
derivative and comparing coefficients. Finally we multiply the
truncated sum at $v=1$ by $e^{s_1}$, which has modulus at most one.
This argument takes no logarithm of a possibly zero factor.

Power sums with absolute errors $\zeta=e^{-C_8L}$ suffice. In the
preceding formal identity, such errors replace $F$ through degree $K$
by $F\exp(E(v))$, where $|E|\le\zeta(1+\log K)$ on the unit circle.
There $|F|\le e^{L+1}$. Cauchy's formula on that circle bounds the
change of each coefficient by $O(e^{L+1}\zeta(1+\log K))$. The
error in $s_1$ fits in the same accuracy budget after exponentiation.
One can use exact rational arithmetic for the truncated recurrence
on dyadic power sums: when all power sums have denominator $2^m$,
a common denominator for degree $j$ divides $j!\,2^{mj}$.
By the recurrence, the numerators have a number of bits polynomial
in $L$.

It remains to compute the power sums. Expanding the binomial gives
\[
 s_h=\sum_{a=0}^h(-1)^{h-a}\binom ha e^{i a y_\ell}
                    \sum_i p_i^h\,\mathrm e(a\ell\alpha_i).
\]
Hence only polynomially many (in $L$) ordinary nonuniform Fourier sums
are needed, with maximum frequency $KM$; the same holds for
$D_\ell$. We recall a direct implementation of the grid expansion
that underlies fast nonuniform transforms~\cite{DR}. For bounded
coefficients $c_i$ and frequencies $|j|\le B_0$, use a power-of-two
grid size $B\asymp B_0+1$. Write $B\{\alpha_i\}=b_i+d_i$ with integer
$b_i$ and $0\le d_i<1$. Accumulate the grid arrays
$\sum_{i:b_i=b}c_i d_i^u$ and transform each by FFT. Multiplying
frequency $j$ by $(2\pi i j/B)^u/u!$ and summing over
$u=0,\ldots,O(L)$ approximates $\sum_i c_i\mathrm e(j\alpha_i)$:
the Taylor remainder for each site is bounded by
$e^{2\pi B_0/B}(2\pi B_0/B)^{u+1}/(u+1)!$.

Finally, choose the precision to allow for the factor $2^K$ in the binomial
expansion and for the sum of $n$ inputs. Rounding phases and FFT
butterflies with $O(L)$ fractional guard bits, with a sufficiently
large multiplier, suffices. Indeed, an unnormalized transform
amplifies entry errors by at most its length, and errors of the
twiddle factors accumulate over logarithmically many stages on
entries of polynomial magnitude. This proves the asserted work bound.

\subsection{Conditioning on the cardinality}
This subsection treats the uniform $k$-subset law, by extracting the
coefficient of a fixed cardinality through Fourier inversion on a short
cyclic group. We state the extraction for arbitrary biases, since
Subsection~\ref{rw:mc-coeff} reuses it.

Let $X_i$ be independent bits with biases $q_i$, let $\theta_i$ be real
phases, and write $K=\sum_iX_i$, $\mu=\sum_iq_i$, $V=\sum_iq_i(1-q_i)$
and $F(k)=\E[e^{i\sum_i\theta_iX_i}\one_{\{K=k\}}]$. Fix $J\ge1$ (until the end of this subsection $J$, $\Pi$ and $j$ are
local to the extraction, unrelated to the number of scales and the
period range above, and $K$ is the cardinality, not the degree of the
previous subsection) and let
$I$ be the set of integers $k$ with $|k-\mu|\le C_{10}(\sqrt{VJ}+J)$;
for a suitable absolute $C_{10}$, Bernstein's inequality gives
$\Pr(K\notin I)\le2e^{-4J}$. Let $\Pi$ be a power of two with
$|I|\le\Pi<2|I|$. Inversion on $\mathbb Z/\Pi$ of
$\phi(j)=\prod_i(1-q_i+q_ie^{i\theta_i}\mathrm e(j/\Pi))
=\sum_{k'}F(k')\mathrm e(jk'/\Pi)$ returns $\sum_{k'\equiv k}F(k')$.
For $k\in I$ all other terms have $k'\notin I$, and
$|F(k')|\le\Pr(K=k')$. Hence, for every $k\in I$,
\begin{equation}\label{sp:cardext}
  F(k)=\Pi^{-1}\sum_{j\in\Gamma}\phi(j)\,\mathrm e(-jk/\Pi)+O(e^{-3J}),
\end{equation}
with $\Gamma=\mathbb Z/\Pi$, and also with the following set $\Gamma$
of $O(J)$ indices, independent of $k$ and of the number of sites.

As in the previous subsection, $|\phi(j)|\le\exp(-V+\Re[B\,\mathrm
e(j/\Pi)])$ with $B=\sum_iq_i(1-q_i)e^{i\theta_i}$. If an additive-$1$
approximation of $V-|B|$ exceeds $3J+1$, every term is at most
$e^{-3J}$ and $\Gamma$ is empty. If $V\le10J$, then $\Pi=O(J)$ and
$\Gamma=\mathbb Z/\Pi$. Otherwise $|B|\ge V-3J-2\ge V/2$, so at circular
distance $x$ from $-\arg B$ the damping is at least
$|B|(1-\cos x)\ge Vx^2/\pi^2$. We keep the $j$ with $2\pi j/\Pi$ within
$\pi\sqrt{3J/V}$ of $-\arg B$, enlarged by two grid spacings after
rounding the center; this arc is listed directly. The center is
computed from an approximation of $B$ with absolute error below a
small constant; since $|B|\ge V/2$ and $\Pi=O(V)$, this moves $\arg B$
by less than one grid spacing, which the enlargement absorbs. The omitted terms
are at most $e^{-3J}$, and
$|\Gamma|=O(1+\Pi\sqrt{J/V})=O(J)$. Each retained $\phi(j)$ is a product
as in the previous subsection, with the offset $2\pi j/\Pi$ common to
all sites (split into the two complementation groups as described
there), and $B$ is one nonuniform Fourier sum.

For the uniform slice, the cases $k=0,n$ are deterministic, and
complementation reduces $k>n/2$ to one minus a lower-tail query at $W-T-1$ on
the smaller complement. Take all $q_i=p=k/n$, so that $\mu=k\in I$,
and put $d=\Pr(\Bin(n,p)=k)$; since $k$ is a mode, $d\ge1/(n+1)$.
The characteristic value at phases $\theta_i=2\pi\ell\alpha_i$ is
$F(k)/d$, and $d$ is $F(k)$ at all $\theta_i=0$. As $|F(k)|\le d$, it
suffices to compute both to absolute accuracy $\xi/[10(n+1)]$. We use
\eqref{sp:cardext} with $J=C_9\log((n+M+2)/\xi)$. The $r$-th retained
index of every row $|\ell|\le M$ enters one batched call of the
product algorithm, padding rows with fewer indices. These $O(J)$ calls,
polylogarithmic in number, cost $\tO(n+M)$ per scale. This proves the soft bound of
Theorem~\ref{sp:additive}.

\paragraph{Panel quadrature.}
Later sections use the following quadrature for real-line integrals.
Let $f$ be analytic and bounded by $A$ within distance $w$ of a real
interval. Subdivide the interval into panels of length at most $w/4$,
and on each panel integrate the polynomial interpolant at $N_q$ equally
spaced nodes. Cauchy's derivative bound and the real interpolation
remainder, applied separately to the real and imaginary parts, give
uniform interpolation error $O(A4^{-N_q})$. The total absolute
quadrature weight is at most the panel length times $(2e)^{N_q}$: bound
each Lagrange numerator by the panel length to the power of its
degree, and its denominator using the node spacing and the two
factorials. Hence, for $N_q$ logarithmic in the accuracy, inverse-polynomial
errors suffice in all function evaluations and node positions.

\subsection{A strictly linear input scan}\label{sp:strict}
We now explain how to remove the logarithmic factors in $n$ when
$W\le(n+\varepsilon^{-1}+2)^C$, which gives the literal bound of
Theorem~\ref{sp:additive}. The key point is that all scales share the
common period $\vartheta$, so the phases at every scale are bit shifts of one
fixed-point number per site. Bucketing the sites by a window of these
bits, with constant-degree moments of the remaining fractional part,
then suffices at every scale.

The soft bound suffices if $n\le \varepsilon^{-(1+\delta/2)}$, by absorbing fixed
powers of logarithms. Otherwise choose a small constant
$\gamma>0$ depending on $\delta$. For all sufficiently large $n$,
the largest required integer frequency, including the powers in
the product algorithm, satisfies $KM\le n^{1-8\gamma}$. Any fixed
additional powers of $\log(n+2)$ can be allowed in this inequality
by the initial choice of $\gamma$. There are $O_C(\log(n+2))$
scales. Crucially, they use a common $\vartheta$: put $\beta_i=H w_i/\vartheta$,
so that all phases are $\beta_i/2^j$. Use fixed-point versions of the
$\beta_i$ with $C_{12}\log_2(n+2)$ fractional bits. Their error in
any requested ordinary sum is at most
$2\pi n KM\,2^{-C_{12}\log_2(n+2)}$, which can be made smaller
than any prescribed fixed inverse polynomial. Under the polynomial
magnitude assumption, modular integer arithmetic and division give
these values in constant original-word work per item.

Let $\chi=\lfloor(1-2\gamma)\log_2 n\rfloor$ and
$d=\max(1,\lfloor\gamma\log_2 n\rfloor)$. Choose upper bit
positions $U$ at intervals $d$, so that each scale $j$ satisfies
$U-d<j\le U$ for some position $U$. There are $O_{C,\delta}(1)$ such
positions. For each position set $l=U-\chi$ and write
\[
 b_i=\lfloor\beta_i/2^l\rfloor\bmod 2^\chi,\qquad
 d_i=\{\beta_i/2^l\},\qquad B_j=2^{j-l}.
\]
Negative bit positions use the stored fractional digits. Each
$B_j$ is a power of two, with $n^{1-4\gamma}\le B_j\le 2^\chi$
for large $n$. A phase at integer frequency $a$ has the exact
fixed-point representation $\mathrm e(a(b_i+d_i)/B_j)$, since
$B_j$ divides $2^\chi$. For each of the constantly many positions,
we store in bucket $b_i$ the moments $\sum_{i:b_i=b'}d_i^h$ through
a fixed degree $D_0$. For fixed $D_0$, the moment entries are sums of
dyadics with $O_{C,\delta}(1)$ words. Direct indexing,
integer products and addition build all these arrays in
$O_{C,\delta}(n)$ word operations.

At scale $j$, fold the buckets modulo $B_j$ and apply ordinary FFTs
to their moment arrays. The Taylor expansion of
$\mathrm e(a d_i/B_j)$ now has an argument of order $n^{-4\gamma}$
for every $|a|\le KM$. Consequently a \emph{constant} degree
$D_0=D_0(C,\delta)$ gives any required fixed inverse-polynomial
total error, including the sum of $n$ remainders and the $2^K$
coefficient loss. The inverse precision demanded in the product
algorithm and the cardinality extraction is polynomial in $n+\varepsilon^{-1}+2$, with an
absolute fixed exponent. All transforms and the subsequent numerical
work cost $(2^\chi+M)\log^{O(1)}(n+2)=O_\delta(n)$ across all scales.
With a common $p$, only unweighted site power sums are needed: the
factors $p^h$ and $e^{iay_\ell}$ are inserted after summation. This
also handles the cardinality extraction and completes the proof of
Theorem~\ref{sp:additive}.

\subsection{Exponentially weighted thresholds}
Later sections need, besides single distribution functions, the
weighted sums $\sum_{t\le T}h_t\,\nu(\{t\})$ with exponential weights
$h_t$, for all thresholds $t\le T$ at once. The next lemma reduces such
a sum to distribution functions by summation by parts, and evaluates
the resulting sums of Fourier phases in logarithmic time.

\begin{lemma}[Exponentially weighted thresholds]\label{ft:geom}
Let $T\ge1$ be an integer, $\alpha\ge0$, and
$P(x)=\pi_0+\pi_1x+\pi_2x^2$ a real polynomial such that
$h_t=P(\alpha(t-T))e^{\alpha(t-T)}$ is nondecreasing in
$t\in\{0,\dots,T\}$, with $h_0\ge0$ and $h_T=1$.
\begin{enumerate}[label=(\alph*)]
\item For every finite measure $\nu$ on $\{0,1,2,\dots\}$, with
  $F_\nu(t)=\nu([0,t])$,
  \[\textstyle\sum_{t=0}^{T}h_t\,\nu(\{t\})
     =F_\nu(T)-\sum_{t=0}^{T-1}(h_{t+1}-h_t)F_\nu(t),\]
  and the coefficients $h_{t+1}-h_t$ are nonnegative with sum at most one.
\item Let $|z|=1$, $q=e^{-\alpha}\bar z$, $c_a=\pi_a(-\alpha)^a$ and
  $S_a(q,N)=\sum_{d=1}^{N}d^aq^d$. Then
  \[\textstyle\sum_{t=0}^{T-1}(h_{t+1}-h_t)z^t
    =z^{T-1}\Bigl(1+\sum_{a=0}^2c_aS_a(q,T-1)\Bigr)
     -z^{T}\sum_{a=0}^2c_aS_a(q,T).\]
\item For $|q|\le1$, $0\le a\le2$, an integer $N\ge0$ and $\eta>0$,
  $S_a(q,N)$ is computed to absolute error $\eta$ by $O(\log(N+2))$
  arithmetic operations on fixed-point numbers with
  $O(\log^2(N+2)+\log(1/\eta))$ bits, starting from such an
  approximation of $q$.
\end{enumerate}
\end{lemma}
For example, $P=1$ and $P(x)=1+x^2$ satisfy the hypothesis for every
$\alpha\ge0$, since $(1+x^2)e^x$ has derivative $(1+x)^2e^x$.
\begin{proof}
(a) Put $F_\nu(-1)=0$. Since $\nu(\{t\})=F_\nu(t)-F_\nu(t-1)$,
summation by parts gives the identity, using $h_T=1$. The differences
are nonnegative by monotonicity, and they sum to $h_T-h_0\le1$.

(b) Substitute $d=T-t$, so that $h_{T-d}=P(-\alpha d)e^{-\alpha d}$
and $z^{T-d}=z^T\bar z^{\,d}$. The terms $h_{t+1}z^t$ give
$z^{T-1}\sum_{e=0}^{T-1}P(-\alpha e)q^e$, whose term $e=0$ is
$P(0)=h_T=1$; the terms $h_tz^t$ give $z^T\sum_{d=1}^{T}P(-\alpha d)q^d$.
Expanding $P(-\alpha d)=\sum_ac_ad^a$ gives the formula. No division by
$z$ or by $q$ occurs.

(c) For integers $L_1,L_2\ge0$, the binomial translation formula gives
\[\textstyle
  S_a(q,L_1+L_2)=S_a(q,L_1)+q^{L_1}\sum_{b=0}^{a}\binom ab
       L_1^{a-b}S_b(q,L_2).
\]
We follow the binary digits of $N$ from the top, keeping the tuple
$(q^L,S_0,S_1,S_2)$ at the current length $L$, and pass from $L$ to
$2L$ (take $L_1=L_2=L$) and from $L$ to $L+1$ (take $L_2=1$, where
$S_b(q,1)=q$). Each of the $O(\log(N+2))$ steps uses $O(1)$
operations. All powers of $q$ have modulus at most one and
$|S_a(q,L)|\le(N+1)^3$, so each step amplifies the errors of its
inputs by at most a fixed polynomial in $N+2$ and adds its own
rounding error. Over $O(\log(N+2))$ steps the amplification is
$(N+2)^{O(\log(N+2))}$, and $O(\log^2(N+2))$ guard bits beyond
$\log_2(1/\eta)$ suffice.
\end{proof}

\section{The displaced tilt and a conditional local estimate}\label{dt:section}
This section collects the probabilistic tools of Sections~\ref{pw:section}
and~\ref{rw:linear} that go beyond the balanced tilt of
Subsection~\ref{s48:tilting}. The first is a \emph{displaced tilt}: a
parameter $\lambda$ under which the capacity lies between one and three
tilted standard deviations above the tilted mean. Unlike the balanced
tilt, it is found deterministically, by bisection on tilted means and
variances (Lemma~\ref{dt:search}), and its price $C_\lambda/Z$ is
still polynomial in $n$ (Lemma~\ref{dt:price}). Both lemmas rest on a
Berry--Esseen anti-concentration estimate (Lemma~\ref{dt:anti}). The
second tool, in Subsection~\ref{rw:pwlocal}, is a local estimate
conditional on several cardinalities. Appendix~\ref{app:structured}
uses the displaced tilt again.

\subsection{Anti-concentration and the price of a displaced tilt}
\begin{lemma}[Anti-concentration at a displaced tilt]\label{dt:anti}
Let $w_1,\dots,w_n\ge0$, $W=\sum_iw_i$ and $0\le T<W/2$. For $\lambda\ge0$
let $X_1,\dots,X_n$ be independent with $X_i\sim\Bern(p_i)$,
$p_i=(1+e^{\lambda w_i})^{-1}$, under the law $Q=Q_\lambda$, and let
$S=\sum_iw_iX_i$, $m(\lambda)=\E_QS$ and $s(\lambda)^2=\Var_QS$. Suppose
\begin{equation}\label{sp:tiltposition}
   s(\lambda)\le T-m(\lambda)\le3s(\lambda).
\end{equation}
Then $\lambda>0$ and $s(\lambda)>0$. Put $u_i=\lambda w_i$, $\tau=\lambda T$,
$\sigma=\lambda s(\lambda)$, $x=\lambda S-\tau$,
\[ r=\E_Q\bigl[e^x\one_{\{x\le0\}}\bigr],\quad m_0=\E_Q\frac1{1+x^2},\quad
   \rho=\sum_i\E_Q|u_i(X_i-p_i)|^3,\quad \beta=\frac\rho{\sigma^2},\quad
   b=1+\min(\sigma,\beta). \]
Then $r\le m_0$, $\sigma\le\sqrt n$, and for an absolute constant $C_{22}$
\begin{equation}\label{rw:anti}
   \frac{m_0}{r}\le C_{22}e^{C_{22}b},\qquad
   \frac1r\le C_{22}(1+\sigma)e^{C_{22}b}.
\end{equation}
Moreover, for every $B\ge3$, with $N_B=\#\{i:u_i>B\}$,
\begin{equation}\label{sp:tilttail}
   \rho\le B\sigma^2+N_BB^3e^{-B}.
\end{equation}
\end{lemma}
\begin{proof}
At $\lambda=0$ the mean $W/2$ exceeds $T$, so \eqref{sp:tiltposition}
forces $\lambda>0$; as some $w_i>0$, also $s(\lambda)>0$ and $\sigma>0$.
Write $U=\lambda S=\sum_iu_iX_i$, so that $x=U-\tau$.
Let $C_{\mathrm{BE}}$ be the absolute constant in the Berry--Esseen
inequality for sums of independent, not necessarily identically
distributed, centered summands~\cite{Berry}: the distribution
functions of $(U-\E U)/\sigma$ and of a standard normal differ
by at most $C_{\mathrm{BE}}\rho/\sigma^3=C_{\mathrm{BE}}\beta/\sigma$
uniformly. Let $\varphi_3=e^{-9/2}/\sqrt{2\pi}$ be the standard
normal density at $3$, and $C_{\mathrm{ac}}=\max(1,4C_{\mathrm{BE}}/\varphi_3)$.

\emph{Step 1: always $Q(\tau-6\sigma\le U\le\tau)\ge7/18$, and
hence $r\ge\frac7{18}e^{-6\sigma}$.}
By \eqref{sp:tiltposition},
$a=\tau-\E U\in[\sigma,3\sigma]$. Cantelli's one-sided
inequality gives $Q(U-\E U>a)\le\sigma^2/(\sigma^2+a^2)\le1/2$,
and Chebyshev's inequality gives $Q(U<\E U-3\sigma)\le1/9$.
Hence $Q(\E U-3\sigma\le U\le\tau)\ge7/18$. Since
$\E U-3\sigma\ge\tau-6\sigma$, this event lies in
$\{\tau-6\sigma\le U\le\tau\}$, on which $e^x\ge e^{-6\sigma}$.

\emph{Step 2: if $\sigma\ge\kappa:=C_{\mathrm{ac}}(1+\beta)$, then
$Q(\tau-\kappa<U\le\tau)\ge\varphi_3\kappa/(2\sigma)$, every
interval of length one has $Q$-probability at most
$(1+2C_{\mathrm{BE}}\beta)/\sigma$, and consequently}
\[
  r\ge\frac{\varphi_3\,\kappa\,e^{-\kappa}}{2\sigma},\qquad
  m_0\le\frac{5(1+2C_{\mathrm{BE}}\beta)}{\sigma}.
\]
Since $\kappa\le\sigma$, the standardized interval
$[(\tau-\kappa-\E U)/\sigma,(\tau-\E U)/\sigma]$ has
length $\kappa/\sigma$ and lies in $[0,3]$, where the normal density
is at least $\varphi_3$. Applying Berry--Esseen at both endpoints,
$Q(\tau-\kappa<U\le\tau)\ge(\varphi_3\kappa-2C_{\mathrm{BE}}\beta)/\sigma
\ge\varphi_3\kappa/(2\sigma)$, because $\kappa\ge4C_{\mathrm{BE}}\beta/\varphi_3$.
On this event $e^x\ge e^{-\kappa}$, which gives the bound on $r$.
Likewise an interval of length one in $U$ becomes an interval of
length $1/\sigma$ after standardizing, of normal mass at most
$1/(\sigma\sqrt{2\pi})\le1/\sigma$, so its $Q$-probability is at most
$(1+2C_{\mathrm{BE}}\beta)/\sigma$. Summing over the intervals $(k-1,k]$,
$k\in\mathbb Z$, in the variable $x$ against the suprema of $1/(1+x^2)$ on
them, whose sum is $2\sum_{j\ge0}(1+j^2)^{-1}<5$, gives the bound on
$m_0$.

\emph{Step 3: the bounds on $r$, $\sigma$ and \eqref{rw:anti}.}
The function $(1+x^2)e^x$ increases on $x\le0$, with derivative
$(1+x)^2e^x$, so $e^x\one_{\{x\le0\}}\le1/(1+x^2)$ and $r\le m_0$.
Next, $u_i^2p_i(1-p_i)\le u_i^2e^{-u_i}\le4e^{-2}<1$, so
$\sigma^2\le n$. For \eqref{rw:anti}, suppose first $\sigma<\kappa$.
Then $\sigma<C_{\mathrm{ac}}(1+\beta)$ and trivially
$\sigma<C_{\mathrm{ac}}(1+\sigma)$, so $\sigma\le C_{\mathrm{ac}}b$; by Step~1 and
$m_0\le1$ both ratios are at most $\frac{18}7e^{6C_{\mathrm{ac}}b}$.
Otherwise $\sigma\ge\kappa$ forces $\beta<\sigma$, so $b=1+\beta$ and
$\kappa=C_{\mathrm{ac}}b\ge1$. By Step~2, $1/r\le(2/\varphi_3)\sigma
e^{C_{\mathrm{ac}}b}$, and, using $(1+2C_{\mathrm{BE}}\beta)/\kappa\le
1+2C_{\mathrm{BE}}$, $m_0/r\le10(1+2C_{\mathrm{BE}})\varphi_3^{-1}e^{C_{\mathrm{ac}}b}$.
Thus $C_{22}=\max(6C_{\mathrm{ac}},10(1+2C_{\mathrm{BE}})/\varphi_3)$ works.

\emph{Step 4: \eqref{sp:tilttail}.}
The sites with $u_i\le B$ contribute at
most $B\sigma^2$ to $\rho$, since
$\E_Q|u_i(X_i-p_i)|^3\le u_i\cdot u_i^2p_i(1-p_i)$.
The other sites each contribute at most $u_i^3 e^{-u_i}\le B^3e^{-B}$,
since $p_i(1-p_i)\le e^{-u_i}$ and $u^3e^{-u}$ decreases for $u\ge3$.
\end{proof}

\begin{lemma}[Price of a displaced tilt]\label{dt:price}
Let $T\ge0$ and $w_1,\dots,w_n\ge0$ be integers, $W=\sum_iw_i$ and
$Z=\#\{A\subseteq[n]:\sum_{i\in A}w_i\le T\}$. For $\lambda\ge0$ let
$C_\lambda=e^{\lambda T}\prod_i(1+e^{-\lambda w_i})$, and let
$m(\lambda)$ and $s(\lambda)^2$ be the mean and variance of
$\sum_iw_iX_i$ for independent $X_i\sim\Bern((1+e^{\lambda w_i})^{-1})$.
\begin{enumerate}[label=(\alph*)]
\item If $T<W/2$ and $\lambda$ satisfies \eqref{sp:tiltposition}, that is,
  $s(\lambda)\le T-m(\lambda)\le3s(\lambda)$, then
  $\lambda\ge4(W/2-T)/\sum_iw_i^2\ge2/W^2$ and
  \begin{equation}\label{sp:tiltpoly}
     C_\lambda/Z\le(n+2)^{C_{\mathrm{tilt}}}
  \end{equation}
  for an absolute constant $C_{\mathrm{tilt}}$.
\item If $T\ge W/2$, then $C_0/Z=2^n/Z\le2$.
\end{enumerate}
\end{lemma}
\begin{proof}
(a) By convexity, $1/(1+e^x)\ge1/2-x/4$ for $x\ge0$, so
$m(\lambda)\ge W/2-\lambda\sum_iw_i^2/4$. Since
\eqref{sp:tiltposition} gives $T\ge m(\lambda)$, and $W-2T\ge1$ for
integers with $T<W/2$, we get
$\lambda\ge4(W/2-T)/\sum_iw_i^2\ge2/W^2$.

For \eqref{sp:tiltpoly} we use the notation of Lemma~\ref{dt:anti}.
By \eqref{s48tilt:identity}, $Z=C_\lambda r$. Choose
$B=C\log(n+2)$ with an absolute $C\ge3$ large enough that
$nB^3e^{-B}\le1$. Then $N_BB^3e^{-B}\le1$, and \eqref{sp:tilttail}
gives $\min(\sigma,\beta)\le B+1$: for $\sigma<1$ use $\sigma$ itself,
and for $\sigma\ge1$ use $\beta=\rho/\sigma^2\le B+1$. Since
$\sigma\le\sqrt n$, \eqref{rw:anti} gives
$C_\lambda/Z=1/r\le C_{22}(1+\sqrt n)e^{C_{22}(B+2)}\le(n+2)^{C_{\mathrm{tilt}}}$.

(b) Complementation maps the infeasible subsets injectively into the
feasible ones, since $w(A)>T$ implies $W-w(A)<W-T\le T$. Hence
$Z\ge2^{n-1}$.
\end{proof}

\subsection{Finding a displaced tilt in linear time}
The search bisects in the rescaled parameter $y=\lambda T$. Part~(b)
of the next lemma implements it with a single input scan and a
sublinear amount of further work, for arbitrary original words; the
weights are rounded once, and the slack in the stopping rule absorbs
the rounding.

\begin{lemma}[Finding a displaced tilt]\label{dt:search}
Let $T\ge1$ and $0\le w_1,\dots,w_n\le T$ be integers with
$W=\sum_iw_i>2T$. Let $p(u)=(1+e^{u})^{-1}$, and for $\lambda\ge0$ let
$m(\lambda)=\sum_iw_ip(\lambda w_i)$,
$s(\lambda)^2=\sum_iw_i^2p(\lambda w_i)(1-p(\lambda w_i))$,
$C_\lambda=e^{\lambda T}\prod_i(1+e^{-\lambda w_i})$ and
$f(y)=(m(y/T)+2s(y/T))/T$.
\begin{enumerate}[label=(\alph*)]
\item Given an oracle that returns $f(y)$ to absolute error
  $1/[40(n+2)]$ at dyadic $y\in[0,10(n+1)]$, bisection with
  $O(\log(n+2))$ oracle calls finds a dyadic $y\in(0,10(n+1))$ with
  $O(\log(n+2))$ bits such that $\lambda=y/T$ satisfies
  \eqref{sp:tiltposition}, that is, $s(\lambda)\le T-m(\lambda)\le3s(\lambda)$.
\item Fix $A\ge1$ and $0<\gamma<1$. If the $w_i$ and $T$ are given as
  original words, a deterministic algorithm finds $y$ as in~(a) and
  $\log C_{y/T}$ to absolute error $(n+2)^{-A}$ in $O_{A,\gamma}(n)$
  word operations. It reads the input in a single scan; all further
  work takes $O_{A,\gamma}(n^\gamma\log^{O(1)}(n+2))$ operations.
\end{enumerate}
\end{lemma}
\begin{proof}
(a) Put $z_i=w_i/T\in[0,1]$ and $v=p(1-p)$, so that
$f(y)=\sum_iz_ip(yz_i)+2(\sum_iz_i^2v(yz_i))^{1/2}$. First,
$f(0)=W/(2T)+2s(0)/T>1$. At $y_1=10(n+1)$ we have $y_1m/T\le n/e$ and
$y_1s/T\le\sqrt n$, since $u/(1+e^u)\le1/e$ and $u^2e^{-u}\le4e^{-2}<1$
for $u\ge0$; hence $f(y_1)\le(n/e+2\sqrt n)/(10(n+1))<1$. Next, $f$ is
$n$-Lipschitz on $[0,y_1]$. Indeed, $|p'|=v\le1/4$, so the derivative of
$m/T$ is at most $n/4$ in absolute value; and $|v'|=v|1-2p|\le v$, so
the derivative of $(s/T)^2$ is at most $(s/T)^2$ in absolute value, and
that of $s/T$ at most $(s/T)/2\le\sqrt n/4$.

We keep a bracket $[y_-,y_+]$ with $f(y_-)>1>f(y_+)$, initially
$[0,y_1]$. Let $\tilde f$ be the oracle value at its midpoint $y$. If
$|\tilde f-1|\le1/[20(n+2)]$, we stop and return $y$. Otherwise
$f(y)-1$ has the sign of $\tilde f-1$, and $y$ replaces $y_-$ or $y_+$
accordingly. Once the bracket has length at most $1/[20n(n+2)]$, its
midpoint lies within half this length of a solution of $f=1$ in the
bracket, so $|f(y)-1|\le1/[40(n+2)]$ by the Lipschitz bound, and the
search stops. Hence it makes $O(\log(n+2))$ calls and returns a dyadic
$y\in(0,y_1)$ with $O(\log(n+2))$ bits and $f(y)=1+e$, where
$|e|\le3/[40(n+2)]$.

Such a $y$ satisfies \eqref{sp:tiltposition}. Indeed, write $m,s$ at
$\lambda=y/T$ and $p_i=p(\lambda w_i)\le1/2$. Since
$p_i(1-p_i)\ge p_i^2$, the Cauchy--Schwarz inequality gives
$s^2\ge\sum_iw_i^2p_i^2\ge m^2/n$, that is, $m\le\sqrt n\,s$. Hence
$T(1+e)=m+2s\le(\sqrt n+2)s$, and $s\ge0.9\,T/(\sqrt n+2)\ge|e|T$. As
$T-m=2s-eT$, this gives $s\le T-m\le3s$.

(b) Put $\ell(u)=\log(1+e^{-u})$ and
$\eta=(n+2)^{-A-4}/10^4$. A trial at $y\le y_1$ needs the three sums
\[
 \frac{m}{T}=\sum_iz_ip(yz_i),\qquad
 \frac{s^2}{T^2}=\sum_iz_i^2v(yz_i),\qquad
 \log C_{y/T}-y=\sum_i\ell(yz_i).
\]
We compute each of them to absolute error $2\eta$. Before taking the
square root, replace the approximation to $s^2/T^2$ by its maximum
with zero. This cannot increase its absolute error. Since
$|\sqrt a-\sqrt b|\le\sqrt{|a-b|}$ for $a,b\ge0$, this gives $f(y)$ to
error $2\eta+2\sqrt{2\eta}\le5\sqrt\eta\le1/[40(n+2)]$, so (a) applies
to the true weights. The third sum gives $\log C_{y/T}$ to error
$2\eta\le(n+2)^{-A}$. For small $n$, below a threshold depending on
$A$ and $\gamma$, we evaluate the sums directly. Each summand has the
form $\phi(z_i)$ with $\phi(z)=z^jg(yz)$, $j\in\{0,1,2\}$ and
$g\in\{p,v,\ell\}$.

\emph{Rounding.} Let $F$ be the least integer with
$2^F\ge10(n+2)^2/\eta$, so $F=O_A(\log(n+2))$. Put
$N_i=\lfloor2^Fw_i/T\rfloor$ and $\tilde z_i=2^{-F}N_i$; this needs a
single original-word quotient after scaling the numerator, which
occupies $O_A(1)$ words. On $[0,1]$ each $\phi$ has derivative at most
$2+y\le10(n+2)$ in absolute value, since $0\le g\le1$ and
$|g'|\le1/2$ for $g\in\{p,v,\ell\}$ (as $|p'|=v\le1/4$, $|v'|\le v$
and $|\ell'|=p\le1/2$) and $y\le10(n+1)$. As $|\tilde z_i-z_i|\le2^{-F}$, replacing $z_i$ by
$\tilde z_i$ changes each sum by at most $10n(n+2)2^{-F}\le\eta$. It
remains to evaluate the sums for the $\tilde z_i$ to error $\eta$, at
each of the $O(\log(n+2))$ trials and at the returned $y$.

\emph{Bands and bins.} Sites with $N_i=0$ contribute $\phi(0)$ and are
only counted. For $N_i\ge1$ let $k_i$ be the position of its leading
bit, the \emph{band} of the site, and put $x_i=N_i/2^{k_i}\in[1,2)$.
The band is found in $O_{A,\gamma}(1)$ word operations: split the
$F+1$ bits of $N_i$ into chunks of $\lceil\gamma\log_2(n+2)\rceil$
bits, and look up the leading bit of the highest nonzero chunk in a
table of size $O(n^\gamma)$. Let $G$ be the power of two with
$G\le n^\gamma<2G$, and split $[1,2)$ into $G$ \emph{bins} of width
$1/G$; the bin of $x_i$ is read off by a shift of $N_i-2^{k_i}$. For
every band, bin and $0\le h\le D$, with a fixed $D=D(A,\gamma)$ chosen
below, the scan accumulates $\sum_i(x_i-c)^h$ over the sites in the bin,
where $c$ is the bin centre. The displacements are dyadic with
$O_A(\log(n+2))$ bits, so each site costs $O_{A,\gamma}(1)$ word
operations. There are $O_{A,\gamma}(n^\gamma\log(n+2))$ stored moments.

\emph{Evaluating a trial.} On band $k$ put $\mu_k=y2^{k-F}$, so that
$\phi(\tilde z_i)=2^{j(k-F)}x_i^jg(\mu_kx_i)$, where $2^{j(k-F)}\le1$.
Since $0\le g(u)\le e^{-u}$ for $u\ge0$, the bands with
$\mu_k>L_0=\log(3n/\eta)$ may be omitted, at total cost at most $\eta/3$.
On the remaining bands $\mu_k\le L_0$. Let $\rho_0=1/[4(1+L_0)]$. For
complex $\xi$ within $\rho_0$ of a real $c\in[1,2]$ we have
$|\Im(\mu_k\xi)|\le1/4$ and $\Re(\mu_k\xi)\ge0$. Hence
$e^{\pm\mu_k\xi}$ have positive real parts, so $|p(\mu_k\xi)|\le1$ and
$|1-p(\mu_k\xi)|\le1$, while $1+e^{-\mu_k\xi}$ has real part at least
one and modulus at most two; we use the branch of the logarithm that is
real on the real axis. Thus $|\xi^jg(\mu_k\xi)|\le16$ on these discs.
For a site at distance at most $1/(2G)$ from its bin centre, Cauchy's
estimate bounds the Taylor remainder after degree $D$ by
$32(2(1+L_0)/G)^{D+1}$, once $2(1+L_0)/G\le1/2$. As $L_0=O_A(\log(n+2))$
and $G>n^\gamma/2$, a fixed $D=D(A,\gamma)$ makes the sum of these
remainders over all sites at most $\eta/3$ for all $n$ above the
threshold. The Taylor coefficients at the bin centres are computed in
short precision, so that the arithmetic adds at most $\eta/3$, and are
combined with the stored moments. This takes
$O_{A,\gamma}(n^\gamma\log^{O(1)}(n+2))$ operations per trial, and the
$O(\log(n+2))$ trials and the final normalizer stay within this bound.
\end{proof}

\subsection{A local estimate conditional on several cardinalities}\label{rw:pwlocal}
The following lemma is the probabilistic core of the second tilt in
Section~\ref{pw:section}. Consider
independent Bernoulli variables split into groups, and condition on
all group counts taking their means. The lemma states that a
Cauchy-type weight $1/(1+\gamma^2(Y-a)^2)$ of a linear statistic is
dominated, up to a factor exponential in $\gamma b$, by the one-sided
exponential payoff $e^{\gamma(Y-a)}\one_{\{Y\le a\}}$. Here $b$
bounds the deviations of the coefficients from their group centers. In
Subsection~\ref{rw:pwrecover} this lemma bounds the total positive
mass \eqref{pw:sumprice} paid by the Fourier estimate.

We give the estimate with absolute constants in the window length.
This point matters because in Section~\ref{pw:section} the number of
groups is chosen as a function of $\delta$. The threshold $\varpi_0$ in the lemma depends on $D_1$ and on the number $m$ of groups.

\begin{lemma}\label{pw:local}
Take independent Bernoulli variables with biases $q_i$, group counts
$K_g$, integer means $k_g=\E K_g$ and variances
$V_g=\sum_{i\in g}q_i(1-q_i)$. For real $d_i$, put $v_i=q_i(1-q_i)$,
$c_g=\sum_{i\in g}v_i d_i/V_g$, $Y=\sum_i d_i X_i$,
$M=\sum_i q_i d_i$ and $H^2=\sum_i v_i(d_i-c_{g(i)})^2$.
Fix $D_1>0$. Suppose $H>0$, $b>0$, $|d_i-c_{g(i)}|\le b$ and
$b/H+\sum_gV_g^{-1/2}\le\varpi_0$, for an explicit
$\varpi_0=\varpi_0(D_1,m)>0$ given in the proof. Write $Q^*$ for conditioning
on $K_g=k_g$ for all $g$; then $\Pr(K_g=k_g\text{ for all }g)\ge\frac12\prod_g(2\pi V_g)^{-1/2}$. For $|a-M|\le D_1H$ and every $\gamma>0$,
\begin{equation}\label{pw:localprice}
 \E_{Q^*}\frac1{1+\gamma^2(Y-a)^2}
 \le C_1 e^{C_1(1+\gamma b)}
       \E_{Q^*}\!\left[e^{\gamma(Y-a)}
                             \one_{\{Y\le a\}}\right].
\end{equation}
For $|a-M|\le D_1H$, $Q^*(a-H\le Y\le a)\ge c_1>0$; hence, by monotonicity,
$Q^*(Y\le a)\ge c_1$ for every $a\ge M-D_1H$. The constants
$c_1,C_1$ are explicit and depend only on $D_1$.
\end{lemma}
\begin{proof}
The idea is to prove a conditional local limit theorem for $Y$ with
explicit error terms, and then to compare both sides of
\eqref{pw:localprice} with intervals of length comparable to
$b+1/\gamma$. Throughout, write $\zeta_i=d_i-c_{g(i)}$. Then
$\sum_{i\in g}v_i\zeta_i=0$ for every group, $|\zeta_i|\le b$, and
$H_g^2=\sum_{i\in g}v_i\zeta_i^2$ satisfies $H_g^2\le b^2V_g$ and
$\sum_gH_g^2=H^2$. Put
\[
  \varpi=\frac bH+\sum_g V_g^{-1/2},\qquad
  G=\prod_g(2\pi V_g)^{-1/2}.
\]

\paragraph{Step 1: Fourier representation and one-site facts.}
Since $Y-M=\sum_i\zeta_i(X_i-q_i)+\sum_gc_g(K_g-k_g)$, on the
conditioning event $Y-M=\sum_i\zeta_i(X_i-q_i)$. For real $s$ and
$y\in[-\pi,\pi]$ put
\[
 \psi_g(s,y)=\E\exp\Bigl(i\sum_{i\in g}(s\zeta_i+y)(X_i-q_i)\Bigr),
 \qquad
 \Phi(s)=(2\pi)^{-m}\!\int_{[-\pi,\pi]^m}\prod_g\psi_g(s,y_g)\,dy .
\]
Because $\sum_{i\in g}q_i=k_g$, orthogonality of the characters
$e^{iy(K_g-k_g)}$ and independence of the groups give
$\Phi(s)=\E[e^{is(Y-M)}\one_{\{K=k\}}]$. Hence
$\phi(s):=\E_{Q^*}e^{is(Y-M)}=\Phi(s)/\Phi(0)$, where
$\Phi(0)=\prod_g\Pr(K_g=k_g)>0$. For one site with bias $q$, $v=q(1-q)$
and real $\theta$ we use two elementary facts.
\begin{itemize}
\item[(F1)] $|1-q+qe^{i\theta}|^2=1-2v(1-\cos\theta)
 \le e^{-2v(1-\cos\theta)}$.
\item[(F2)] For $|\theta|\le1$ the function
 $f(\theta)=\log(1-q+qe^{i\theta})-iq\theta$ (principal branch; the
 real part of the argument is at least $\cos1$) satisfies
 $|f(\theta)+v\theta^2/2|\le v|\theta|^3/4$. Indeed $f(0)=f'(0)=0$,
 $f''(0)=-v$, and
 $f'''(\theta)=-ive^{i\theta}(1-q-qe^{i\theta})/(1-q+qe^{i\theta})^3$
 has modulus at most $v((1+\cos1)/2)^{-3/2}\le 3v/2$ by (F1), so
 Taylor's theorem with remainder applies.
\end{itemize}
Put $D_g(s)=\sum_{i\in g}v_ie^{is\zeta_i}$ and
$\alpha_g=\arg D_g(s)$. By (F1),
\begin{equation}\label{rw:psibound}
 |\psi_g(s,y)|\le e^{-\sum_{i\in g}v_i(1-\cos(s\zeta_i+y))}
 =e^{-(V_g-|D_g(s)|)}\,e^{-|D_g(s)|(1-\cos(y+\alpha_g))}.
\end{equation}
For $|s|b\le1$ we also have the following two bounds.
\begin{itemize}
\item[(D1)] Since $\sum_{i\in g}v_i\zeta_i=0$ and
 $|e^{ix}-1-ix|\le x^2/2$, we get
 $|D_g(s)-V_g|\le s^2H_g^2/2\le V_g/2$. Hence
 $|D_g(s)|\ge\Re D_g(s)\ge V_g/2$, and
 $|\Im D_g(s)|=|\sum_{i\in g}v_i(\sin(s\zeta_i)-s\zeta_i)|\le|s|^3bH_g^2/6$.
\item[(D2)] $V_g-|D_g(s)|\ge(2/\pi^2)s^2H_g^2$. To see this, use the
 pair-difference identity
 $V_g^2-|D_g(s)|^2=\sum_{i,j\in g}v_iv_j(1-\cos(s(\zeta_i-\zeta_j)))$, the
 bound $1-\cos x\ge 2x^2/\pi^2$ for $|x|\le\pi$ (here
 $|s(\zeta_i-\zeta_j)|\le2$), and
 $\sum_{i,j\in g}v_iv_j(\zeta_i-\zeta_j)^2=2V_gH_g^2$. This gives
 $V_g^2-|D_g|^2\ge(4/\pi^2)s^2V_gH_g^2$; divide by
 $V_g+|D_g|\le2V_g$.
\end{itemize}
Finally, $1-\cos y\ge2y^2/\pi^2$ on $[-\pi,\pi]$ gives, for $v>0$,
\begin{equation}\label{rw:Ibound}
 \Upsilon(v):=\frac1{2\pi}\int_{-\pi}^{\pi}e^{-v(1-\cos y)}\,dy
 \le\frac1{2\pi}\int_{\mathbb R}e^{-2vy^2/\pi^2}\,dy
 =\frac\pi2\,(2\pi v)^{-1/2}.
\end{equation}

\paragraph{Step 2: a quantitative Gaussian approximation.}
We show that for $A\ge1$ with $A^3\varpi\le1$ and $|s|H\le A$,
\begin{equation}\label{rw:Napprox}
 \bigl|\Phi(s)-Ge^{-s^2H^2/2}\bigr|\le G\,\mathcal E,\qquad
 \mathcal E=C_m'\bigl(A^3\varpi\,e^{A^3\varpi}+e^{-A^2/(4\pi^2)}\bigr),
 \quad C_m'=6m(\pi/\sqrt2)^m .
\end{equation}
Let $\mathcal B=\{y:|y_g|\le AV_g^{-1/2}\ \forall g\}$; it lies in
the torus because $AV_g^{-1/2}\le A\varpi\le1$. Write
$Ge^{-s^2H^2/2}=(2\pi)^{-m}\int_{\mathbb R^m}e^{-\mathcal Q/2}\,dy$
with $\mathcal Q=s^2H^2+\sum_gV_gy_g^2$, and split the difference into
three parts.

\emph{Inside $\mathcal B$.} Put $\theta_i=s\zeta_i+y_{g(i)}$. Then
$|\theta_i|\le Ab/H+AV_g^{-1/2}\le A\varpi\le1$, and
$\sum_iv_i\theta_i^2=\mathcal Q$ because
$\sum_{i\in g}v_i\zeta_i=0$. By (F2), $\prod_g\psi_g=e^{-\mathcal Q/2+\rho}$
with
\[
 |\rho|\le\tfrac14\sum_iv_i|\theta_i|^3
 \le\sum_iv_i\bigl(|s|^3|\zeta_i|^3+|y_{g(i)}|^3\bigr)
 \le |s|^3bH^2+\sum_gV_g|y_g|^3\le A^3\varpi ,
\]
using $(x+y)^3\le4(x^3+y^3)$. Hence
$|\prod_g\psi_g-e^{-\mathcal Q/2}|\le|\rho|e^{|\rho|}e^{-\mathcal Q/2}
\le A^3\varpi e^{A^3\varpi}e^{-\mathcal Q/2}$. Integrated over
$\mathcal B$ and normalized by $(2\pi)^{-m}$, this contributes at most
$GA^3\varpi e^{A^3\varpi}$.

\emph{The Gaussian outside $\mathcal B$.} By the union bound and
$\Pr(|\mathcal N(0,1)|>A)\le2e^{-A^2/2}$, we have
$(2\pi)^{-m}\int_{\mathbb R^m\setminus\mathcal B}e^{-\mathcal Q/2}\le
2mGe^{-A^2/2}$.

\emph{The torus outside $\mathcal B$.} Here some $|y_g|>AV_g^{-1/2}$.
By (D1), with $|s|b\le A\varpi\le1$ and
$|s|^3bH_g^2\le A^3b/H\le A^3\varpi\le1$, we get $|D_g|\ge V_g/2$ and
$|\alpha_g|\le|\Im D_g|/\Re D_g\le 1/(3V_g)\le AV_g^{-1/2}/2$ (using $\arctan x\le x$ for $x\ge0$ and $\Re D_g>0$), since
$V_g\ge\varpi^{-2}\ge1$. So for $AV_g^{-1/2}\le|y|\le\pi$ the
distance from $y+\alpha_g$ to $2\pi\mathbb Z$ is at least $|y|/2$.
By \eqref{rw:psibound},
$|\psi_g(s,y)|\le e^{-V_gy^2/(4\pi^2)}$ there, and a Gaussian tail
bound gives
$(2\pi)^{-1}\int_{AV_g^{-1/2}\le|y|\le\pi}|\psi_g|\le
2\sqrt2\pi(2\pi V_g)^{-1/2}e^{-A^2/(4\pi^2)}$. For every other group
$h$, \eqref{rw:psibound} and \eqref{rw:Ibound} give
$(2\pi)^{-1}\int_{-\pi}^{\pi}|\psi_h|\le \Upsilon(|D_h|)\le(\pi/\sqrt2)(2\pi
V_h)^{-1/2}$. Taking the union over the exceptional group $g$, this
part is at most $4m(\pi/\sqrt2)^mGe^{-A^2/(4\pi^2)}$.

Adding the three parts proves \eqref{rw:Napprox}. Now suppose also
$\mathcal E\le1/2$. At $s=0$, \eqref{rw:Napprox} gives
$|\Phi(0)-G|\le G\mathcal E$, so $\Phi(0)\ge G/2$ and, for $|s|H\le A$,
\begin{equation}\label{pw:effective}
\begin{aligned}
 \bigl|\phi(s)-e^{-s^2H^2/2}\bigr|
 &\le\frac{|\Phi(s)-Ge^{-s^2H^2/2}|+e^{-s^2H^2/2}|G-\Phi(0)|}{\Phi(0)}\\
 &\le4\mathcal E
 =C_m\bigl(A^3\varpi\,e^{A^3\varpi}+e^{-A^2/(4\pi^2)}\bigr),
\end{aligned}
\end{equation}
with the explicit constant $C_m=4C_m'=24m(\pi/\sqrt2)^m$.

\paragraph{Step 3: a Gaussian majorant.}
Suppose $\Phi(0)\ge G/2$. Then, for $|s|\le1/b$,
\begin{equation}\label{rw:majorant}
 |\phi(s)|\le M_me^{-2s^2H^2/\pi^2},\qquad M_m=2(\pi/\sqrt2)^m .
\end{equation}
Indeed, by \eqref{rw:psibound}, rotation invariance of the integral
over a full period, (D2), (D1) and \eqref{rw:Ibound},
\[
 |\Phi(s)|\le\prod_g e^{-(V_g-|D_g|)}\Upsilon(|D_g|)
 \le\prod_g e^{-2s^2H_g^2/\pi^2}\,\frac\pi{\sqrt2}(2\pi V_g)^{-1/2}
 =G(\pi/\sqrt2)^me^{-2s^2H^2/\pi^2},
\]
and we divide by $\Phi(0)\ge G/2$.

\paragraph{Step 4: explicit thresholds.}
We now fix all parameters explicitly. Let
\[
 g_0=\frac{2\sin^4(1)}{\sqrt{2\pi}}e^{-(D_1+1)^2/2},\qquad
 \eta=\frac{g_0}4,
\]
so $\eta\le(2\pi)^{-1/2}/2<1/5$, and let $L=L(D_1)\ge2$ be the least
integer with $8/(3\sqrt{2\pi}L^2)\le g_0/4$. Let $A=A(D_1,m)\ge1$ be the
least integer such that
\[
 \int_{|u|>A}\bigl(M_me^{-2u^2/\pi^2}+e^{-u^2/2}\bigr)\,du\le\eta
 \qquad\text{and}\qquad C_me^{-A^2/(4\pi^2)}\le\frac{\eta}{4A},
\]
and set
\[
 \varpi_0(D_1,m)=\min\Bigl\{A^{-3},\ \frac{\eta}{4eC_mA^4},\
 \frac1{8L}\Bigr\}.
\]
Assume $\varpi\le\varpi_0$. Then $e^{A^3\varpi}\le e$, so
$4\mathcal E\le\eta/(4A)+\eta/(4A)$ in Step~2. In particular
$\mathcal E\le1/2$, and \eqref{pw:effective} gives
\begin{equation}\label{rw:window}
 |\phi(s)-e^{-s^2H^2/2}|\le\frac{\eta}{2A}\qquad(|s|H\le A).
\end{equation}
Moreover $\Phi(0)\ge G/2$, so \eqref{rw:majorant} holds for
$|s|\le 1/b$. Finally $b/H\le1/(8L)$. The constants
$c_1,C_1$ obtained below are built from $g_0$ and $L$ only, so they
depend only on $D_1$.

\paragraph{Step 5: probabilities of intervals.}
We now pass from these frequency estimates to conditional
probabilities. Assume $\varpi\le\varpi_0$, let $l\ge4b$, and let
$\xi\sim\mathcal N(M,H^2)$. Let $F\in L^1$ be continuous with
transform $\widehat F(s)=\int F(x)e^{-isx}\,dx$ bounded and supported
in $[-4,4]$. For real $x$, Fourier inversion gives
\[
 \E_{Q^*}F\Bigl(\frac{Y-x}l\Bigr)
 =\frac l{2\pi}\int_{|s|\le4/l}\widehat F(ls)\,\phi(s)\,e^{is(M-x)}\,ds,
\]
and the same formula holds for $\xi$ with $e^{-s^2H^2/2}$ in place of
$\phi$. All frequencies here satisfy $|s|\le1/b$. On $|s|H\le A$ we
use \eqref{rw:window}, and on $|s|H>A$ we use \eqref{rw:majorant}
and the choice of $A$. This gives
$\int_{|s|\le4/l}|\phi-e^{-s^2H^2/2}|\le2\eta/H$, and hence
\begin{equation}\label{rw:transfer}
 \Bigl|\E_{Q^*}F\Bigl(\frac{Y-x}l\Bigr)-\E F\Bigl(\frac{\xi-x}l\Bigr)\Bigr|
 \le\frac l{2\pi}\,\|\widehat F\|_\infty\,\frac{2\eta}H .
\end{equation}
We use the kernel $\Psi(x)=(\sin x/x)^4$, with $\Psi(0)=1$. It
satisfies $0\le \Psi\le\min(1,x^{-4})$ and $\Psi\ge\sin^4(1)$ on
$[-1,1]$. Its transform is supported in $[-4,4]$, being a
fourfold convolution of the transform $\pi\one_{[-1,1]}$ of
$\sin x/x$, and $|\widehat \Psi|\le\int \Psi=2\pi/3$.

\emph{Upper bound.} The density of $\xi$ is at most
$(2\pi)^{-1/2}H^{-1}$ and $\eta<1/5$, so \eqref{rw:transfer} gives
\begin{equation}\label{rw:upper}
 \E_{Q^*}\Psi\Bigl(\frac{Y-x}l\Bigr)
 \le\frac{\sqrt{2\pi}+2\eta}3\,\frac lH\le\frac lH
 \qquad(x\in\mathbb R).
\end{equation}

\emph{Lower bound.} The function
$\Psi_L(u)=(1-u^2/L^2)\Psi(u)$ lies below $\one_{\{|u|\le L\}}$.
Its transform $\widehat\Psi_L=\widehat\Psi+L^{-2}\widehat\Psi''$ is
supported in $[-4,4]$. Since $\Psi(u)\le\min(1,u^{-4})$, we have
$\int u^2\Psi(u)\,du\le8/3$ and
$\|\widehat\Psi_L\|_\infty\le 2\pi/3+8/(3L^2)<3$.
For $|a-M|\le D_1H$ and $4b\le l\le H/(2L)$ take $x=a-Ll$.
The density of $\xi$ on $[x-l,x+l]$ is at least
$(2\pi)^{-1/2}H^{-1}e^{-(D_1+1)^2/2}$. Thus the contribution of this
interval to $\E\Psi_L((\xi-x)/l)$ is at least $3g_0l/(4H)$.
The negative contribution has magnitude at most
$8l/(3\sqrt{2\pi}L^2H)\le g_0l/(4H)$ by the density upper bound.
By \eqref{rw:transfer}, replacing $\xi$ by $Y$ changes this
expectation by at most $3\eta l/(\pi H)\le g_0l/(4H)$. Applying the
pointwise minorant gives
\begin{equation}\label{rw:onesided}
 Q^*(a-2Ll\le Y\le a)\ge c_3\frac lH,\qquad c_3=\frac{g_0}4 .
\end{equation}

\paragraph{Step 6: proof of \eqref{pw:localprice} and of the last assertion.}
Let $l_0=4b+1/\gamma$ and $l=\min\{l_0,H/(2L)\}$. Then $l\ge4b$,
because $H/(2L)\ge4b$ by $b/H\le1/(8L)$, and
$\gamma l\le\gamma l_0\le4(1+\gamma b)$. By \eqref{rw:onesided}, the
right side of \eqref{pw:localprice} satisfies
\[
 \E_{Q^*}\bigl[e^{\gamma(Y-a)}\one_{\{Y\le a\}}\bigr]
 \ge e^{-2L\gamma l}c_3\frac lH\ge e^{-8L(1+\gamma b)}c_3\frac lH .
\]
For the Cauchy expectation, suppose first that $l=l_0$, so
$\gamma l\ge1$. Since $1+t^2\le3(1+x^2)$ for $|t-x|\le1$ (as $t^2\le2x^2+2$) and
$\Psi\ge\sin^4(1)$ on $[-1,1]$, we have $(1+x^2)^{-1}\le
\frac3{2\sin^4(1)}\int\Psi(x-t)(1+t^2)^{-1}\,dt$. Take $x=(Y-a)/l$.
Fubini, \eqref{rw:upper} at the points $a+lt$ and
$\int(1+t^2)^{-1}dt=\pi$ bound the Cauchy expectation by
$\E_{Q^*}(1+(Y-a)^2/l^2)^{-1}\le C_Ul/H$, where
$C_U=3\pi/(2\sin^4(1))$.
If instead $l=H/(2L)$, the Cauchy expectation is at most $1=2Ll/H$. In
both cases it is at most $\max\{C_U,2L\}\,l/H$. Comparing with the
display above proves \eqref{pw:localprice} with
$C_1=\max\{8L,\ \max\{C_U,2L\}/c_3\}$.

For the last assertion, let $|a-M|\le D_1H$. Then \eqref{rw:onesided}
with $l=H/(2L)\ge4b$ gives
$Q^*(a-H\le Y\le a)\ge c_1:=c_3/(2L)$. If only $a\ge M-D_1H$, apply
this at $\min\{a,M+D_1H\}$ and use monotonicity.
\end{proof}

\section{All capacities at polynomial weight magnitude}\label{pw:section}
In this section we allow an arbitrary list of polynomially bounded
weights and prove Theorem~\ref{pw:main}, which gives a relative
approximation at every capacity. The difficulty is that a single
exponential tilt can put too much mass on a large infeasible atom
close to the threshold. The idea is to group sites with similar tilted
weights, to retain the counts of chosen sites in each group, and to
apply a second change of measure to the deviations of the tilted
weights within groups. The number of groups is fixed once the desired
running-time exponent is specified.

The proof combines several earlier tools. The displaced tilt of
Lemma~\ref{dt:anti} and its price (Lemma~\ref{dt:price}) control
the first change of measure, and Lemma~\ref{dt:search} finds this
tilt. A small set of exceptional sites is handled by the coarse parity
trees of Lemma~\ref{ex:coarse}. The group counts are
inverted and summed over integer points of polytopes by the methods
of Subsection~\ref{mc:region}, and the threshold itself is evaluated by
the Fourier construction of Section~\ref{sp:section}, with the input
scan of Section~\ref{sp:strict}. The key ingredient is the local
estimate of Lemma~\ref{pw:local} (Subsection~\ref{rw:pwlocal}), which bounds the price of the second
tilt conditionally on all group counts. The construction is reused
later: Lemma~\ref{pw:param} records how it depends on $\log W$, and
Section~\ref{fin:section} runs it at each baseline.

The section is organized as follows.
Subsection~\ref{rw:pwtrim} trims the instance and splits off the
exceptional sites. Subsection~\ref{rw:pwtilt} defines the second tilt
on regions of group counts, and Subsection~\ref{rw:pwrecover} recovers
the original threshold and bounds the total mass of the resulting
measures. Subsection~\ref{mc:region} develops Fourier estimates
summed over the integer points of a polytope of group counts,
Subsection~\ref{rw:pwchar} uses them to compute the required
characteristic values, and Subsection~\ref{rw:pwprec} treats finite
precision and the strictly linear input term.

\begin{theorem}\label{pw:main}
Write $R=n+\varepsilon^{-1}+2$. For every fixed $C,\delta>0$, if the
total weight $W=\sum_iw_i$ satisfies $W\le R^C$, a relative $\varepsilon$ approximation to
$Z$ for any given capacity $T$ takes $O_{\delta,C}(n+\varepsilon^{-(1+\delta)})$ word
operations with success probability at least $2/3$.
\end{theorem}

\subsection{Trimming by tilted occupancy}\label{rw:pwtrim}
This subsection prepares the instance for the second tilt. After an
initial tilt, sites of large tilted weight are discarded and the rest
are grouped by tilted weight. Sites and groups that would spoil the
hypotheses of Lemma~\ref{pw:local}, namely groups of small variance
and sites far from their group center, are moved to an exceptional set
$E$ of polylogarithmic tilted occupancy. The exceptional sites are
then handled by the coarse parity trees of
Lemma~\ref{ex:coarse}, and \eqref{pw:rootmass} bounds the total
mass of their root entries.

We may take $T<W/2$ by Theorem~\ref{sp:additive}. Use the displaced
tilt, that is, $\lambda$ satisfying \eqref{sp:tiltposition} of
Lemma~\ref{dt:anti}, writing $u_i=\lambda w_i$,
$\tau=\lambda T$, $p_i=(1+e^{u_i})^{-1}$, $v_i=p_i(1-p_i)$,
$P(\lambda)=\prod(1+e^{-u_i})$, $C_\lambda=e^\tau P(\lambda)$ and $J=\lceil\log R\rceil$.
Lemma~\ref{dt:price} gives $C_\lambda/Z\le R^{C_3}$ with an
absolute $C_3$, and Lemma~\ref{dt:search}(b) finds such a tilt
deterministically with a strictly linear input scan. Only the tilt is
taken from that lemma here; scalar normalizers are computed to
inverse-polynomial accuracy in $R$ as in Subsection~\ref{rw:pwprec}.

Next we discard the sites with $u_i>C_4J$. Since
$0\le e^{\lambda S-\tau}\one_{\{\lambda S\le\tau\}}\le1$, the feasible sets containing site $i$ number
at most $C_\lambda p_i\le C_\lambda e^{-u_i}$ (see after \eqref{s48tilt:identity}), so this costs at most
$n C_\lambda e^{-C_4J}$ in the count. We take an absolute $C_4$ sufficiently large
and retain the same tilt. All subsequent losses that are
inverse-polynomial relative to $C_\lambda$ can pay for this price
and for the required relative error.

Fix $0<\beta\le1$, to be chosen at the end. Group the remaining
$u_i$ by intervals of width $\beta J$; the number of groups $m\le
2+C_4/\beta$ is then fixed. For a current group write
$V_g=\sum_{i\in g}v_i$, $a_g=\sum_{i\in g}v_i u_i/V_g$ and
\[
 d_i=u_i-a_g,\quad
 \varrho^2=\sum_i v_i d_i^2,\qquad \mu_g=\sum_{i\in g}p_i .
\]
Here $\varkappa$ is a sufficiently small absolute constant. We repeatedly
apply the following two tests until neither applies:
\begin{itemize}
\item move any group with $V_g<(J+m)^{10}$ to an exceptional set $E$;
\item if $\varrho>0$, move a single site with $|d_i|>\varkappa \varrho/\sqrt J$ to
  $E$, and recompute the affected center.
\end{itemize}

We claim that the total $\sum_{i\in E}p_i$ is polylogarithmic.

\emph{Single-site removals.} A site removal reduces the within-group
sum of weighted squared deviations by at least $v_i d_i^2$, since
recentering minimizes that sum. To turn this into a bound we need the
range of $\varrho^2$. A positive $\varrho^2$ is bounded below by an inverse
polynomial: a group with two distinct weights has two $u_i$ separated
by at least $\lambda$, and each retained $v_i$ is inverse polynomial.
(Here $\lambda\ge2/W^2\ge2R^{-2C}$ by Lemma~\ref{dt:price}.) The upper bound is $O(nJ^2)$. The resulting telescoping
decrease in $\log \varrho^2$ bounds the sum of $v_i$ over the single
removals by $O_C(J^2/\varkappa^2)$; the final removal, which may bring
$\varrho^2$ to zero, is accounted for separately.

\emph{Whole-group removals.} These cost at most $m(J+m)^{10}$.
Finally $p_i\le 2v_i$, which converts both bounds into bounds on the
$p_i$.

Restrict the degree in $E$ to $K=O(\sum_E p_i+(C_3+10)J)$ with
a sufficiently large fixed multiplier. The product-law Chernoff
bound \eqref{s48tilt:trunc} then loses at most $\varepsilon Z/100$ in feasible count. Build the
coarse trees of Lemma~\ref{ex:coarse} for these
sites up to degree $K$. A root entry has mass $\varphi'>0$ and lower
representative $b'$ in physical weight, with conditional true weight
$V_E\ge b'$. Write $F_{\mathrm{core}}$ for the true counting measure of the
remaining core and $P_{\mathrm{core}},P_E$ for the two product normalizers. Using
both signs, we have the following positive bounds:
\begin{align}
 \sum \varphi'F_{\mathrm{core}}(\le T-b')&\le 2e^{1/16} Z, \notag\\
 \sum \varphi'e^{-\lambda b'}&\le 2e^{1/16}P_E .                 \label{pw:rootmass}
\end{align}
Both follow from Lemma~\ref{ex:coarse}(c), applied for each $j\le K$
and summed, with $f(x)=F_{\mathrm{core}}(\le T-x)$ and with
$f(x)=e^{-\lambda x}$ respectively, since $\sum_{j\le K}\int e^{-\lambda x}\,dF_j\le P_E$.

Fix an absolute $D>C_3+5$ for the following truncations.
By \eqref{pw:rootmass}, errors bounded for one entry by
$R^{-D} \varphi'e^{t_0}P_{\mathrm{core}}$, where $t_0=\lambda(T-b')$, cost
$O(R^{-D}C_\lambda)$ in total across all entries, irrespective of
signs. Entries with $b'>T$ contribute zero. If the core is empty, its
point mass is treated as a constant core below.

\subsection{A second tilt on count regions}\label{rw:pwtilt}
From now on we work with one exceptional root entry at a time, so that
$t_0=\lambda(T-b')$ is the tilted capacity available to the core. We
first discard atypical group counts and atypical residuals. We then
introduce a one-parameter family of product laws $Q^{(l)}$ that tilt the
core along the normalized deviations while keeping the expected group
counts fixed. Each count vector is assigned to the tilt whose residual
mean best matches the remaining capacity. Finally we check that, for
each count vector, the hypotheses of Lemma~\ref{pw:local} hold with
absolute constants.

For now suppose $\varrho>0$, put $Y=\sum_{\mathrm{core}} d_i X_i$, $Y_0=\sum_{\mathrm{core}} p_i d_i$
and $z_i=d_i/\varrho$, and write $U=\lambda S_{\mathrm{core}}=a\cdot k+Y$ at group
counts $k=(K_g)_g$. Under the initial core law,
$\sum v_i z_i^2=1$, $\sum_{i\in g}v_i z_i=0$ and
$\max|z_i|\le\varkappa/\sqrt J$. Keep the box
$|k_g-\mu_g|\le C_5\sqrt{V_g J}$ and the additional strip
\begin{equation}\label{pw:mainstrip}
  -C_5\sqrt J \le h(k):=(t_0-Y_0-a\cdot k)/\varrho
                         \le C_5\sqrt J+D J/\varrho .
\end{equation}
Choose $C_5$ sufficiently large in terms of $D$. We justify the
discards outside the box and the strip. Bernstein's inequality bounds
the probability outside the box or outside $|Y-Y_0|\le C_5\varrho\sqrt J$
by $O_m(R^{-D})$, increasing $C_5$ if necessary. Inside the latter
event, counts below the lower endpoint of \eqref{pw:mainstrip} cannot
fit the core cutoff. Above the upper endpoint we have
$U-t_0\le-DJ$, giving payoff at most $e^{-DJ}$. Multiplication by
$\varphi'e^{t_0}P_{\mathrm{core}}$ and \eqref{pw:rootmass} justify these discards for the
joint problem, since the exceptional excess is nonnegative.

For a parameter $l$ define a product law $Q^{(l)}$ on the core by
\[
     \frac{q_i}{1-q_i}=e^{-u_i+\alpha_g-l z_i},\qquad
                         \sum_{i\in g}q_i=\mu_g .
\]
The intercepts need not make $\mu_g$ integers. We use a mesh of
values of $l$ that includes $l=0$, has spacing at most one, and covers
$[-\min(\varrho,C_6\sqrt J),\,C_6\sqrt J]$, including both endpoints.
The constants $C_5,D$ determine a sufficiently large absolute
$C_6$, and we take $\varkappa C_6$ sufficiently small. Since each
$\alpha_g$ lies between the extremes of $l z_i$ in its group,
all new variances are comparable to the original $v_i$ by
absolute factors. Define
\[
       M(l)=\varrho^{-1}\bigl(\sum_i q_i d_i-Y_0\bigr),\qquad
       \gamma=1+l/\varrho\ge0.
\]
Clearly $M(0)=0$, and differentiating the constraints gives
$-M'(l)=\sum_g\sum_{i\in g}q_i(1-q_i)(z_i-\bar z_g^{(l)})^2$, where
$\bar z_g^{(l)}$ is the $q_i(1-q_i)$-weighted mean of the $z_i$ in $g$. By minimization over the
group centers, it is comparable to one. These comparison constants
can be fixed in advance on $|l z_i|,|\alpha_g|\le1/4$.

Assign each $h(k)$ to the nearest value $M(l)$, with a fixed tie
rule. The resulting regions are bounded polytopes, obtained by slicing
the box and \eqref{pw:mainstrip} with parallel hyperplanes. Since
$M(0)=0$, $-M'$ is comparable to one and $C_6$ is large in terms of
$C_5$ and $D$, we have $M(C_6\sqrt J)<-C_5\sqrt J$ and
$M(-C_6\sqrt J)>(C_5+D)\sqrt J$, and the latter bounds the upper
endpoint of \eqref{pw:mainstrip} when $\varrho\ge\sqrt J$. Hence
$|h(k)-M(l)|\le C_7$, with an absolute $C_7$, in every region except
the region of the endpoint $l=-\varrho$ when $\varrho\le C_6\sqrt J$ (the saturated
region). That region may extend above $M(-\varrho)$, always has
$h(k)\ge M(-\varrho)-C_7$, and has $\gamma=0$.

For the local estimate only, we change each intercept again so that
the actual count $k_g$ becomes its independent mean. Conditioning on
the counts is invariant under these new group intercepts. We now
check that the resulting change in the normalized residual mean is
$o(1)$, uniformly in the entire count box. Put
$\sigma_g^2=\sum_{i\in g}v_i z_i^2$, so $\sum \sigma_g^2=1$. At zero intercept
the linear change in the group count under $-lz_i$ vanishes. The
Taylor remainder and the intercept derivative therefore give
\[
       |\alpha_g|\le C_8 l^2 \sigma_g^2/V_g .
\]
The additional count-fitting intercept change $\eta_g$ has magnitude
$O(\sqrt{J/V_g})$. At the first intercept, the derivative of the
group residual mean in normalized units is bounded by
$C(|l|\sigma_g^2+|\alpha_g|\sqrt{V_g}\sigma_g)$. This follows by subtracting
$\sum v_i z_i=0$ and using the Lipschitz change of logistic
variance in a bounded log-odds interval. The quadratic remainder
in $\eta_g$ is $O(\eta_g^2\sqrt{V_g}\sigma_g)$. Summing the linear and
quadratic changes gives $O_m(J/\sqrt{\min V_g})=o(1)$.

We now verify the hypotheses of Lemma~\ref{pw:local}, applied to the
refitted law with the same $d_i$; there $q_i,v_i,V_g$ refer to that
law and the cutoff of that lemma (called $a$ there) is $t_0-a\cdot k$. The resulting
$H$ of Lemma~\ref{pw:local} obeys $H\asymp \varrho$, by minimization and
variance comparability. Take
$b=2\max|d_i|\le 2\min(\beta J,\varkappa \varrho/\sqrt J)$.
Thus $b/H\to0$ uniformly, and the fitted counts have variances
tending to infinity. In a nonsaturated region the core cutoff
$t_0-a\cdot k$ is within an absolute $D_1H$ of the fitted
residual mean. In a saturated region it is at least that mean
minus $D_1H$. All bounds determining $D_1$ are absolute for
large $R$ at each fixed $m$. Since $b/H$ and $\max_gV_g^{-1/2}$ are
bounded by explicit negative powers of $J=\Theta(\log R)$, the
threshold $\varpi\le\varpi_0(D_1,m)$ of Lemma~\ref{pw:local} holds for
all $R\ge R_0(D_1,m)$ with an explicit $R_0$. Instances with $R<R_0$
have $n,\varepsilon^{-1}<R_0$ and are counted exactly in
$O(n)+O_{\delta,C}(1)$ operations, so no ineffective constant is needed.

\subsection{Recovering the original threshold}\label{rw:pwrecover}
We now write the contribution of each root entry and count region
exactly in terms of the tilted law $Q^{(l)}$. The threshold indicator
becomes a payoff of bounded variation integrated against a positive
measure $\nu$. Lemma~\ref{pw:local} bounds the total mass of these
measures by \eqref{pw:sumprice}, and the Fourier threshold
construction of Section~\ref{sp:section} then gives the estimate. The
characteristic values it needs are computed in
Subsection~\ref{rw:pwchar}.

Fix an exceptional root entry and a mesh region $\mathcal C$.
Define
\begin{align*}
 L_0&=\sum_g\alpha_g\mu_g+(l/\varrho)(t_0-Y_0),\\
 B_l&=\frac{\exp(\gamma t_0-L_0)}{\prod_{\mathrm{core}}(1-q_i)},\qquad
 W_l(k)=\exp\!\left[-\sum_g\alpha_g(k_g-\mu_g)+l h(k)\right].
\end{align*}
At a core configuration of weight $U$, the counting mass equals its
$Q^{(l)}$ probability times $B_l W_l(k)e^{\gamma(U-t_0)}$. This follows
by multiplying out the site odds. Uniformly on the retained region,
$|\log W_l|\le O_m(J)$. For nonsaturated regions this uses
$|h(k)|=O(\sqrt J)$; on the saturated one, $l=-\varrho$,
$\varrho\le C_6\sqrt J$ and \eqref{pw:mainstrip} give the same result.
We also need a bound on the scalar prefactor:
\begin{equation}\label{pw:prefactor}
 \frac{B_l}{e^{t_0}P_{\mathrm{core}}}
  =\E_{Q^{(0)}}\exp\!\left[\sum_g\alpha_g(K_g-\mu_g)
                            -l(Y-Y_0)/\varrho\right]\le R^{C_9}.
\end{equation}
Indeed, the linear combination is centered and its single-site
coefficients are uniformly bounded. Its log moment is therefore
$O(\sum V_g\alpha_g^2+l^2)=O(J)$, by the centered Bernoulli moment
bound and the estimate on $\alpha_g$. In particular, polynomial
absolute errors after scaling by $B_l$ are affordable.

Let $x=\gamma(U+\chi-t_0)$ where $\chi=\lambda(V_E-b')\ge0$ is the
exceptional excess. Draw the exceptional path conditionally on the
root, and the core independently under $Q^{(l)}$. If $\gamma>0$, define a
positive measure $\nu$ on the original physical total $S=S_{\mathrm{core}}+V_E$
with weight
\[
        \one_{\mathcal C}(k) W_l(k)\,e^{-\gamma\chi}
                                  G(x),\qquad G(x)=(1+x^2)^{-1}.
\]
Then the root and region contribute, before their parity sign,
\begin{equation}\label{pw:exact}
       \varphi' B_l\int e^x(1+x^2)\one_{\{S\le T\}}\,d\nu(S),
             \qquad x=\gamma\lambda(S-T).
\end{equation}
In a saturated region with $\gamma=0$, take $G=1$ and use the same
formula with $x=0$. The exponential $e^{-\gamma\chi}$ exactly
cancels the exceptional part of the change of measure in
\eqref{pw:exact}.

The total positive mass paid for these measures satisfies
\begin{equation}\label{pw:sumprice}
               \sum_{\text{roots, regions}} \varphi'B_l\|\nu\|
                       \le C_{10} e^{C_{11}\beta J} Z
\end{equation}
with an absolute $C_{11}$, for sufficiently large $R$ at fixed $m$.
To prove this, first note that $e^{-v}G(y+v)\le G(y)$ for $v\ge0$,
since $(1+x^2)e^x$ is increasing; this
removes the exceptional excess. For each fixed $k$ in a nonsaturated region, apply
Lemma~\ref{pw:local} to the core alone. Its cost is
$C e^{C\gamma b}$, and $\gamma b\le 2\beta J+2C_6\varkappa$. In the
saturated region the one-sided assertion of the lemma gives a
constant cost. Multiply the resulting bound by the $Q^{(l)}$ count
probability and by $\varphi'B_l W_l(k)$. By the exact odds identity, this is
at most the stated factor times $\varphi'$ times the actual number of core
subsets with these counts and weight at most $T-b'$. Sum over the
disjoint count regions and then apply \eqref{pw:rootmass}. Note that
the local comparison uses absolute constants for each fixed $m$, so
no $m$ appears in the exponent multiplying the group width.

The payoff in \eqref{pw:exact} has bounded variation on the real
line, by an absolute constant: $e^x(1+x^2)$ increases from zero
to one for $x\le0$, its derivative being $e^x(x+1)^2$.
Consequently the Fourier threshold construction of
Section~\ref{sp:section}, with the common period and internal
accuracy parameter $\varepsilon'=\varepsilon e^{-C_{11}\beta J}/C_{12}$, gives expected
absolute alias error $O(\varepsilon'\|\nu\|)$ for each term before its
scalar prefactor. Use summation by parts and the geometric sums of
Lemma~\ref{ft:geom}. The physical weights in the parity trees
need at most $O((K+1)(W+1))$ as an upper bound, which is polynomial.
Summing the expected absolute errors and applying Markov's
inequality to \eqref{pw:sumprice} gives relative error at most
$\varepsilon/3$, with any prescribed fixed high probability, by increasing
$C_{12}$. The deterministic numerical errors described below can be
made $O(\varepsilon Z/100)$ by \eqref{pw:prefactor}, \eqref{pw:rootmass} and
the bound on $|\log W_l|$.

\subsection{Fourier estimation inside a polytope of multiplicities}\label{mc:region}
For $m$ nonempty classes of sizes $n_g$, let $k_g$ count the selected
sites in class $g$ and set $R=n+\varepsilon^{-1}+2$. For each integer
vector $k$, count inversion gives the Fourier mass of configurations
with those multiplicities under the tilted product law. We multiply
this mass by the region weight and sum over the integer vectors in
the prescribed polytope. This gives the unnormalized weighted sum
required in Subsection~\ref{rw:pwchar}, even for regions of very
small probability. Equation~\eqref{pw:sumprice} controls the
aggregate error over regions. Subsection~\ref{rw:mc-coeff} gives
the single-class coefficients, and Subsection~\ref{rw:mc-phase}
uses the polynomial lattice-point sums of Section~\ref{sec:fdsum}
to evaluate the resulting phases.

\subsubsection{Fourier coefficients indexed by a multiplicity}\label{rw:mc-coeff}
We first treat a single class in isolation. The goal is to express
the Fourier mass of the class restricted to each fixed multiplicity
as a short exponential sum in that multiplicity; this form can later
be summed over a polytope of multiplicities. The idea is to invert on
a small cyclic group that separates the typical multiplicities, and
to keep only the few Fourier indices at which the product is not
negligible.

Consider a group of $n_g$ independent sites with biases $p_i\in[0,1]$.
Write $K_g=\sum_i X_i$, $\mu=\sum_i p_i$ and
$V=\sum_i p_i(1-p_i)$. Let
$\theta_{ri}=2\pi r\xi_i$ for $|r|\le M_1$.
We need \emph{all} the complex masses
\[
  F_r(k)=\E\!\left[
    e^{i\sum_i\theta_{ri}X_i}\one_{\{K_g=k\}}\right]
\]
in a form that can be summed over $k$. Note that always
$|F_r(k)|\le\Pr(K_g=k)$.

Take $J_1=C_{1,m}\log R$ with a sufficiently large multiplier for the
prescribed inverse-polynomial tolerance. Apply the cardinality
extraction~\eqref{sp:cardext} to the biases $p_i$ and phases
$\theta_{ri}$. Keep the counts in its window $I$, intersected
with $[0,n_g]$. The probability outside this window is at most $2e^{-4J_1}$. With the power of
two $\Pi$ chosen there, every $k\in I$ satisfies
\begin{equation}\label{mc:rankchar}
 F_r(k)=\Pi^{-1}\sum_{j\in\Gamma_r}
    \prod_i(1-p_i+p_i e^{i\theta_{ri}}\mathrm e(j/\Pi))\,
                         \mathrm e(-jk/\Pi)+O(e^{-2J_1}).
\end{equation}
The set $\Gamma_r$ has $O(J_1)$ indices, independently of $n_g$ and
of $k$.

Some implementation remarks follow. If a computed bias is used, it
should first be rounded to total variation error $e^{-4J_1}$ over all
sites. The product algorithm of Section~\ref{sp:section} evaluates the
coefficients for all rows in $\tO(n_g+M_1)$ time; in each batched call,
the angular offset $2\pi j/\Pi$ is specified separately for each row.
Unused entries may be padded with coefficient zero.

\subsubsection{Summing the phases over integer points}\label{rw:mc-phase}
Expanding products of the expressions \eqref{mc:rankchar} over the
classes produces phases that are linear in the multiplicity vector
$k$. This subsection shows how to sum such phases, weighted by a
polynomial, over the integer points of a polytope to
inverse-polynomial absolute accuracy without enumerating them.
Informally, we introduce one extra integer variable that records the
integer part of the phase, replace the exponential of the remaining
fractional phase by a Taylor polynomial, and sum the resulting
polynomial exactly by the lattice-point summation of
Section~\ref{sec:fdsum}. The only error is the Taylor error.

Let $\mathcal C$ be a rational polytope described by short
inequalities and contained in a box $c_g\le k_g\le c_g+N$ with short
integers $c_g$ and $N\ge0$ (for multiplicity polytopes $c=0$ and $N=n$), and let $f(k)$ be
a rational polynomial of polylogarithmic degree and coefficient bit
length. Let $\omega_g,\omega_0$ be rational turns of polylogarithmic
bit length. Reduction modulo one then gives an approximation, with
the absolute error stated below, of
\begin{equation}\label{mc:phasesum}
      \sum_{k\in\mathcal C\cap\mathbb Z^m}
           f(k)\,\mathrm e(\omega_0+\textstyle\sum_g\omega_g k_g).
\end{equation}

Translating $k$ by $c$ changes only $\omega_0$, $f$ and the (still short) inequalities of $\mathcal C$, so we assume $c=0$.
We reduce the individual turns to $[0,1)$ and clear their common
denominator $q$. We then introduce a single integer variable $z$
subject to
\[
  0\le q(\omega_0+\textstyle\sum_g\omega_g k_g-z)\le q-1 .
\]
Exactly one such $z$ exists for each $k$, and $0\le z\le(m+1)N$.

On $0\le x\le1$, the real and imaginary parts of $\mathrm e(x)$ admit
rational Taylor polynomials of degree $O(J_1)$, each with error at most
$e^{-3J_1}/\sqrt2$; this follows from the factorial remainder bound
together with a short rational approximation of $\pi$. Their
combination therefore approximates $\mathrm e(x)$ with complex error
at most $e^{-3J_1}$. We apply the fixed-dimensional polynomial
lattice-point summation of Section~\ref{sec:fdsum} to the two
resulting polynomials in $(k,z)$; it evaluates their sums over the
integer points exactly. The computed value of \eqref{mc:phasesum} thus
differs from the true value only by the Taylor error, summed over at
most $(N+1)^m$ integer points. If $|f(k)|\le F_0$ on the polytope, the
absolute error is therefore at most $F_0(N+1)^m e^{-3J_1}$. The work
is polylogarithmic for fixed $m$. Because the polynomial summation is
exact rational arithmetic, cancellation inside the integer-point
computation cannot increase this error.

To bound the bit lengths, let $Q$ be the denominator of the rational
approximation of $2\pi$ and let $t=O(J_1)$ be the Taylor degree. After
substitution of $x=\omega_0+\sum_g\omega_g k_g-z$, the coefficient
denominators of the resulting polynomial divide $(qQ)^t t!$. In fixed dimension there are $(t+1)^{O(m+1)}$ monomials,
and their numerators have bit lengths polynomial in $t+\log(qQ)$
by the multinomial formula. Multiplication by the short polynomial
$f$ preserves these bounds, so exact rational summation handles
cancellation at polylogarithmic bit cost.

\subsection{Computing the characteristic values}\label{rw:pwchar}
It remains to compute, at each external physical frequency, the
characteristic values of the measures $\nu$ to inverse-polynomial
absolute accuracy. We write the Cauchy weight $G$ as a Fourier
integral in an auxiliary variable $y$, and we split the integrand into
an exceptional factor and a core factor. The core factor is computed
by count inversion and the lattice phase sums of
Section~\ref{mc:region}. The integral over $y$ is restricted to an
adaptively located set where the core factor is not negligible. The
work shared between root entries is identified, and the
case of a constant core is treated at the end.

\paragraph{Splitting the integrand.}
In a nonsaturated region use
$G(x)=\frac12\int_{\mathbb R}e^{-|y|}e^{iyx}\,dy$. At external
physical frequency $f$, separate the exceptional factor
\[
 E_f(y)=\E\!\left[
             \mathrm e(f V_E)e^{-(1-iy)\gamma\chi}\right].
\]
It is bounded by one, also on $|\Im y|\le1/2$. The core factor is
\[
 C_f(y)=\E_{Q^{(l)}}\!\left[
 \one_{\mathcal C}(k)W_l(k)\,\mathrm e(f S_{\mathrm{core}})
                              e^{iy\gamma(U-t_0)}\right].
\]
Let $\mathcal J=C_{13,m}\lceil\log R\rceil$, with the multiplier
sufficiently large for all inverse-polynomial absolute tolerances.
Truncate the integral to $|y|\le\mathcal J$ and treat its two
halves separately.

\paragraph{Count inversion.}
For each group choose a power-of-two period
$\Pi_g\asymp\sqrt{V_g\mathcal J}+\mathcal J$ that contains its
typical count interval without collisions (this interval contains the
box $|k_g-\mu_g|\le C_5\sqrt{V_gJ}$ of Subsection~\ref{rw:pwtilt}
once $C_{13,m}$ is large in terms of $C_5$), with a sufficiently
large absolute width. Aliases at any retained count have total
probability $O(e^{-4\mathcal J})$ by Bernstein's inequality, since
any different count mapping to it lies far outside the mean
interval. This remains an absolute error bound for every real
characteristic phase. Use discrete inversion of those counts, that is,
\eqref{mc:rankchar} with biases $q_i$, period $\Pi_g$, $\mathcal J$ in place of $J_1$
and site phases $2\pi fw_i+y\gamma d_i$.
For the residual phases set
\[
  B_g(y)=\sum_{i\in g} q_i(1-q_i)
      \exp\!\left(2\pi if w_i+iy\gamma d_i\right),
   \qquad V'_g=\sum_{i\in g}q_i(1-q_i).
\]
At a fixed $y$, retain only $O(\mathcal J)$ count angles per group
around $-\arg B_g(y)$, as in \eqref{sp:cardext}, or discard the
whole row when $V'_g-|B_g(y)|$ is large. The inversion is normalized
by $1/\prod \Pi_g$, so the omitted mass per count is
$O(e^{-2\mathcal J})$. The shift $a_g k_g$ is applied after inversion.

For each retained tuple, the sum over $k\in\mathcal C\cap\mathbb Z^m$
is $W_l(k)$ times a character. Its turns are obtained by adding
$y\gamma a_g/(2\pi)$ to the inverse count phases, together with an
external constant phase $-y\gamma t_0/(2\pi)$. Taylor-approximate the
exponential of the affine $\log W_l(k)$ on its bounded range, and
apply the phase lift \eqref{mc:phasesum}.

\paragraph{Precision budget.}
On each retained region $\mathcal C$, Subsection~\ref{rw:pwrecover}
gives $|\log W_l(k)|\le B_m\log R$ for a fixed $B_m$.
Put $\varsigma=e^{-2\mathcal J}$. Each one-group coefficient in the count
inversion has absolute error at most $3\varsigma$, including numerical
evaluation, since \eqref{sp:cardext} with $\mathcal J$ in place of $J$
gives $O(e^{-3\mathcal J})$. Telescoping the product of $m$ coefficients of exact
modulus at most one gives error at most $3m\varsigma(1+3\varsigma)^{m-1}$ per
count vector, before multiplication by $W_l(k)$.

At a real node, round the turns $y\gamma a_g/(2\pi)$ and
$-y\gamma t_0/(2\pi)$ and the $m+1$ coefficients of the affine
function $\log W_l(k)$ to dyadics with absolute error at most
$\varsigma/(1+mn)$ each, using $O_m(\mathcal J+\log(n+1))$ fractional
bits. Since $0\le k_g\le n$, the total turn and the affine argument each
change by at most $\varsigma$. Thus the character
changes by at most $2\pi\varsigma$, and the scalar exponential by at most
$2R^{B_m}\varsigma$ on $\mathcal C$. Approximate the exponential of the
rounded affine argument by a rational Taylor polynomial $P_W$ with
uniform absolute error at most $\varsigma$ on
$[-B_m\log R-1,B_m\log R+1]$. The factorial remainder gives degree
$O_m(\mathcal J)$ and $|P_W(k)|\le 2R^{B_m}$ on $\mathcal C$.
In \eqref{mc:phasesum} use the unit-phase Taylor approximation with
complex error at most $\varsigma$, also of degree $O(\mathcal J)$.
Its error after multiplication by $P_W(k)$ is at most $2R^{B_m}\varsigma$
per point. There are at most $R^m$ points and $O_m((1+\mathcal J)^m)$
retained tuples with normalized coefficients of modulus at most one.
Adding these errors to the count-inversion error therefore gives
$O_m((1+\mathcal J)^m R^{m+B_m}e^{-2\mathcal J})$ at each node.
In a saturated region $y=0$ alone suffices. The precision multiplier
may depend on $m$, while $C_{11}$ in \eqref{pw:sumprice} is absolute.

\paragraph{What is shared between roots.}
Fix a mesh tilt $l$, so that the biases $q_i$ and
$\gamma=1+l/\varrho$ are fixed. The root representative determines
$t_0=\lambda(T-b')$, which enters the core factor through the phase
$-y\gamma t_0/(2\pi)$, the polytope $\mathcal C$ and the weight
$W_l$. At each frequency $f$ the damping function $D_f(y)$ below
depends only on $l$ and the core data, so all roots use the same
quadrature nodes for this $(l,f)$. At each $(l,f,y)$ we compute the
group sums $B_g(y)$, their power sums and the inverse count
coefficients once for all roots. Each root and region then uses its
own weighted lattice sum \eqref{mc:phasesum}. The common mesh has
$O(1+\sqrt J)$ tilts, already counted in the region bound below.
The total cost of the root-dependent part is therefore at
most
\[
 (\#\text{roots})\cdot(\#\text{regions})\cdot(\#f)\cdot(\#y)
 \cdot\mathcal J^{O(m)}\cdot\operatorname{poly}(\mathcal J)
 \le (K+1)^{O(1)}J^{O_m(1)}\cdot \varepsilon'^{-1} ,
\]
since the root list of a coarse tree has $O((K+1)^4J)$ entries per
state and there are $O((K+1)^2)$ states (by the list-length count in
the proof of Lemma~\ref{lem:basegeneral}), there are $J^{O_m(1)}$
regions, nodes and retained tuples, and there are $O(\varepsilon'^{-1}J^{O(1)})$
external frequencies. With $K$
polylogarithmic this is $\varepsilon'^{-1}\log^{O_m(1)}R$, which is within the stated
bound by the choice of $\beta$ in Subsection~\ref{rw:pwprec}.

\paragraph{Locating the significant range of $y$.}
To control the cost of the integral, consider the nonnegative function
\[
  D_f(y)=\sum_g\frac{(V'_g)^2-|B_g(y)|^2}{2V'_g}.
\]
It estimates the minimum count-angle damping to within a factor
of two. Put $L_1=\max(1,2\gamma\max|d_i|)=O(J+1)$ and
$S_1^2=\sum_g\sum_{i\in g}q_i(1-q_i)
(\gamma d_i-\overline{\gamma d}_g)^2\asymp(\gamma \varrho)^2$,
with group variance-weighted means for the current $Q^{(l)}$.
The pair-difference expansion, now with external phase differences,
gives $D_f''\ge S_1^2-L_1^2 D_f$ and
$|D_f'''|\le L_1 S_1^2$. Partition each half of the truncated
range into cells of length at most $1/(10L_1)$, using
$O(1+\mathcal J L_1)$ cells. We distinguish a large-variance and a
small-variance branch.

\emph{Large variance, $S_1^2>C_{14}L_1^2\mathcal J$.} Then a cell
containing any point of $D_f\le 3\mathcal J$ has $D_f''\ge0.8 S_1^2$.
Indeed, $C_{14}\ge30$ gives $D_f''\ge0.9S_1^2$ at the low point, and
the third derivative bound permits a decrease of at most $0.1S_1^2$
across the cell. At the constrained minimum $y_*$, the first-order term
in the Taylor lower bound is nonnegative, even at a cell endpoint.
Thus $D_f(y)\ge0.4S_1^2|y-y_*|^2$, and every low point lies within
$\sqrt{7.5\mathcal J}/S_1$ of $y_*$. We locate $y_*$ by bisecting the
cell. At each step, estimate $D_f'$ at a midpoint to absolute error
$\omega=S_1\sqrt{\mathcal J}/10$. If the estimated magnitude exceeds $2\omega$, its
sign selects the half containing the constrained minimum. Otherwise
strong convexity gives distance at most $3\omega/(0.8S_1^2)$ to that
minimum. Stop also when the bracket width is at most
$\sqrt{\mathcal J}/S_1$. Enlarging the computed neighborhood by the
preceding radius covers all low points, in $O(\log(2+S_1/L_1))$ steps
per cell. Cells with no low point may return any such interval.
Outside the retained parts, the inversion just described bounds
$|C_f(y)|$ by $R^{O_m(1)}e^{-2\mathcal J}$.

\emph{Small variance.} In the small-variance branch, keep each full
cell.

\paragraph{Quadrature.}
On a complex strip of width $h_0=c/(1+L_1+\gamma \varrho)$, the true core
integrand has magnitude at most $R^{O_m(1)}$. Indeed, after
restricting to $\mathcal C$ the nonconstant count weight is bounded by
$R^{O_m(1)}$, and $|t_0-a\cdot k-\E_{Q^{(l)}}Y|=O(\varrho)$ in a
nonsaturated region by its construction. The centered residual has
log moment $O(h_0^2\gamma^2 \varrho^2)$ on that strip, since
$h_0\gamma\max|d_i|$ is bounded. Dropping the count restriction after
these absolute estimates gives the claimed strip bound. Use panels of
length at most $h_0/4$ inside retained cells. There are
$O(1+\sqrt{\mathcal J})$ of them per cell in both variance regimes.
Use $N_q=\lceil\mathcal J/4\rceil$ nodes per panel in the quadrature of
Section~\ref{sp:section}. The panels have total length at most
$2\mathcal J$, so their interpolation error is
$O(\mathcal J R^{O_m(1)}4^{-N_q})$ and their total absolute quadrature
weight is at most $2\mathcal J(2e)^{N_q}=O(\mathcal J e^{\mathcal J/2})$.
The latter factor preserves exponential decay of the node error above.
Evaluate the exceptional factor and the remaining node arithmetic
to absolute error at most $e^{-2\mathcal J}$, allowing for the
$R^{B_m}$ bound on the core factor when multiplying the two. Thus a
sufficiently large multiplier in $\mathcal J$ gives any fixed
inverse-polynomial absolute accuracy.

At one cell with center $y_0$, the quadrature nodes $y_f$ used at
frequency $f$ have offsets $y_f-y_0=O(1/L_1)$. For core power sums of degree $\rho=O(\mathcal J)$, the extra
factors in the ordinary site sums are
$e^{ijy_0\gamma d_i}e^{ij(y_f-y_0)\gamma d_i}$ with $|j|\le\rho$.
Taylor-expand the second factor through $O(\mathcal J)$ terms, since
its argument is $O(\mathcal J)$. This separates row and site, with
coefficient amplification at most $e^{O(\mathcal J)}$, using
$d_i/\max|d_i|$ as the expansion variable. The same procedure computes
the group dampings, their derivatives, and the count-angle products by
the entire-function method of Section~\ref{sp:section}.

\paragraph{Biases above one half.}
In the entire-function method, a second-tilt bias $q_i$ can now exceed
$1/2$, so the norm bound used there, which has constant two for
biases at most $1/2$, must be replaced. Write $\theta_i$ for the phase
of site $i$ and $D_f^*=\sum_i q_i(1-q_i)(1-\cos\theta_i)$. Since $u_i\ge0$,
$|\alpha_g|\le1/4$ and $|l z_i|\le1/4$, the odds are at most
$e^{1/2}$. Consequently
\[
 \sum_i|q_i(e^{i\theta_i}-1)|^2
  =2\sum_i\frac{q_i}{1-q_i}\,q_i(1-q_i)(1-\cos\theta_i)
  \le 2e^{1/2}D_f^* .
\]
On rows with $D_f^*\le L+1$ (where $L$ is the damping cutoff of the
entire-function method, not the constant of Lemma~\ref{pw:local}),
the entire-function bound on $|v|=2$ is therefore
$\exp(4e^{1/2}(L+1))$. Increasing the constant in the Taylor degree
gives the required accuracy with the same asymptotic work. The
large-damping test uses the original modulus bound $e^{-D_f^*}$.

For $E_f(y)$ use Lemma~\ref{ex:coarse}(d) with
$\zeta=(1-iy)\gamma\lambda$, whose real part
$\gamma\lambda(1+\Im y)$ is nonnegative for $\gamma\ge0$ and
$|\Im y|\le1/2$. Internal factors for increments $\Delta$ with
$\gamma\lambda\Delta>C_{15,m}J$ have modulus at most
$e^{-C_{15,m}J/2}$; replacing them by zero is an error of that size at
a group, which propagates additively as in the proof of~(d). Raw groups use the same expansion about $y_0$,
now with $|(y_f-y_0)\gamma\lambda\Delta|=O_m(J)$ on the retained
sites. The tree takes only polylogarithmic work per frequency after
evaluation of its raw leaf sums. Consequently all integrals, including
their adaptive bisections, cost $\tO_{m,C}(n+\varepsilon'^{-1})$ in the soft
model.

Suppose the core is constant inside every group ($\varrho=0$). At each root,
use the count box (Bernstein's inequality bounds the mass outside it as
in Subsection~\ref{rw:pwtilt}, since $e^{U-t_0}\le1$ on feasible
configurations) and the linear strip $t_0-DJ\le a\cdot k\le t_0$
under $Q^{(0)}$. Relative to $e^{t_0}P_{\mathrm{core}}$, the positive core counting
measure there has weight $e^{a\cdot k-t_0}$. Its original-weight
characteristic is computed by count inversion and the same
fixed-dimensional phase sum. Multiply by the undamped characteristic
of the exceptional path and apply the physical threshold. Its
normalized mass after the scalars is bounded by \eqref{pw:rootmass}
directly. Counts below the strip cost at most $e^{-DJ}e^{t_0}P_{\mathrm{core}}$;
counts above the strip cannot fit. This also treats an empty core.

\subsection{Finite precision and the strict input term}\label{rw:pwprec}
This subsection completes the proof of Theorem~\ref{pw:main}. We first
explain how all decisions are made with short dyadic numbers, so that
no equality between transcendental numbers is ever tested. We then
choose $\beta$, and show that when $n$ dominates, the input scan can
be organized, as in Section~\ref{sp:strict}, so that it takes
$O_{\delta,C}(n)$ operations.

\paragraph{Finite precision.}
The physical magnitude assumption ensures that all raw weights,
count periods, reciprocals of nonzero residual scales, and
frequencies before phase reduction have polynomial magnitude
in $R$, with constants allowed to depend on $m,C$. Consequently,
one may implement trimming with short positive dyadics $\tilde v_i$
of arbitrarily high fixed inverse-polynomial absolute accuracy.
Use their exact moments, testing a zero group span by its original
endpoints. Integrality of the weights at a positive span, the lower
bound on the retained $v_i$, and the search lower bound on
$\lambda$ give the polynomial lower bound on $\varrho$ also for these
moments. These moments give rational centers exactly; the true
covariance centering equations then hold with arbitrarily small
inverse-polynomial slack. This slack changes $M(0)$, the intercept
estimates and the local comparison by $o(1)$ in normalized units.

Use short dyadic approximations for the residual scale $\varrho$,
the tilt parameters and the cuts. For a positive residual scale, use
the same dyadic $\varrho$ to define $z_i=d_i/\varrho$ and
$\gamma=1+l/\varrho$, with $d_i=u_i-a_g$ for the chosen centers.
The saturated endpoint $l=-\varrho$ then has $\gamma=0$ exactly.
The intercepts, $\mu_g$, $M(l)$ and the endpoints of the strips can
all be used with inverse-polynomial slack. Make a single rational
partition with a fixed assignment at each endpoint. Either adjacent
tilt remains within the absolute margin used in the local comparison.
The typical box and the two outside tail bounds allow constant
buffers. These conventions avoid decisions about equality between
transcendental numbers.

For a nonconstant core, analyze the law $Q=Q^{(l)}$ with exact
logistic probabilities and exact intercepts for the chosen tilt and
centers, so \eqref{pw:exact} holds. Apply Lemma~\ref{pw:local} to the
law $Q_k$ obtained by refitting the intercepts so that
$\E_{Q_k}K_g=k_g$ exactly, as in Subsection~\ref{rw:pwtilt}.
The group intercept changes $\eta_g$ leave
the law conditional on $K=k$ invariant. By the change of measure
and Jensen's inequality,
\[
  Q(K=k)=Q_k(K=k)\,\E_Q e^{\sum_g\eta_g(K_g-k_g)}
     \ge Q_k(K=k)\,e^{-\sum_g\eta_g(k_g-\mu_g)} .
\]
The refit bound $|\eta_g|=O(\sqrt{J/V_g})$ and the count box make
the last exponential at least $e^{-O_m(J)}$. Lemma~\ref{pw:local}
bounds $Q_k(K=k)$ below by an inverse polynomial, so
$d=Q(K=k)\ge R^{-O_m(1)}$ for all retained $k$.

If $\tilde q_i$ are the implemented probabilities, a coordinatewise
coupling bounds the total variation error of the product law by
$\Delta_q=\sum_i|\tilde q_i-q_i|$. For $\Delta_q<d$, the two laws
conditional on $K=k$ differ in total variation by at most
$2\Delta_q/d$, by comparing the numerators and denominators of the
conditional probabilities. The Fourier computation uses unconditioned
expectations with the count restriction, for which the coupling
error is at most twice $\Delta_q$ times the supremum of the integrand.
Those integrands are polynomially bounded in $R$. Taking the
probability approximations sufficiently accurate therefore fits
both comparisons within any prescribed inverse-polynomial budget.

Use short logarithms for scalar factors and use \eqref{pw:prefactor}
to budget absolute errors.
Quadrature weights and phase-lift calculations allow arbitrary
fixed inverse-polynomial accuracy at polylogarithmic cost.

Choose $\beta$ so small that $C_{11}\beta<
\delta/[40(1+\delta)]$. This choice uses the absolute constant
in \eqref{pw:sumprice}. If $n\le \varepsilon^{-(1+\delta/2)}$, the soft
bound already fits the claimed time. Otherwise $R=O(n)$, and
all ordinary site frequencies at a physical scale, including
cycle lengths and powers in the product, are at most
$n^{1-8\eta_*}$ for some fixed $\eta_*=\eta_*(\delta)>0$. The rest of the
proof treats this case.

We describe the shared input preprocessing explicitly. The initial
tilt is found by Lemma~\ref{dt:search}(b). After sorting in a
fixed number of radix passes, all group deletions are from the
ends of sorted intervals, or remove entire groups. Only $m$ pairs of
endpoints need be tried at a step. Use running dyadic moments
of orders zero, one and two. The sigmoid and variance for the
initial $u_i\in[0,C_4J]$ are evaluated in constant work per
input, from a table of $O(n^{\eta_*}\log R)$ cells with fixed-degree
Taylor coefficients. Its width $O(n^{-\eta_*})$ gives arbitrarily
high fixed inverse-polynomial accuracy, by choosing the degree
once. Initial sorting, trimming and all these scans take
$O_{\delta,C}(n)$.

\paragraph{Core sites: phase windows and physical bins.}
Use the common phase windows of Section~\ref{sp:strict} for all
external frequencies and scales. For core sites in group $g$, add
one physical coordinate $(u_i-a_g)/A_g\in[-1,1]$, where $A_g$
is its exact maximal deviation; constant groups are stored in
one bin. Form fixed-degree mixed moments of the phase displacement
and the physical displacement within a grid of $O(n^{\eta_*})$ physical
bins. All site multipliers in a cell calculation have derivatives
through any fixed order bounded by a polynomial in $\log R$
times at most a fixed polynomial in $R$. More explicitly, the
logistic log odds have bounded slope contribution $|l|A_g/\varrho$
and baseline slope $A_g=O(J)$. Powers of the bias of degree
$O(\mathcal J)$, bounded powers of $d_i/\max|d_i|$, the fixed
phase $e^{ijy_0\gamma d_i}$, and variance or normalizer factors
are bounded by $O(1)$ or $\log^{O(1)}R$ on complex discs of
radius $1/\log^{O(1)}R$ around real bin locations, increasing
the power in the radius as needed. This follows from
$|y_0|\le\mathcal J$, $\gamma A_g=O(J)$ and the separation of
the real logistic argument from its complex poles. Cauchy's
formula then gives the requisite fixed-order local expansions.
The same physical moments alone evaluate all the intercept searches
and group moments.

\paragraph{Exceptional raw leaves.}
The trees of Lemma~\ref{ex:coarse} need the raw-leaf averages
$a_{I,\ell}$ of part~(d) of that lemma, over polylogarithmically many
rank intervals $I$ of $E$, with minimum weight $b_I$ and cycle lengths
$\ell\le K$. The trimming bounds only $\sum_{i\in E}p_i$, so $|E|$
itself may be polynomial. Even so, each average is a sum over $I$ of a
multiplier that depends on the site only through $w_i$. The phases use
the common phase windows of Section~\ref{sp:strict} on the original
$w_i$; a cycle length merely multiplies the requested frequency by
$\ell$. We separate the sites of $E$ by raw group $I$ and by the dyadic
magnitude band of $\Delta_i=w_i-b_I$, with one more case for $\Delta_i=0$; the
band is found in constant work as in the scan of
Lemma~\ref{dt:search}. In a
nonzero band write $\Delta_i=2^tx_i$ with $1\le x_i<2$, and store
fixed-degree moments of the displacement of $x_i$ within one of
$O(n^{\eta_*})$ equal physical bins, mixed with the phase displacement
moments. The scaled argument $\gamma\lambda\ell2^tx_i$ is either
negligible after damping or $O_m(J)$, as described above. In the first
case the multiplier can be set to zero with any fixed
inverse-polynomial error; a band straddling the numerical cutoff can
retain the entire band with a doubled cutoff. In the second case a
fixed-degree Taylor expansion on a bin gives that precision, by
Cauchy's formula as for the core sites. So all these sums, at all frequencies
and $y$-nodes, come from the same per-site mixed moments that are
accumulated once in the input scan, exactly as for core groups. In
the soft bound one may instead use a nonequispaced transform
\cite{DR} over $E$ at all required frequencies, at cost
$(|E|+\#f)\log^{O(1)}R$ per node.

\paragraph{Total cost.}
Each input contributes to a fixed number of arrays. Their total
length, including physical bins and the polylogarithmic group,
band and case counts, is $n^{1-\eta_*}\log^{O_{m,C}(1)}R$.
All row-dependent Taylor expansions above have at most
$e^{O_m(J)}$ coefficient amplification. Therefore fixed-degree moment
expansions at mesh $O(n^{-\eta_*})$, together with the negative-power
ratio of site frequencies to the phase grid, suffice after
summing over items. The remaining transforms, phase sums and
quadratures require $O_{\delta,C}(n)$ operations, including their
polylogarithmic repetitions. All retained scalar inputs use a
constant number of original or short words in the item scan.
This proves the running time and completes the proof of
Theorem~\ref{pw:main}.

\section{A near-linear algorithm for small inputs}\label{bc:section}
This section proves the following proposition, which settles the
small-input regime of Theorem~\ref{fin:main}.
\begin{proposition}\label{bc:main}
Let $T$, $1\le w_1,\ldots,w_n\le T$ and $0<\varepsilon<1$ be as in
Theorem~\ref{fin:main}. For each fixed $\delta>0$, relative $\varepsilon$
counting at all capacities takes
\[
       O_\delta\bigl(n\log^{C_\delta}(n+2)+\varepsilon^{-(1+\delta)}\bigr)
\]
word operations, with success probability at least $2/3$, for a
constant $C_\delta$. In particular it has running time
$\tO_\delta(n+\varepsilon^{-(1+\delta)})$ without a weight or density promise.
The literal bound \eqref{eq:target} follows for every instance
with $n\le \varepsilon^{-(1+\delta/2)}$.
\end{proposition}
Theorem~\ref{cg:structure} of Section~\ref{rw:hidden} reduces the problem to a core with a
logarithmic-rank representation plus a list of exceptional sites, of
which at most a polylogarithmic number $K$ may be selected. The error
from the exceptional sites must be measured against the feasible
count. To achieve this, we compute, for each of a small number of
residual capacities, a table approximating \emph{all} smaller prefix
counts of the core, with precision relative to the count at that
residual capacity. We then integrate each table against the
corresponding exceptional sums. The three ingredients are: prefix
counts for weights of polylogarithmic bit length
(Lemma~\ref{bc:cdf}, an extension of Theorem~\ref{pw:main}); a
decoding step that returns these tables to the original scale
(Lemma~\ref{bc:decode}); and the combination with the exceptional
sites through the parity trees of Section~\ref{sec:extra}
(Proposition~\ref{bc:mixed}). Lemmas~\ref{bc:cdf} and~\ref{bc:decode}
and Proposition~\ref{bc:mixed} are reused in
Section~\ref{fin:section}.

\subsection{Many prefix counts for short integer weights}
Theorem~\ref{pw:main} assumes that the weight sum satisfies
$W\le R^C$. We first isolate the places where its proof uses this
hypothesis. Lemma~\ref{pw:param} extends the soft-time algorithm to
arbitrary $W$ by increasing the numerical precision and allowing
polynomial dependence on $\log W$. Lemma~\ref{bc:cdf} uses the
polylogarithmic bit-length case to estimate many prefix counts at
once, all with error relative to a single baseline count.

\begin{lemma}[Dependence of Theorem~\ref{pw:main} on $\log W$]\label{pw:param}
Let $R=n+\varepsilon^{-1}+2$, $J=\lceil\log R\rceil$ and $\Lambda=\log(W+2)+J$. In the
soft-time algorithm of Theorem~\ref{pw:main}, fix the group-width
parameter $\beta$ and allow arbitrary $W$. With the precision choices
below, the same error and success guarantees hold. In particular,
$C_\lambda/Z\le R^{C_3}$, \eqref{pw:localprice}, \eqref{pw:sumprice}
with the factor $C_{10}e^{C_{11}\beta J}$, and the alias bound at
$\varepsilon'=\varepsilon e^{-C_{11}\beta J}/C_{12}$ retain their
constants. The deleted occupancy is
$\sum_{i\in E}p_i=O(J\Lambda/\varkappa^2+m(J+m)^{10})$.
The degree cap $K$, tree sizes and states, numbers of regions and
roots, number of physical scales, and working bit lengths are all
bounded by fixed powers of $\Lambda$. The cost is
\[
     \bigl(n+\varepsilon^{-1}e^{C_{11}\beta J}\bigr)\Lambda^{O_{\beta}(1)} .
\]
\end{lemma}
\begin{proof}
We track the dependence on $W$ in the soft-time algorithm of
Section~\ref{pw:section} and its Fourier computations.
\begin{enumerate}[leftmargin=*,label=(\alph*)]
\item \emph{Price.} Lemma~\ref{dt:price} gives
$C_\lambda/Z\le(n+2)^{C_{\mathrm{tilt}}}$ at every weight magnitude.
The weights here need not be original words, so the tilt is found by
the bisection of Lemma~\ref{dt:search}(a), whose oracle accuracy does
not depend on $W$. Each oracle value and $\log C_\lambda$ are sums of
$n$ terms evaluated with $\Lambda^{O(1)}$-bit arithmetic, so the search
costs $n\Lambda^{O(1)}$.
\item \emph{Trimming.} A positive $\varrho^2$ is at least
$\min_i v_i\cdot\lambda^2/4$. Indeed, two distinct integer weights in a
group have $u$-values at least $\lambda$ apart, so one of them deviates
from the center by at least $\lambda/2$. Here $v_i\ge
e^{-C_4J}/4$, and $\lambda\ge2/W^2$ by Lemma~\ref{dt:price}.
Hence $\log(1/\varrho^2)=O(\Lambda)$, while $\varrho^2\le nJ^2C_4^2$. The
telescoping argument then bounds the single-site deleted occupancy by
$O(J\cdot\log(\varrho^2_{\max}/\varrho^2_{\min})/\varkappa^2)=O(J\Lambda/\varkappa^2)$.
Whole-group removals contribute at most $O(m(J+m)^{10})$ to
$\sum_Ep_i$. Thus $K=O(\sum_Ep_i+J)=O_\beta(J\Lambda+J^{10})$,
and the tree and state counts are polynomial in $K+J$ by
Lemma~\ref{lem:basegeneral}.

\item \emph{Scales and accuracy.} Let $L_{\mathrm{sc}}$ be the number
of physical scales at internal accuracy $\varepsilon'$, using
$(K+1)(W+1)$ as a support bound to include exceptional paths with
repeated labels. The dyadic scales of Section~\ref{sp:section} give
$L_{\mathrm{sc}}=O(\Lambda)$. The common random period has range of
length $\Theta(L_{\mathrm{sc}}/\varepsilon')$, and each scale uses
$\varepsilon'^{-1}\Lambda^{O(1)}$ frequencies. Summing
\eqref{sp:alias} gives expected error $O(\varepsilon'\|\nu\|)$.

The Fourier coefficients have bounded one-norm at each scale, and
the summation-by-parts coefficients have total absolute weight at most
two. Evaluate each characteristic value, including the mass at
frequency zero, to absolute error at most $R^{-A}/(1+L_{\mathrm{sc}})$,
in the units of $\nu$ itself. Since $|W_l|\le R^{B_m}$ and the
prefactor bound below give $\|\nu\|\le R^{O_m(1)}$, this needs only
$O_m(\log R)$ further bits of working precision; the exponent $A$ is
fixed after $B_m$ and $C_9$.
Allocate the same order of arithmetic error per scale, including the
rounding of the period using the derivative bound in
Section~\ref{sp:section}. Summing over scales gives deterministic error
$O(R^{-A})$ per measure. There are $O(1+\sqrt J)$ mesh values per root.
By \eqref{pw:prefactor} and \eqref{pw:rootmass}, the sum of the
prefactors $\varphi'B_l$ is at most
$O((1+\sqrt J)R^{C_9}C_\lambda)$. The constant-core case uses
$e^{t_0}P_{\mathrm{core}}$ in place of $B_l$, and in the fair branch the measure
can be normalized to a probability law. The polynomial price bound
in (a) therefore lets us take a fixed $A$ large enough for a total
deterministic error at most $\varepsilon Z/100$.

\item \emph{Quadrature and bit lengths.} In Subsection~\ref{rw:pwchar}
take
\[
 \mathcal J=\left\lceil C_m
       \bigl(J+\log(2+L_{\mathrm{sc}})+\log(K+2)\bigr)\right\rceil.
\]
Use this enlarged $\mathcal J$ also in the dyadic rounding tolerance
$e^{-2\mathcal J}/(1+mn)$ and in both Taylor tolerances
$e^{-2\mathcal J}$ of Subsection~\ref{rw:pwchar}. For the constant core
the scalar affine argument is $a\cdot k-t_0\in[-DJ,0]$ on the retained
strip, so the same estimates apply there. With
$N_q=\lceil\mathcal J/4\rceil$ quadrature nodes per panel, the total
error from count inversion, phase and scalar approximation, integration
tails, interpolation and tree evaluation is at most a fixed polynomial
in $R,K,\mathcal J$ times $e^{-c\mathcal J}$ for a fixed $c>0$.
Choosing $C_m$ sufficiently large in terms of this polynomial and $A$
makes the bound at most $R^{-A}/(1+L_{\mathrm{sc}})$ at every frequency.
When a group has count variance $O(\mathcal J)$, its entire grid
has $O(\mathcal J)$ angles and can be retained. For the exceptional
recursion, use a sufficiently large multiple of $\mathcal J$ as the
cutoff for the scaled damping argument $\gamma\lambda\Delta$.
The retained offset expansions then have arguments $O(\mathcal J)$
and use $O(\mathcal J)$ Taylor terms.

The physical sums and cutoffs have $O(\Lambda)$ bits, and (b) gives
$\varrho\ge 2^{-O(\Lambda)}$ whenever $\varrho>0$. The exact moments of dyadic
variance approximations in Subsection~\ref{rw:pwprec} can therefore
use a fixed power of $\Lambda$ guard bits, chosen large enough
for the divisions by $\varrho$ and the multiplications by physical
increments. Reduce external turns modulo one before evaluating
their phases. The working precision, rational count-polytope data,
and the numbers of cells, nodes and tree entries are all polynomial
in $\Lambda$. Each batched computation processes at most $n$ sites
and $\varepsilon'^{-1}\Lambda^{O(1)}$ frequencies per scale, with
$\Lambda^{O_\beta(1)}$ work per site or frequency over all the
auxiliary calls and scales.
\item \emph{Local and price constants.} The proofs of
Lemma~\ref{pw:local} and of \eqref{pw:sumprice} use only the scaled
weights $u_i\le C_4J$, the group width $\beta J$, and $V_g\ge(J+m)^{10}$.
None of these involves $W$, and the effective form \eqref{pw:effective}
gives the same explicit constants.
\end{enumerate}
Collecting (a)--(e) gives the stated cost.
\end{proof}

For the next lemma, let $F(x)$ denote the number of subsets of a given
list of nonnegative integer weights with total weight at most $x$. In the lemma, $R$ bounds the total number
of items plus $1/\varepsilon$ plus two, up to a fixed constant.

\begin{lemma}\label{bc:cdf}
Suppose the integer weights and a nonnegative baseline cutoff $v$
have at most $\log^C R$ bits, for fixed $C$. Given $L$ integer
cutoffs $0\le x\le v$, one can estimate each $F(x)$ to additive
error $\varepsilon F(v)$ simultaneously with probability $1-R^{-a}$ for
any fixed $a>0$. For every fixed $\theta>0$ the work is
\[
       (n+L+\varepsilon^{-1}R^\theta)\log^{O_{C,\theta,a}(1)}R
\]
when $L$ is polynomial in $R$. The same work up to a
polylogarithmic factor suffices for a polylogarithmic number of
adaptive rounds, with polynomially many queries in total and
$L$ the total number of queries.
\end{lemma}
\begin{proof}
The plan is to run the algorithm of Theorem~\ref{pw:main} once, at the
baseline $v$, and to read off all prefix counts $F(x)$ from the same
characteristic values. Two points need care: the error must be
controlled relative to $F(v)$ for every $x\le v$ (Step~3), and the
$L$ endpoints must be evaluated in additive rather than
multiplicative time (Steps~4 and~5).

\paragraph{Step 1: preprocessing and the dense case.}
Let $z$ be the number of zero weights and retain the $n_v$ sites
with positive weights at most $v$. If $F_{\mathrm{ret}}(x)$ counts
feasible subsets of these sites, then $F(x)=2^z F_{\mathrm{ret}}(x)$
for every $0\le x\le v$. In both branches we compute retained counts
with error relative to $F_{\mathrm{ret}}(v)$ and multiply the answers
by $2^z$. If $n_v=0$, return $2^z$ for every query.

When $v$ is at least half the total retained weight, complementation
gives $F_{\mathrm{ret}}(v)\ge2^{n_v-1}$. Use the fair-law expansion
of Section~\ref{sp:section} on these $n_v$ sites, retaining the
characteristic values once per scale and period. Steps~4--6 apply
with payoff one. Multiplying the resulting prefix probabilities by
$2^{n_v}$ converts additive probability error $\varepsilon/2$ into
count error at most $\varepsilon F_{\mathrm{ret}}(v)$. For the rest
of the proof write $F$ for the retained count and keep the ambient
$n$ in the precision and running-time bounds.

\paragraph{Step 2: running Theorem~\ref{pw:main} at the baseline.}
Otherwise, run the soft (not strictly linear) construction of
Theorem~\ref{pw:main} at capacity $v$, building the ordinary site
arrays of Section~\ref{sp:section} for each query of parameters and
omitting the strictly linear input preprocessing. Since
$W\le n2^{\log^CR}$, Lemma~\ref{pw:param} applies with
$\Lambda=O(\log^CR+J)$. We take $\beta=\min(1,\theta/C_{11})$, so that
$e^{C_{11}\beta J}=O(R^\theta)$ because $J\le\log R+1$. Thus the error,
probability and price bounds of Section~\ref{pw:section} hold with the
constants of that lemma, all sizes, counts and bit lengths listed there
are polylogarithmic in $R$, and all characteristic values at one choice
of period cost $(n+\varepsilon^{-1}R^\theta)\log^{O_{C,\theta}(1)}R$.

\paragraph{Step 3: one baseline controls all prefixes.}
Keep the root and region partitions at $v$ for all $x\le v$.
Every discarded nonnegative path contribution to a prefix at $x$
is bounded by its contribution at $v$. In particular, the bounds
before the parity subtractions in Section~\ref{pw:section} apply
to every such prefix. Reducing their error budgets by any fixed
factor has the same cost. For each retained measure in
\eqref{pw:exact}, replace its payoff by
\[
     h_s\one_{\{s\le x\}},\qquad
     h_s=e^{\alpha(s-v)}(1+\alpha^2(s-v)^2),\quad \alpha=\gamma\lambda .
\]
Here $s$ is the physical integer sum, $\alpha\ge0$, and $h_s$
increases to one on $s\le v$. The resulting identity for each prefix
follows from the same pointwise change of measure. At fixed $x$,
summation by parts writes the prefix as
$h_x F_\nu(x)-\sum_{s=0}^{x-1}(h_{s+1}-h_s)F_\nu(s)$, where
$F_\nu$ is the cumulative of that retained measure; the total absolute
coefficient in this expression is at most two. Therefore the uniform
pointwise expected alias estimate \eqref{sp:alias}, integrated
against the positive measures and summed using \eqref{pw:sumprice},
gives expected absolute error $O(\varepsilon F(v))$ for each specified $x$,
with an arbitrarily small fixed multiplier. Here
and below, deterministic truncation and numerical errors are
allocated in the same units. The case $\alpha=0$ is simply
$F_\nu(x)$.

\paragraph{Step 4: evaluating many endpoints in additive time.}
We now explain why evaluating $L$ endpoints requires work additive
in $L$ and the number of frequencies. Fix a scale $\rho$ and a period
$\vartheta$, and set $z_\ell=\mathrm e(-\ell/(\vartheta\rho))$. For a single frequency,
the contribution of the endpoint is proportional to
\[
        h_x z_\ell^x-\sum_{s=0}^{x-1}(h_{s+1}-h_s)z_\ell^s .
\]
For $\alpha>0$, finite geometric sums of degrees zero through two
write this as a term independent of $x$ plus
$z_\ell^x e^{\alpha(x-v)}$ times a polynomial of degree at most two
in $x$, with coefficients depending on $\ell$. Indeed, write
$z=z_\ell$ and $q=z^{-1}e^{-\alpha}$. The displayed expression equals
$h_0+(1-z^{-1})\sum_{j=1}^x h_j z^j$. Solve for the quadratic
$G$ in $G(j)-qG(j-1)=1+\alpha^2(j-v)^2$ by comparing coefficients of
descending powers. By telescoping, the last sum is
$e^{\alpha(x-v)}z^xG(x)-e^{-\alpha v}G(0)$.
This form avoids large positive exponentials. The constant CDF
term has coefficient $h_0$. At $\alpha=0$ only $z_\ell^x$ remains.
Thus, per scale and retained case, a constant number of evaluations
of trigonometric polynomials at the points $\{-x/(\vartheta\rho)\}$ suffice.
For a polynomial of bandwidth $M$, split each evaluation point
as $(j+d)/B$ on a grid $B\asymp M+1$, with $0\le d<1$.
By FFT, evaluate the polynomials with coefficients multiplied by
$(2\pi i\ell/B)^t/t!$ at all grid points. Then Taylor-sum with
the powers $d^t$ at each requested point. To set the Taylor degree,
let $A_1$ bound the coefficient one-norm of each separated
trigonometric polynomial for a single measure $\nu$, before applying
the root and region prefactor in \eqref{pw:exact}. With $c=2\pi M/B=O(1)$, the error through
degree $t$ at each grid displacement is at most
$A_1 e^c c^{t+1}/(t+1)!$. The remaining factors
$x^j e^{\alpha(x-v)}$, $0\le j\le2$, have magnitude at most
$(1+v)^2$. For a per-term tolerance $0<\eta<1$ it suffices to take
$t=O(1+\log(2+A_1)+\log(1+v)+\log(1/\eta))$ with a sufficiently
large constant. The guard-bit bounds in Step~5 make this degree
polylogarithmic in $R$, so the work is $(L+M)\operatorname{polylog}R$.

\paragraph{Step 5: numerical precision.}
We check that the separated coefficients can be computed with
polylogarithmic precision. In the finite tilt construction, including saturation, a
positive $\alpha$ can be chosen exactly as a positive rational of
polylogarithmic bit length. Since
$|1-z_\ell^{-1}e^{-\alpha}|\ge1-e^{-\alpha}$, their reciprocals cost at most polylogarithmically many
guard bits. The factors $v,x$ have polylogarithmic bit lengths.
The recurrence for $G$ uses at most three such divisions, and there
are $O(M+1)$ separated coefficients per polynomial. Hence
$\log(2+A_1)=\log^{O(1)}R$. Exponentials with sufficiently negative
arguments may be truncated at the working precision.
Use a common stored approximation of each characteristic value in
all endpoint terms. Its own error is amplified by at most two
in the endpoint expression before the geometric sums are separated.
The higher arithmetic precision used to evaluate these sums and
FFTs controls cancellation between their terms.

\paragraph{Step 6: amplification and adaptive rounds.}
Use $O(\log R)$ independent periods and take the median at each query.
The individual constant-probability bounds, amplified and combined by
a union bound over all queries, give the stated guarantee. The
baseline partitions and any tilt-search success event can be prepared
with an inverse-polynomial failure budget. For adaptive queries, use
fresh period repetitions at each round, conditional on the earlier
answers.
\end{proof}

\subsection{Returning quantiles to the original scale}
The core weights produced by Theorem~\ref{cg:structure} are of the
form $b\cdot a_i$ with generators $b$ of a constant number of original
words each, so their physical
values are too long for Lemma~\ref{bc:cdf}. Lemma~\ref{bc:decode}
handles this. It encodes the core by the sign compression of
Section~\ref{sec:fdsign}, which preserves the order of all physical
subset sums, computes approximate quantiles in the short encoded
coordinates with Lemma~\ref{bc:cdf}, and maps them back with a
monotone decoder. The output is an explicit table $\mu_u$ of $O(\varepsilon^{-1})$
point masses whose cumulative approximates every prefix count up to
the cutoff $u$.

\begin{lemma}\label{bc:decode}
Suppose integer weights on the core have a supplied representation
$b\cdot a_i\ge0$ in $d\le\log^C R$ generators, each occupying a
constant number of original words, with
$\log(2+\max_i\|a_i\|_\infty)\le\log^C R$. Given a nonnegative
cutoff $u$ of a constant number of original words, let $F_b(x)$ be their prefix count.
One can construct a positive table $\mu_u$ of $O(\varepsilon^{-1})$ nonnegative
original integer coordinates such that, with probability $1-R^{-a}$,
\[
       |\mu_u(\le x)-F_b(x)|\le \varepsilon F_b(u)
                     \qquad(0\le x\le u).
\]
For each fixed $\theta>0$ the cost is
$(n+\varepsilon^{-1}R^\theta)\log^{O_{C,\theta,a}(1)}R$.
\end{lemma}
\begin{proof}
We encode the core in short coordinates, compute approximate quantiles
there, and decode them monotonically.

Compress the vector $z=(b,u)$ by Section~\ref{sec:fdsign}, in
dimension $d+1$, with norm budget
$D\ge 4(d+1)(n+1)(1+\max_i\|a_i\|_\infty)$ and $D\ge10(d+1)$.
Let $\psi_1,\ldots,\psi_s$ be its rows and $M$ its odd radix. The
compressed vector
$\bar z=\sum_{j=1}^s M^{s-j}\psi_j$ preserves signs within the norm
budget. The encoded core weights are $\bar z\cdot(a_i,0)$, and
they are nonnegative. The encoded capacity is $\bar z\cdot(0,1)$,
with the zero vector in the first $d$ coordinates. Hence the ordering
of all physical subset sums, and their comparisons to $u$, are
preserved, including ties. Since $d$ and $\log D$ are
polylogarithmic, all these encoded integers and their sums have
polylogarithmic bit length.

We need a nondecreasing map back to physical coordinates, defined even
at unattained integer indices. For a vector $t$ of one-norm
at most $D$, the balanced base-$M$ digits of $\bar z\cdot t$
are $k_j=\psi_j\cdot t$. By \eqref{fd:digit}, every attainable
physical value with the extended prefix $k_{\le j}$ belongs to
\[
    [c_j(k_{<j})+\varrho_j(k_j-1/8),\
     c_j(k_{<j})+\varrho_j(k_j+1/8)].
\]
The decoder processes the digits in order. Begin with the interval
$[0,W_b]$ of physical core sums. At each digit, clip both endpoints of
the new interval to the preceding interval, projecting to the
nearest endpoint if outside. The resulting child intervals are ordered
for increasing $k_j$, lie inside the parent, and contain all
attainable physical values with that prefix. Since $z$ belongs to the
span of all the rows, at the last step we clip the exact projected
value $c_{s+1}(k_{\le s})$ to the final interval, and round this value
up to an integer. This defines a nondecreasing decoder on the
range of the balanced digit strings that agrees with every actual
core sum. Extend it constantly outside the range if needed.
In the zero-row case all physical values are zero. By the cost bound of
Section~\ref{sec:fdsign}, each query requires
only polylogarithmically many operations on original words and
short rationals.

Write $V$ for the smaller of the encoded capacity and the total encoded weight,
and $A(x)$ for the encoded CDF truncated at $V$ and normalized by
its total mass $F_b(u)$. This mass is positive. Approximate it to
relative error at most $\varepsilon/100$ by Lemma~\ref{bc:cdf} at $V$.
The same lemma, with larger internal accuracy, supplies normalized
CDF values to absolute error $\zeta=\varepsilon/50$ at all requested
endpoints. Let $m=\lceil 8/\varepsilon\rceil$. For the ranks $j/m$,
$1\le j\le m$, bisect simultaneously in the integer interval
$[-1,V]$ to find indices $v_j$. The endpoint conventions are
$A(-1)=0$ and $A(V)=1$: pin the left sentinel to zero and the right
endpoint to one, even when the approximate total is slightly too
small. At each midpoint, move the upper endpoint left when the
computed value there is at least $j/m$, and otherwise move the lower
endpoint right. These are searches for approximate rank brackets;
note that the estimated CDF values themselves need not be monotone.
On the joint accuracy event, the final adjacent endpoints satisfy
\[
           A(v_j)\ge j/m-\zeta,\qquad A(v_j-1)\le j/m+\zeta.
\]
There are $O(\log(V+2))$ adaptive rounds. The empirical CDF of
the $v_j$, counted with repetitions, differs from $A$ by at most
$\zeta+1/m$ everywhere. Indeed, ranks below $A(x)-\zeta$ have $v_j\le x$,
and ranks above $A(x)+\zeta$ have $v_j>x$, using strict inequalities
at the rank boundaries. The nondecreasing decoder preserves this
uniform CDF bound under pushforward, since the preimage of a
lower interval is a lower interval. Finally, put mass equal to the
estimated $F_b(u)/m$ at each decoded index. The normalization error
and the uniform CDF error together give the claim, including when
some encoded indices found by the bisection are unattained.
\end{proof}

\subsection{Combining the core tables with the exceptional sites}
We now combine the core tables of Lemma~\ref{bc:decode} with the
exceptional sites. The exceptional sums are organized by the two
positive parity trees of Lemma~\ref{lem:base}, coarsened as in
Lemma~\ref{ex:coarse}. Each root entry $\omega$ of these trees
leaves a residual capacity $T-b_\omega$ for the core, and we build one
table per such residual capacity. The key estimate \eqref{bc:budget}
shows that the sum of the core counts at all these residual
capacities, weighted by the entry masses, is at most a constant
multiple of the requested count. Hence errors relative to each
residual count add up to a relative error in the final count.

\begin{proposition}\label{bc:mixed}
Let a core as in Lemma~\ref{bc:decode} and arbitrary other
nonnegative weights, each occupying a constant number of original
words, be supplied. For
$K\le\log^C R$, count all feasible subsets selecting at most
$K$ items outside the core. For fixed $\theta>0$ the running time
for relative $\varepsilon$ accuracy with success at least $0.98$ is
$(n+\varepsilon^{-1}R^\theta)\log^{O_{C,\theta}(1)}R$ on the retained input.
\end{proposition}
\begin{proof}
Denote the requested count by $Z_K$.

For the exceptional list, build the two positive parity trees of
Lemma~\ref{lem:base} at degree $K$, keeping track of their difference.
By the repeated-label exchange \eqref{eq:cancel}, their positive sum of
cumulatives through each fixed degree is at most twice the true
exceptional cumulative. Coarsen their vertex lists with the downwards
representatives and cumulative spacing $1/(16\kappa)$ of
Lemma~\ref{ex:coarse}. Here $\kappa=O((K+1)^2)$ is the number of
vertices in each tree, increased to cover both trees if necessary.
For a root entry $\omega$, write $b_\omega$ for its lower
representative and $\varphi_\omega$ for its mass, including the common
slot normalization. Conditional on the entry, the true exceptional
weight $Y_\omega$ is at least $b_\omega$. The coarse construction
preserves path masses. Positive convolution preserves the cumulative
inequalities in \eqref{eq:jump}. The sum of the two coarse cumulatives
is therefore at most $2(1+1/(16\kappa))^\kappa\le2e^{1/16}$ times the
true exceptional cumulative at every cutoff. Integrating against the
core counting measure gives
\begin{equation}\label{bc:budget}
   \sum_{\omega} \varphi_\omega F_b(T-b_\omega)\le 2 e^{1/16} Z_K .
\end{equation}
The sum includes the entries of both signs, with positive masses.
Entries with $b_\omega>T$ contribute zero. The number of remaining
entries, all coarse state and transition counts, and the logarithms of
the integer mass ranges before slot normalization are
polylogarithmic.

Write $u_\omega=T-b_\omega$ for a remaining entry and apply
Lemma~\ref{bc:decode} there at accuracy $\eta=\varepsilon/C_1$.
Let $z_\omega=\varphi_\omega\E F_b(T-Y_\omega)$, and let
$\widetilde z_\omega$ be the same expression using the positive
table $\mu_{u_\omega}$, with both prefix functions extended by zero at
negative cutoffs. Since $Y_\omega\ge b_\omega$, the uniform prefix
guarantee gives
\[
 |\widetilde z_\omega-z_\omega|
     \le\eta\varphi_\omega F_b(u_\omega),\qquad
 0\le\widetilde z_\omega
     \le(1+\eta)\varphi_\omega F_b(u_\omega).
\]

For clarity, we describe how to evaluate an entry against its own
base table within the claimed time. Augment each exception-tree
state by its stored \emph{coarse group index}. At a raw leaf,
partition its original choices by that group's recorded membership.
At an internal vertex, the two coarse child indices and the
original transition determine the parent's coarse group.
Retain only the state of $\omega$ at the root. This positive
finite-state tree now sums the true original leaf weights on
exactly the paths ending at $\omega$. Normalization by the
common slot divisor is performed only at the end. The numbers of
vertices, states per vertex, and transition types are still
polylogarithmic, and the integer path masses have the same bounds as
before. In particular, there is no conditioning on a rare path by
rejection.

Now apply the positive-tree computation in the proof of
Lemma~\ref{lem:base} to this tree and the base table $\mu_{u_\omega}$.
Specifically, the fine CDF groups at each vertex and state have
relative spacing of order $\eta^{1/2}/\kappa$, and the sum of the input
child-pair list sizes is $O(\eta^{-1}\operatorname{polylog}R)$. Estimate each
compression change as in that proof, using the coarse outside tree and
a draw from the base conditional on its allowed prefix. With a
predetermined root state, positive convolution and addition over
the allowed transitions give exactly the same cumulative bounds
at each partial compression. The telescoping probability for a
drawn path still includes all its original transition multipliers.
This gives a relative $\eta$ estimate $\widehat z_\omega$ of
$\widetilde z_\omega$, with exact zero returned if $\widetilde z_\omega=0$,
at cost $(n+\varepsilon^{-1})\operatorname{polylog}R$ per entry once
its table is given.

Let $s_\omega\in\{-1,1\}$ be the parity sign. The generating-function
identity gives $Z_K=\sum_\omega s_\omega z_\omega$. On joint success
of the table and root estimates, the triangle inequality and
\eqref{bc:budget} give
\begin{equation}\label{bc:rooterror}
 \left|\sum_\omega s_\omega\widehat z_\omega-Z_K\right|
 \le \eta(2+\eta)\sum_\omega\varphi_\omega F_b(u_\omega)
 \le 2e^{1/16}\eta(2+\eta)Z_K.
\end{equation}
To compute the signed sum, accumulate the $N$ root estimates in two
sums, one per parity, and subtract. Use relative roundoff at most
$\tau=\varepsilon/[100(N+1)]$ at each addition and at the subtraction.
On joint success the estimates are nonnegative, and \eqref{bc:budget}
and the bound on $\widetilde z_\omega$ give
$\widehat\Sigma:=\sum_\omega\widehat z_\omega\le 2e^{1/16}(1+\eta)^2Z_K$.
In each parity sum the accumulated relative error is at most
$(1+\tau)^N-1$, and the subtraction adds absolute error at most
$\tau(1+\tau)^N \widehat\Sigma$. Thus, since $\eta\le1/2$, the total arithmetic error $E_{\rm sum}$ satisfies
\begin{equation}\label{bc:sumround}
       |E_{\rm sum}|\le\bigl((1+\tau)^{N+1}-1\bigr)\widehat\Sigma
       \le 2(N+1)\tau \widehat\Sigma\le \varepsilon Z_K/10 .
\end{equation}
This takes $O(\log((N+1)/\varepsilon))$
significand bits. When the exponent gap exceeds this precision by
two bits, discarding the smaller term of the signed operation already
meets the relative roundoff bound.

The budget \eqref{bc:budget} and the per-entry tree estimate hold at
every degree cap $K$. For the polylogarithmic
cap here, take inverse-polynomial failure probabilities for the tables
and amplify the root estimates by medians. A union bound gives joint
success probability at least $0.98$.
Taking $C_1$ sufficiently large proves the proposition.
\end{proof}

We can now assemble the proof of the main result of this section.

\begin{proof}[Proof of Proposition~\ref{bc:main}]
We first check correctness and then the running time.

Handle the trivial case of Section~\ref{s48:model}.
Use $R=n+\lceil 1/\varepsilon\rceil+2$ with a fixed smaller internal error.
Theorem~\ref{cg:structure} supplies the data needed by
Proposition~\ref{bc:mixed}. The one-sided loss of the former is at most
$\varepsilon/100$, and the relative error of the latter in the truncated
count can be made at most $\varepsilon/4$. Their composition gives the desired relative error. By the union
bound over the success events ($0.97$ and $0.98$), the success probability is at least $2/3$.

For each fixed $\theta>0$, the work is
$(n+\varepsilon^{-1}R^\theta)\log^{A_\theta}R$. To obtain the stated bound, choose
$0<\theta<\delta/(8(1+\delta))$. If $n\le \varepsilon^{-(1+\delta/2)}$, both
terms, including all logarithms, are $O_\delta(\varepsilon^{-(1+\delta)})$.
Otherwise $R=O(n)$ and $\varepsilon^{-1}\le n^{1/(1+\delta/2)}$, so the second
term, including its logarithms, is $O_\delta(n)$. This leaves
$O_\delta(n\log^{C_\delta}(n+2))$ for the first term and proves
the proposition. \end{proof}

\section{Linear input time}\label{rw:linear}
This section removes the polylogarithmic factor on $n$ from
Proposition~\ref{bc:main} in the large-input regime.
Subsection~\ref{lcg:section} learns a fixed-rank core in linear input
time, and Subsection~\ref{fin:section} assembles the proof of
Theorem~\ref{fin:main}.

\subsection{Learning a fixed-rank core in linear input time}\label{lcg:section}
This subsection proves Theorem~\ref{lcg:main}, a version of the
structural reduction of Theorem~\ref{cg:structure} whose input scan
is strictly linear. The logarithmic rank in
Theorem~\ref{cg:structure} is affordable when every site operation
may cost a power of a logarithm, but not when the scan of the input
must be strictly linear. We therefore work with a much smaller pilot
sample, search for a coarser admissible rounding mesh, and recover
fixed-dimensional integer coordinates by direct rounding. The price
for the fixed dimension is a larger exceptional degree, a small power
$K=O_a(n^a)$ instead of a polylogarithm.

The ingredients are the following. The existence of a hidden
progression of fixed rank comes from Lemma~\ref{iv:weighted} and
Proposition~\ref{iv:inner}. The pilot approximations use the pivot
step of the proof of Theorem~\ref{cg:structure}. A fixed-rank
progression is recognized on the pilot by Lemma~\ref{lcg:recognize}
below, which is based on the sign compression of
Section~\ref{sec:fdsign} and LLL. The full input is then classified with
the projection rows of Section~\ref{sec:fdsign}. Theorem~\ref{lcg:main} is
the structural input to the proof of Theorem~\ref{fin:main} in
Section~\ref{fin:section}.

\begin{theorem}\label{lcg:main}
Fix $0<a<1$ and suppose $2\le \varepsilon^{-1}\le n$. In $O_a(n)$ original-word
operations one can return a rounded integer instance, a partition
into a core and an exceptional list $E$, integer generators and
coordinates for the core, and a cap $K=O_a(n^a)$. The core has
rank $O_a(1)$ and its coordinates are bounded by $n^{O_a(1)}$.
With probability at least $0.95$, the returned feasible subsets
using at most $K$ exceptional items constitute a subfamily of the
original feasible subsets of size at least $(1-\varepsilon/10)Z$.
Each generator, each exceptional weight, and the capacity occupy
a constant number of original words, with the constant allowed to depend on $a$.
The returned weight of each exceptional site $i$ is $c\lceil Mw_i/g\rceil$
for positive integers $c,M,g$, returned by the algorithm and each
occupying a constant number of original words.
\end{theorem}

\subsubsection{Recognizing a small progression on a sample}
The following lemma is the recognition step used on the pilot sample.
Under the promise that some given integers are the values of a linear
form at short integer vectors in a space of fixed dimension $t$, it
finds an explicit representation with at most $t$ generators taken from
the input and with polynomially bounded coordinates. Its cost per value
is polylogarithmic.

\begin{lemma}\label{lcg:recognize}
Fix $t,C$. Let $x_1,\ldots,x_m$ be integers occupying a fixed number
of original words each, with $m\le n$. Suppose a rational linear form
on $\mathbb Q^t$ takes the values
$x_i$ on unknown integer vectors $\alpha_i$ with
$\|\alpha_i\|_\infty\le H=n^C$. In
$(m+1)\log^{O_{t,C}(1)}(n+2)$ original-word operations one can find
at most $t$ input values $b_j$, one positive integer $\mu$, and
integer vectors $u_i$ with $\mu x_i=b\cdot u_i$ for every $i$.
The denominator and coordinates are bounded by $n^{O_{t,C}(1)}$.
Without the promise the same procedure either returns a representation
within these bounds or reports failure.
\end{lemma}
\begin{proof}
The proof has three steps. First we build a test for short integer
relations among at most $t+1$ values. Then we use it to grow a basis
of input values incrementally, recording rational coordinates of each
$x_i$. Finally we replace the rational coordinates by integer
coordinates over a single short denominator.

We first describe a relation test on $h\le t+1$ values $z_j$.
Given a positive budget $B$, the test returns a nonzero relation of
infinity norm at most $c_hB$ whenever a relation of norm at most $B$
exists. Every returned relation is verified exactly. Take the
integer $c_h=2^{h+1}(h+1)$. Sign-compress $z$ as in
Section~\ref{sec:fdsign}, with one-norm budget $h c_hB$. Then apply
LLL to the rank-$h$ row lattice
$\{(v,N\bar z\cdot v):v\in\mathbb Z^h\}\subseteq\mathbb Z^{h+1}$,
with integer $N>c_hB$.
If a relation of norm at most $B$ exists, it gives a lattice vector
showing that the first minimum is at most $\sqrt h B$.
The first reduced vector then has norm at most
$2^{(h-1)/2}\sqrt h B<c_hB$, so its last coordinate vanishes.
Since sign compression preserves signs, the relation transfers back
to $z$. The test returns the first $h$ entries of the first reduced
vector if they are nonzero, within the bound, and orthogonal to $z$.
In all remaining cases it reports failure. At fixed $t$ its time is
polynomial in $\log(B+2)$.

Start with an empty ordered basis. Choose integer thresholds
$B_1>\cdots>B_{t+1}>(t+1)!(H+1)^{t+1}$, decreasing them
by a fixed factor larger than $2\max_{h\le t+1} c_h$ at each step.
When the basis has $j$ values, test the tuple formed by adjoining the
next $x_i$ at budget $B_{j+1}$. When $j>0$, the current basis admits
no nonzero relation of norm at most $B_j$, by the test that last
enlarged it. Hence any returned relation uses $x_i$ with a nonzero
coefficient. In that case store the resulting rational coordinates
of $x_i$ in this basis. If no relation is returned, append $x_i$ to
the basis and pad the previous coordinates by zero. Reject if the new
dimension exceeds $t$. Under the promise this never happens: every
$t+1$ hidden coordinate columns have a nonzero integer dependence of
norm at most $(t+1)!(H+1)^{t+1}$, obtained by taking minors on a
minimally dependent subfamily. This proves the claimed dimension
bound. The padding of older coordinates may be postponed until the
basis is finalized.

It remains to obtain one short denominator. Consider any pass in
which $q_i x_i=b\cdot u_i$, $1\le q_i\le B$, and
$\|u_i\|_\infty\le B$ for every item. Initially $B$ is bounded by
a fixed multiple of $B_1$. In the hidden coordinate space let
$V$ be the span of the vectors
\[
          q_i\alpha_i-\sum_j u_{ij}\alpha(b_j).
\]
The physical linear form vanishes on $V$. We claim that there exists
an integer matrix $\Gamma$ with kernel $V$ over $\mathbb Q$, with at most
$t$ rows and entries bounded by $(2(t+1)^2BH)^{2t}$.
Indeed, choose independent spanning columns of $V$ and a nonsingular
minor on their rows. Express every remaining row of this column
matrix in terms of the selected rows by Cramer's rule, clearing
the same minor in each resulting equation. These equations give
the rows of $\Gamma$. If the span is zero we use the identity matrix, and
if it is everything we use the empty matrix. The entries of all
$\Gamma\alpha_i$ are then bounded by $H_1=tH(2(t+1)^2BH)^{2t}$.

There are now two cases. If the columns $\Gamma\alpha(b_j)$ are
independent, one nonzero minor of their matrix is a common
denominator for the reduced rational coordinates $u_i/q_i$, again by
Cramer's rule. If these columns are dependent, a nonzero relation
among them is also a physical relation among the $b_j$, since the
physical form factors through $\Gamma$. In both cases the absolute value
of the minor, or the norm of the relation, is bounded by
$B'=(t+1)!(H_1+1)^{t+1}$.

The algorithm needs neither the hidden vectors nor $\Gamma$. It computes
the least common multiple of all reduced coordinate denominators,
stopping as soon as it exceeds $B'$. If the least common multiple
stays within $B'$, the algorithm clears it and verifies all the
resulting identities. Otherwise it tests the basis for a relation at
budget $B'$. Under the promise this test succeeds, by the preceding
dichotomy. A nonzero coefficient of the relation is used to eliminate
one basis value, and the relation is substituted into all
coordinates. After a common denominator for the coefficients of
each item is taken, their new numerators and denominators are at
most $2(t+1)c_t B B'$ (as $c_h$ increases with $h$).
We then repeat with that bound in place of $B$. Each elimination
lowers the dimension, so there are at most $t$ such steps. Thus all
budgets are fixed powers of $n$, and all arithmetic on them has
polylogarithmic cost. The procedure rejects on any failed test or
bound. For the empty basis, verify that every $x_i$ is zero and
return the empty coordinate vectors with $\mu=1$ if the check succeeds.
\end{proof}

\subsubsection{Choosing a rounding mesh and scanning the full input}
We now prove Theorem~\ref{lcg:main}. The proof first identifies,
for the analysis only, a scale $\Delta$ at which a fixed-rank hidden
progression exists. A sparse random pilot sample then
yields a polylogarithmic list of candidate meshes $g$, which are
tested with Lemma~\ref{lcg:recognize}. The accepted mesh is used to
round and classify all sites in constant time each. Finally a union
bound over the few possible data-dependent choices controls the
exceptional occupancy.

\begin{proof}[Proof of Theorem~\ref{lcg:main}]
It suffices to prove the theorem for rational $a$, since any smaller
fixed positive rational gives a cap of order at most $n^a$. All
fixed constants below may depend on $a$.

\paragraph{Step 1: the tilt and the density ratio.}
If $T\ge W$, return every site as a core site of weight zero (rank
zero), capacity zero, $E=\emptyset$ and $K=0$; this preserves all
$2^n$ feasible sets. Assume $T<W$ from now on. Use a displaced tilt for $T<W/2$ and the fair law for $T\ge W/2$.
By Lemma~\ref{dt:price}, $C_\lambda/Z\le n^{c_0}$ for an absolute $c_0$
and all sufficiently large $n$. Lemma~\ref{dt:search}(b), with a
fixed $0<\gamma<1$, finds the displaced parameter with an $O(n)$ scan
and $o(n)$ further operations.

\paragraph{Step 2: the hidden scale and the hidden progression.}
Next we define a scale $\Delta$, used only in the analysis, at which
the good sites lie close to a progression of fixed rank. Write
$\rho=n^{-(c_0+5)}$, and let $h_*$ and $\Delta=h_*/[16(n+1)]$ be the
integer $h$ and the mesh of Proposition~\ref{iv:inner} for this $\rho$;
thus $h_*$ is the least positive integer with
$Q_\lambda(|2S-2T-1|\le h_*)\ge\rho$. By part~(b) of that proposition,
some atom of $\sum_i\lceil w_i/\Delta\rceil X_i$ has mass at least
$\rho/(17(n+1))\ge n^{-(c_0+7)}$. We
apply Lemma~\ref{iv:weighted} with $R=n$, with that lemma's atom
exponent set to $c_0+7$, and with its outside exponent $\eta$ set to
$a/8\in(0,1)$, where $a$ is the parameter of this proof. It gives a
hidden progression of fixed rank and polynomial coefficient bounds,
with outside occupancy $O_a(n^{a/8})$. Consequently there are bounded
integer coordinate vectors for the good sites, and a common real
linear form taking these coordinates to numbers within $\Delta$ of
the original weights.

\paragraph{Step 3: drawing the pilot.}
Let $L=2^\ell$ be the smallest power of two at least $n^{a/3}$.
Writing $a/3=p'/q'$ in lowest terms, find $L$ by doubling until
$L^{q'}\ge n^{p'}$, in $O_a(\log n)$ word operations. In one pass, mark item
$i$ when the lowest $\ell$ bits of an independent uniform word are
all zero. The mark count has mean $n/L$. Abort if it exceeds
$\lceil4n/L\rceil$, which by a Chernoff bound has probability at most
$e^{-c n/L}$ for an absolute $c>0$.
Accept each marked candidate independently with probability $2p_i$.
Approximate these probabilities to absolute error $2^{-\beta}$ with
$\beta=O(\log n)$ bits and use dyadic uniforms. The total coupling error
over at most $n$ tests is at most $n2^{-\beta}$. Biases at arguments above a sufficiently large multiple of
$\log n$ can be set to zero within the same budget. The ideal sample
before the cap and rounding has independent rates $2p_i/L$, so it
contains only good sites with probability $1-O_a(n^{a/8}/L)$.

\paragraph{Step 4: approximating the pilot values.}
We now compute short approximate representations of all pilot
values. Run the pivot part of the proof of
Theorem~\ref{cg:structure} on this pilot, with
$q=\lceil\log_2(100/\rho)\rceil$ and $F=2^{3q+5}$, and call its
reported discrepancy $H_0$. The independent-pilot quantile event
there has probability at least $0.995$. When $H_0$ is positive, the
uniform interval bound \eqref{cg:price} for lengths strictly less than $H_0/F$,
now with $L$ independent masks of rate $2p_i/L$, is
$2^{-q}+(1-0.005)^L<\rho$ for sufficiently large $n$. The interval
defining $h_*$ has length $h_*$ on the physical $S$ scale, so it
follows that $H_0/F\le h_*$. Define
\[
 \zeta=\frac{\max(1,H_0)}{64(n+1)F}.
\]
Then $\zeta\le\Delta$. The pivot representations, which have error
at most $H_0$, can be supplemented by the generator $\zeta$, rounding
the remaining discrepancy to its nearest multiple. We have thus
computed approximations $\widehat x$ to all pilot values $x$ with
$|x-\widehat x|\le\zeta$, in a common list $b^*$ of $O(\log n)$
rational generators and with polynomial integer coefficients.
All those generators have a common polynomial denominator.

\paragraph{Step 5: a short list of candidate meshes.}
We next produce polylogarithmically many candidate meshes $g$, at
least one of which satisfies \eqref{lcg:goodmesh} below. On a clean
pilot choose coordinate rows spanning its hidden coordinates. There
are $s$ such rows, for a fixed bound on $s$. The minor argument of
Lemma~\ref{iv:weighted} gives a common $1\le d\le n^\theta$ and
$|c_{xj}|\le n^\theta$ such that the original pilot values and the basis
values $x_j$ satisfy
\[
      U=\max_x\left|d x-\sum_{j=1}^s c_{xj}x_j\right|
                      \le (s+1)n^\theta\Delta.
\]
Here $\theta$ is a fixed constant chosen from the hidden bounds, and the
empty pilot uses $U=0$. Replacing the raw values in such a
combination by their approximations incurs error at most
$B_0\zeta$, where $B_0=(s_{\max}+1)n^\theta$. This observation supplies
a short list of candidate positive integers $g$, as follows.

We always include $\max(1,\lceil\zeta\rceil)$. If $U\le2B_0\zeta$,
this candidate satisfies $U/g\le2B_0$ and $g\le\Delta+1$.
Otherwise, using a maximizer for $U$, some combination of the
approximations has absolute value between $U/2$ and $3U/2$.
It has the form $b^*\cdot z$ with $\|z\|_1\le D_*$, for a
known fixed power $D_*$ of $n$ determined by the preceding bounds.

To locate this value without knowing $z$, run the successive-row
sign compression of Section~\ref{sec:fdsign} on $b^*$ at budget $D_*$,
clearing and later undoing its common denominator. If $\psi_j$ is the
first row with $\psi_j\cdot z\ne0$, then all earlier projected digits
vanish, and \eqref{fd:digit} gives
\[
       \varrho_j/2\le |b^*\cdot z|
                 \le (D_*\|\psi_j\|_\infty+1)\varrho_j .
\]
For each row, start at $\varrho_j/4$ and double until the value
exceeds $4(D_*\|\psi_j\|_\infty+1)\varrho_j$. Add the ceiling of each
value to the list, with minimum one. There are polylogarithmically
many candidates, because the logarithm of the ratio of the endpoints
is polynomial in $\log n$, independently of the physical input
length. When $U>2B_0\zeta$, the quantity $U$ is a positive integer
and the list contains a $g$ between $U/4$ and $4U+1$. In either case,
since $\Delta\ge1/[16(n+1)]$, the list contains a candidate with
\begin{equation}\label{lcg:goodmesh}
           U/g\le n^{\theta'},\qquad g\le n^{\theta'}\Delta
\end{equation}
for a fixed $\theta'$ chosen sufficiently large before the pilot is drawn.
The row scales, even before pruning, are combinations of original
words and integers of polylogarithmic bit length. On the pilot
they can therefore be manipulated at polylogarithmic original-word
cost. Discarding all $g>n^{\theta'+3}(W+1)$ preserves \eqref{lcg:goodmesh},
since $h_*\le 2W+2T+1$ in a nontrivial instance.

\paragraph{Step 6: testing the candidates.}
We test each candidate with Lemma~\ref{lcg:recognize}. Fix
$M=\lceil n^{\theta'+3}\rceil$. For each candidate, test the pilot
integers $v(x)=\lceil Mx/g\rceil$ by Lemma~\ref{lcg:recognize}.
Each occupies a constant number of original words since $g\ge1$,
$x\le T$ and $M$ is polynomial in $n$. At a good candidate,
\[
    \left|d v(x)-\sum_j c_{xj}v(x_j)\right|
                     \le M U/g+(s+1)n^\theta .
\]
The expression inside the absolute value is an integer. Therefore,
adding the generator $1/d$ to the generators $v(x_j)/d$ gives the
promise of the lemma, with fixed rank and with a polynomial
coordinate bound known from the displayed inequalities. These bounds
are used throughout the tests. Take the smallest passing $g$, and
return failure if there is none. On the preceding events
$g/M\le\Delta$.

\paragraph{Step 7: rounding all sites, and the loss at the threshold.}
For this $g$ round \emph{all} sites to $v_i=\lceil Mw_i/g\rceil$ and
use the cutoff $V=\lfloor M(2T+1)/(2g)\rfloor$. This is the rounding
of Proposition~\ref{iv:inner}(a) at the mesh $\Delta'=g/M\le\Delta$.
Hence the rounded instance keeps only feasible subsets, and it misses
at most $C_\lambda\rho$ of them. The half unit in the cutoff matters:
without it, subsets with $S=T$ could be lost when $h_*=1$.

\paragraph{Step 8: classifying the full list in constant time per site.}
It remains to decide, for every site, whether it belongs to the
core, and to compute its coordinates. The accepted pilot
representation supplies $\mu,b$ and a fixed power $B$ of $n$ bounding
its integer coefficients. Form the projection rows of
Section~\ref{sec:fdsign} for the fixed-dimensional $b$, with one-norm
budget $D_1\ge\max(1+\dim(b) B,10\dim(b))$. For a site, set $y=\mu v_i$ and
successively compute
\[
       k_j=\operatorname{nint}
          \big((y-c_j(k_{<j}))/\varrho_j\big).
\]
Reject the site if $|k_j|>D_1\|\psi_j\|_\infty$, or if
$|y-c_j(k_{<j})-\varrho_j k_j|>\varrho_j/8$.
At the end, accept exactly when $y=c_{s'+1}(k_{\le s'})$,
where $s'$ is the number of projection rows. By \eqref{fd:digit},
every member of the pilot container passes. Conversely, every
accepted value has the displayed exact linear representation in its
bounded integer digits. In fixed dimension these digits and the Gram
determinants have $O_a(\log n)$ bits. The coefficients of the last
linear form are rational original-word quantities with the common
denominator $\det(\Psi\Psi^\top)$, where $\Psi$ is the matrix of
projection rows. Clearing this denominator and $\mu$ on the entire
instance gives the claimed integer coordinates and generators. Thus
all weights and the capacity are multiplied by
$c=\mu\det(\Psi\Psi^\top)$, and the returned weight of an exceptional
site $i$ is $c\lceil Mw_i/g\rceil$ with $M,g$ from Steps~6--7. Here
$\mu$ and $\det(\Psi\Psi^\top)$ are polynomial in $n$,
$M=\lceil n^{\theta'+3}\rceil$ and $g\le n^{\theta'+3}(W+1)$, so $c$,
$M$ and $g$ occupy a constant number of original words; they are
returned. In the branches that return without rounding, take
$c=M=g=1$. An
empty basis classifies zero exactly. Each operation in this scan
acts on a constant number of original words. In particular, the
division by $\varrho_j$ uses an exact numerator and denominator,
each obtained with short Gram minors or a fixed number of original
weights. Oversized proposed digits are rejected before further use.

\paragraph{Step 9: the exceptional occupancy and the degree cap.}
Finally we bound the outside occupancy for the data-dependent choice
of mesh, and set the cap $K$. The candidate list is determined by at
most $q$ input pivots, by $H_0$, by the fixed tie rules, and by the
row and doubling index. As in Theorem~\ref{cg:structure}, $H_0$ is
either zero or the discrepancy of one input weight against a
polynomially bounded combination of the pivots. Thus, over all
pilots, there are $\exp(\log^{O_a(1)} n)$ possible values of $g$.
For each $g$ the tested pilot container is specified by at most $t$
input indices, the returned $\mu$, and the coefficient bound $B$, all
with polynomial ranges of choices. Every pilot site lies in this
container, and the full scan accepts at least all its members. The
probability that a fixed outside set of occupancy $\tau_{\mathrm{out}}$ is missed by
the pilot is at most $\exp(-2\tau_{\mathrm{out}}/L)$. A union bound therefore shows
that the final exceptional occupancy is at most $n^{a/2}$, except
with probability $o(1)$.

A deterministic cap $K=\lceil C_a n^a\rceil$ then loses at most
$C_\lambda\rho$ in feasible mass: by \eqref{s48tilt:trunc}, applied to
the list $E$ after fixing the pilot that selected it, the loss is at
most $C_\lambda\exp((e-1)n^{a/2}-C_a n^a)\le C_\lambda\rho$ for
sufficiently large $n$.

\paragraph{Step 10: totals.}
Together with the mesh loss of Step~7, the relative loss is at most
$2C_\lambda\rho/Z\le 2n^{-5}\le\varepsilon/10$ for sufficiently large
$n$, since $\varepsilon^{-1}\le n$. The failure events are: more
than $\lceil4n/L\rceil$ marks (Chernoff, $o(1)$); a pilot that is not
contained in the good sites ($O_a(n^{a/8}/L)=o(1)$); the
upper-quantile event of Step~4 ($0.005$); the event that the union of
$L$ masks misses ($(1-0.005)^L<\rho$); the union bound over
containers in Step~9 ($o(1)$); and the finite sampling errors of
Step~3 (allot $0.005$). Their total is at most $0.01+o(1)\le0.05$
for sufficiently large $n$.
Marking candidates costs one linear scan. Evaluating their biases
and processing the pilot costs $O_a(n^{1-a/3}\log^{O_a(1)}n)=O_a(n)$.
The tilt preparation and full classification also take linear time.
Enlarging constants and handling bounded $n$ directly gives the
asserted probability and error for every $n$.
\end{proof}

Section~\ref{fin:section} evaluates the resulting mixed instance
at the larger exceptional degree allowed by Theorem~\ref{lcg:main}.
There the initial input cost is shared among the exceptional states
and all residual capacities.

\subsection{Linear input time at every capacity}\label{fin:section}
This subsection completes the proof of Theorem~\ref{fin:main} in the
regime $n>\varepsilon^{-(1+\delta/2)}$; the complementary regime is covered by
Proposition~\ref{bc:main}. The linear-time reduction of
Section~\ref{lcg:section} leaves two costs to address. First, the
exceptional list now has a large degree cap $K=O_a(n^a)$, and it must
answer rank queries at cutoffs in the returned units after linear
preparation.
Second, different residual capacities require different tilts of the
same fixed-rank core, and the core must not be rescanned for each of
them. We handle the first cost with the rank structure of
Lemma~\ref{fin:ordered} and an explicit count of the exceptional
tree sizes, which extends Proposition~\ref{bc:mixed} to arbitrary
degree. For the second we store moments on intervals common to all
tilts and use a fixed number of random periods for all prefix
queries (Lemma~\ref{fin:baselines}, a shared-baseline version of
Lemma~\ref{bc:cdf}). The final subsection assembles these pieces.

\subsubsection{Access to a large unsorted exceptional list}
The next lemma gives linear-time preparation and
query time $O_\xi(n^\xi+\log(n+2))$, which is small enough for every
query issued by the exceptional trees.

\begin{lemma}\label{fin:ordered}
Fix $\xi>0$ and a constant $0<\tau<1$. In $O_{\xi,\tau}(n)$ word
operations one can organize any list of at most $n$ original-word
integer weights, with preprocessing failure probability at most
$\tau$. On success, an exact rank or an exact selection by rank costs
$O_\xi(n^\xi+\log(n+2))$ word operations. Moreover, given $Q\ge1$
and $0<\zeta<1$ with $Q/\zeta$ at most polynomial in $n$, up to $Q$
uniform draws from specified nonempty rank intervals cost
$O_\xi(n^\xi+\log(n+2))$ word operations each, apart from their
random bits, and all of them are exact except with total probability
at most $\zeta$.
\end{lemma}
\begin{proof}
Lists of bounded length are handled directly. Compute $J_0=\lceil\log_2(n+2)\rceil$ by
doubling. In both cases below, ranks, selections and draws refer to a
fixed total order of the list in which the weights are nondecreasing;
in the second case it is the order by weight and then position, and
the list is not physically sorted.

If $L_{\mathrm{word}}>J_0^4$, then $L_{\mathrm{word}}>\lfloor\log_2 h\rfloor^4$ for the length $h\le n$ of the
list, and the randomized signature sort of Andersson, Hagerup, Nilsson
and Raman~\cite{AHNR} sorts the weights in $O(h)$ expected time.
In their large-word case $L_{\mathrm{word}}>\lfloor\log_2 h\rfloor^4$,
they compute the field widths from the supplied word length in
$O(\log h)$ time and the masks for the single reduction in
$O(\log\log h)$ time. In our model each such mask costs $O(1)$
operations however large $L_{\mathrm{word}}$ is: a pattern repeating a
$p$-bit block $\beta$ is $\beta\cdot\bigl((2^{pq}-1)\operatorname{div}(2^p-1)\bigr)$
with $q=\lfloor L_{\mathrm{word}}/p\rfloor$, and $2^{pq}-1$ is a shift
of the all-ones word. The sorter draws random integers uniformly
from ranges $[0,m)$ with $1\le
m<2^{L_{\mathrm{word}}}$ given at run time. For $m=1$ return zero
directly. For $m\ge2$ use rejection from one random word $r$: with
$q=\lfloor((2^{L_{\mathrm{word}}}-1)-(m-1))/m\rfloor+1=\lfloor
2^{L_{\mathrm{word}}}/m\rfloor$, accept if $\lfloor r/m\rfloor<q$
and return $r-m\lfloor r/m\rfloor$. Every residue has exactly $q$
accepted preimages, so an accepted value is exactly uniform, and a
trial is accepted with probability $mq/2^{L_{\mathrm{word}}}>1/2$.
Each draw therefore costs $O(1)$ expected operations, the expected
running time remains $O(h)$, and the draws are covered by the work cap
below. Keep each weight's position in the list with it. Check in $O(h)$ time that
the output is nondecreasing and, by marking positions in an array,
that it is a permutation of the input; restart if a check fails. Cap
the total work at $C_\tau n$ operations and report failure if the cap
is reached. By Markov's inequality, a suitable constant $C_\tau$
bounds this probability by $\tau$. The sorted array answers ranks by
binary search and selections by indexing.

If $L_{\mathrm{word}}\le J_0^4$, we use Han's deterministic ordered-partition
theorem~\cite{HanPart}: $h$ distinct integer keys are partitioned in
linear time into blocks of size $O(\sqrt h)$, each block lying
entirely below the next. Make the keys distinct by appending the
position in the list in $\lceil\log_2(n+1)\rceil$ low-order bits. For
$b$-bit keys, Han's conservative model uses words of
$b+\lceil\log_2 h\rceil$ bits and the most significant set bit of a
word. We simulate such a word by a constant number of original words.
Arithmetic, shifts and Boolean operations then take $O(1)$
original-word operations, and the most significant set bit of a
multiword value is that of its highest nonzero word plus that word's
offset. On one original word we use the constant-time routine of
Fredman and Willard~\cite[Section 5]{FW}. Its constants are bit
patterns that are periodic or occupy an arithmetic progression of
positions, so each is built by doubling in $O(\log(L_{\mathrm{word}}+1))=O(\log J_0)$
operations. The same holds for any word-length-dependent constant used
by Han's algorithm, and its indirect addressing uses indices of at
most $\lceil\log_2(n+1)\rceil$ bits, which are original words. Thus
the simulation has constant overhead after this preparation.

Apply the theorem inside each large block, for a fixed number of
levels, until the maximum block size is $O_\xi(n^\xi)$. Indeed, the
block size after $j$ levels is at most $C^2n^{2^{-j}}$, for an
absolute $C$ enlarged to cover small blocks. Each level has linear
total length. Store the block maxima and cumulative sizes in block
order. A rank query uses binary search on the maxima and one block
scan. A selection uses binary search on the sizes and linear-time
selection within the one block. Both cost $O(\log(n+2))+O_\xi(n^\xi)$.

A uniform draw from an interval of $h\ge2$ ranks is a uniform
integer in $[0,h)$ followed by one selection. Draw it by rejection:
take $\lceil\log_2 h\rceil$ random bits and retry if the integer is
at least $h$. Each trial is accepted with probability greater than
$1/2$, and an accepted value is exactly uniform. Allow
$k=\lceil\log_2(Q/\zeta)\rceil+1$ trials per draw and report failure
if all $k$ are rejected. Conditional on the preceding history and the
chosen interval, this has probability at most $2^{-k}\le\zeta/(2Q)$.
A union bound over at most $Q$ draws gives failure probability at most
$\zeta$. Since $Q/\zeta$ is polynomial in $n$, generating the rank
costs $O(\log(n+2))$ operations, followed on success by one selection.
For $h=1$ the rank is deterministic.
\end{proof}

Theorem~\ref{lcg:main} returns each exceptional weight in the returned
units, where it may occupy several original words, whereas
Lemma~\ref{fin:ordered} takes single-word keys. We therefore build the
structure on the \emph{original} weights of the exceptional sites,
which are original words, and translate queries and answers. By
Theorem~\ref{lcg:main}, the returned weight of an exceptional site
$i$ is $c\lceil Mw_i/g\rceil$; in its proof, $M$ and $g$ are the
positive integers of Step~7 and $c$ is the common factor cleared on
the entire instance in Step~8.
This is nondecreasing in $w_i$, so the order of
Lemma~\ref{fin:ordered} on the original weights is also a
nondecreasing order of the returned weights, and ranks, selections and
rank intervals refer to the same order in both units. For an integer
cutoff $x$ in the returned units, put $y=\lfloor x/c\rfloor$. For
integers $w$,
\[
 c\lceil Mw/g\rceil\le x \iff \lceil Mw/g\rceil\le y
 \iff Mw\le gy \iff w\le\lfloor gy/M\rfloor .
\]
Hence the rank at $x$ is $0$ if $y<0$, and otherwise the rank in the
original weights at $\min(\lfloor gy/M\rfloor,T)$, which is an
original word; a strict cutoff $x$ is the cutoff $x-1$. A selection or
a uniform draw returns a site $i$, whose returned weight
$c\lceil Mw_i/g\rceil$ is formed on return. Since $c$, $M$, $g$ and
$x$ occupy a constant number of original words, each translation costs
$O(1)$ original-word operations, so the query bound of
Lemma~\ref{fin:ordered} is unchanged.

\paragraph{Size bounds for the exceptional trees.}
Write $J=\lceil\log(n+2)\rceil$. We apply Lemma~\ref{lem:basegeneral}
with the oracle of Lemma~\ref{fin:ordered}, built on the original
exceptional weights and translated as above, so that
$q_E=O_\xi(n^\xi+J)$, and the construction of every exceptional tree
below shares the single linear preparation of that lemma. The quantity
$J=\lceil\log(n+\varepsilon^{-1}+L+2)\rceil$ of Lemma~\ref{lem:basegeneral},
and $J=\lceil\log R\rceil$ of Section~\ref{pw:section} used in
Lemma~\ref{fin:baselines}, agree with the present $J$ up to a constant factor, since
$\varepsilon^{-1}<n$ and $L$ is polynomial in $n$ here.

Proposition~\ref{bc:mixed} is stated for $K\le\log^CR$, but its proof
uses the positive-tree computation of Lemma~\ref{lem:basegeneral} only
through the quantities counted in that proof. Let $\Gamma=(K+J)^{O(1)}$
bound the coarse list length per state. Recording a coarse group
index gives at most $\Gamma$ augmented states per original state and at
most $\Gamma^2$ augmented transitions per original binary transition.
Indeed, the two child indices and the original transition determine
the parent index. These factors are polynomial in $K+J$ with absolute
exponents. At a raw slot, a coarse group is an interval of ranks and
can be queried relative to its first rank.

The extension to arbitrary $K$ also requires accounting for list
access and the sizes of the masses. First, the term $n$ in the per-entry cost
$(n+\varepsilon^{-1})\operatorname{polylog}R$ comes only from access to
the exceptional list. Here every such access is a rank, selection or
draw query to the oracle, so no term proportional to $n$ remains per
entry, and the queries are counted separately below, as in
Lemma~\ref{lem:basegeneral}. Second, the integer path masses, which
involve the factors $1/(\ell^jj!)$ with $j\le K$, now have
$(K+J)^{O(1)}$ bits rather than polylogarithmically many. Arithmetic on
them costs $(K+J)^{O(1)}$ per operation and is absorbed into the
constant $d_0$ below.
There is an absolute constant $d_0\ge d_1$, where $d_1$ is the
constant of Lemma~\ref{lem:basegeneral}, such that the outer coarse
trees have at most $(K+1)^{d_0}J^{d_0}$ root entries, and, for the
fine and correction trees of Proposition~\ref{bc:mixed} and all
coarse roots together, both the work apart from raw-list queries and
the number of those queries are bounded by
\begin{equation}\label{fin:outercost}
       (1+\varepsilon^{-1})(K+1)^{d_0}J^{d_0}
\end{equation}
once the base tables are supplied. This includes the $B$-bit
transition sampling of Lemma~\ref{lem:basegeneral}, with $B=O(J)$.
Charging $q_E$ per query multiplies this bound by at most $1+q_E$.
Here the parameters are at most polynomial in $n$.

\subsubsection{Many baselines on a single polynomial-weight input}
For the core we need prefix-count tables at many baselines, one for
each outer root of the exceptional trees. Calling Lemma~\ref{bc:cdf}
separately at each baseline would charge the input scan once per
baseline. The following lemma shares a single linear scan among up
to $n^d$ baselines. Informally, the expansion of
Section~\ref{pw:section} is run at every baseline, but all
baselines use the same random periods, and all site sums are
reconstructed from one array of local moments stored on pieces of the
sorted input that are common to all baselines. The lemma is applied
in the proof of Theorem~\ref{fin:main} after the core has been
encoded by Lemma~\ref{bc:decode}.

\begin{lemma}\label{fin:baselines}
Fix $0<\sigma<1/2$. There exists $d>0$, depending only on
$\sigma$, with the following property. Suppose $n$ is an upper
bound on the number of sites and $2\le \varepsilon^{-1}\le n^{1-\sigma}$.
One list of nonnegative integer weights of total at most $n^C$
is given, with $C$ fixed, together with at most $n^d$ integer
baseline cutoffs $v$. In $O_{\sigma,C}(n)$ work, with joint
probability at least $0.98$, one can construct for each $v\ge0$
a positive table of $O(\varepsilon^{-1})$ atoms approximating every prefix count
through $v$ to absolute error $\varepsilon F(v)$, with $F$ as before
Lemma~\ref{bc:cdf} for this list. Its atom coordinates may
subsequently be pushed through a supplied monotone decoder, as in
Lemma~\ref{bc:decode}.
\end{lemma}
\begin{proof}
Clip the baselines to the total weight. Zero weights contribute a
common factor. At each $v$ we use the proof of Lemma~\ref{bc:cdf},
keeping only the weights at most $v$ there; we show how to share its
input cost. The argument below uses the ambient $n$ even when an
individual baseline retains few sites. Probability and numerical
budgets relative to $n$ only strengthen the required estimates.

\paragraph{Step 1: a modified deletion cutoff, and transfer of the estimates.}
Choose small fixed $\theta,\beta>0$. In Section~\ref{pw:section},
replace the cutoff for moving a whole group to the inner
exceptional list by $V_g<n^\theta$. For sufficiently large $n$ we have
$n^\theta\ge(J+m)^{10}$, so every group left for the conditional local
estimate meets the original cutoff $V_g\ge(J+m)^{10}$ of
Section~\ref{pw:section}.
All single-site tests
and the group widths $\beta J$ are unchanged. The analysis of
Section~\ref{pw:section} then gives an inner exceptional degree cap
$k'=O_\beta(n^\theta+J^2)$ for each baseline. Indeed, whole-group
deletion removes at most $O_\beta(n^\theta)$ tilted occupancy in total.
For individual deletions that leave $\varrho^2>0$, the decrease in
$\log \varrho^2$ bounds their total occupancy by $O_C(J^2)$, as in
Section~\ref{pw:section}. Here nonzero encoded spans are integral,
and the tilt (Lemma~\ref{dt:price}) and the retained variances have inverse-polynomial lower
bounds at polynomial weight magnitude. If an individual deletion
makes $\varrho^2=0$, it adds at most one to the occupancy and no further
individual deletion is possible. The Chernoff truncation and the
box and strip errors therefore apply. The local bound
\eqref{pw:localprice} holds verbatim, since its proof uses only
$u_i\le C_4J$, the group width $\beta J$ and $V_g\ge(J+m)^{10}$
(Lemma~\ref{pw:param}(e)). The degree cap enters the derivation of
\eqref{pw:sumprice} only through the sum over roots, which
\eqref{pw:rootmass} bounds at every degree. Hence \eqref{pw:sumprice}
holds at each baseline with the same $C_{10}$ and $C_{11}$. Baselines
at least half the retained total are handled by the fair expansion of
Step~1 of Lemma~\ref{bc:cdf}, with no local bound or price.

\paragraph{Step 2: prefix moments for the tilt and normalization sums.}
We now show how every baseline obtains its scalar sums in sublinear
time from one shared preprocessing. Radix-sort the polynomial weights
once, in $O_C(n)$ operations. Prepare prefix sums of their powers
through a fixed degree $D$, to be chosen for any required fixed
inverse-polynomial precision. This initially costs $O_{C,D}(n)$. For
a trial tilt $\lambda$ and a contiguous interval of indices, the sums
of $w^j$ times the logistic, its variance, or
$\log(1+e^{-\lambda w})$, for fixed $j$, are then available in
$O(n^\xi J^{O(1)})$ operations, where $\xi=\sigma/100$. To see this,
partition $0\le\lambda w\le C_1 J$ into cells of width at most
$n^{-\xi}$, and find their index endpoints by binary search. Taylor
expansion to fixed degree on each cell, followed by subtraction of
prefix moments, sums the polynomial exactly up to the rounding of
the coefficients. The functions are analytic, with uniformly bounded
derivatives of every fixed order on this real half-line. Take
$C_1,D$ large enough for the desired error after the factor
$n^{Cj+1}$. Beyond $C_1J$ the decaying terms are negligible. At
$\lambda=0$ the required scalar sums are computed directly. This
implements the displaced tilt bisection and the initial group and
normalization sums for each baseline, including the trial-dependent
capacity normalization.

For the variance weights used in trimming, fix one dyadic piecewise
polynomial in the integer $w$ at the resulting tilt. Its coefficients
in raw powers can be rounded at an arbitrarily large fixed multiple
of $J$ bits, since $w\le n^C$ and the degree is fixed. Exact dyadic
sums of this polynomial and its evaluations at deleted endpoints
then agree throughout the loop, and they approximate the true variances $p_i(1-p_i)$
to any fixed inverse-polynomial tolerance. Use the span and
finite-precision rules of Section~\ref{pw:section}.

\paragraph{Step 3: the cost of trimming.}
We bound the number of individual deletions at each baseline. Apply
the group test whenever it applies, so that before
an individual deletion all current groups have $V_g\ge n^\theta$. In such a
group the largest and smallest $p_i(1-p_i)$ differ by a factor at most
$e^{\beta J}$, since $|(\log(p(1-p)))'|\le1$ in the scaled weight.
As the group has size at most $n$, each such removal costs at least
$c n^{\theta-1-\beta}$ in tilted occupancy. Consequently there are at most
$O_{\beta,C}(n^{1-\theta+\beta}J^2)$ individual removals per baseline. All
of them are endpoints of at most $m=O(1/\beta)$ ordered intervals, so
the three moments are updated by evaluating only the removed
endpoint. An entire group is moved by updating the interval
description. Fair baselines, after the weights larger than $v$ are
removed, need only their retained prefix. Empty inputs are handled
directly. Apart from the single input scan and the radix sort, the
total preparation so far is at most
\[
       n^d\bigl(n^\xi+n^{1-\theta+\beta}\bigr)J^{O_{\beta,C,\xi}(1)} .
\]

\paragraph{Step 4: intervals common to all baselines.}
We now cut the sorted input into pieces that are compatible with the
partitions of all baselines at once. At each baseline the regular
groups are intervals of the sorted input, and the inner exceptional
list is a union of $O(m)$ intervals. Its concatenated sorted ranks
can be translated to ordinary indices in $O(m)$ operations. Construct
its coarse parity tree using the rank rule for slot leaves in Lemma~\ref{lem:basegeneral}. Collect,
from every baseline, all regular interval endpoints, retained input
endpoints, boundaries of raw leaf groups, and both ends of each
light-prefix singleton. Divide the original sorted list at all these
positions. Within each resulting nonempty segment $\Sigma$ use the
coordinate $w-\min\Sigma$. Split its positive values into dyadic
bands $D_\Sigma\le w-\min\Sigma<2D_\Sigma$, and keep zero as a separate band.
Because $W\le n^C$ there are only $O_C(J)$ bands per segment.
Every raw leaf group is a union of the resulting pieces, and each
site belongs to exactly one piece. Contiguous endpoints are found by
binary search, and segment and band indices can be assigned in a
single ordered scan. All root coordinates arising before
normalization have polynomial physical magnitude, since degrees and
encoded weights are polynomial. The root masses may have large
numerators, whose construction costs a fixed polynomial in $k'+J$,
by Steps~3 and~8 of the proof of Lemma~\ref{lem:basegeneral}.

\paragraph{Step 5: one choice of periods for all the tables.}
We use the same random periods at every baseline, and show that a
fixed number of them suffices for all baselines and endpoints
simultaneously. Use a common upper bound on the physical weights of
paths, for example $(n+2)^2(W+1)$. In the expansion of
Section~\ref{sp:section}, use the same scales and random period at
every baseline, with internal accuracy parameter
$\varepsilon'=\varepsilon e^{-C_{11}\beta J}n^{-\xi}/C_{12}$. Choose $H$, the number of scales,
and the Fourier truncations for this common parameter and bound. For
each specified baseline $v$ and integer $x\le v$, the expected
absolute alias error after summing the positive masses is
$O(n^{-\xi}\varepsilon F(v))$, by \eqref{sp:alias}, \eqref{pw:sumprice} and the
variation estimate of Lemma~\ref{bc:cdf}. Truncation and numerical
errors may be set below $\varepsilon F(v)/(100C_3)$ deterministically, for any
fixed $C_3\ge1$. By Markov's inequality, for a fixed pair $(v,x)$ one
period has error above $\varepsilon F(v)/C_3$ with probability
$p_0\le cn^{-\xi}$, where $c$ depends on $C_3$. Take the median of
$2t-1$ independent periods, with the fixed integer
$t>(C+d+2)/\xi$. The median fails at $(v,x)$ only if at least $t$
periods fail there, which has probability at most
$2^{2t-1}p_0^t\le(4c)^tn^{-t\xi}$. A union bound over the at most
$(1+n^C)n^d$ pairs gives total failure probability
$O(n^{C+d-t\xi})=O(n^{-2})<0.01$ for large $n$. Thus the error is at
most $\varepsilon F(v)/C_3$ simultaneously at every integer endpoint of every
baseline. The baseline partitions do not depend on
these periods, and this union bound permits the reuse of the periods
throughout all adaptive quantile bisections. The same convention is
used for the fair CDF expansion.

\paragraph{Step 6: the shared scan.}
We now describe the single scan that stores all site information.
For each period form the common bit windows of
Section~\ref{sp:strict}, with its parameter $\gamma_0=3\xi$.
There are $O_{C,\xi}(1)$ windows, since the number of bits in
the common physical range is $O_C(J)$. In each piece, put the
physical coordinate $(w-\min\Sigma)/D_\Sigma$ into a grid of width at
most $n^{-\xi}$ on $[1,2]$, using a constant coordinate on its
zero band. At each physical bin and phase bucket, store the
mixed moments of the two local displacements through a fixed
degree. The phase displacement is the fractional part defined
in Section~\ref{sp:strict}. Both displacements can use fixed-point
values with $O_{C,\xi,\beta}(J)$ bits at any chosen fixed
inverse-polynomial precision. Every site updates constantly many
moments, for a fixed number of periods and windows, so the full
input preparation takes linear time. Allocation of bins and later
visits to unused bins are charged separately in Step~8.

We verify that the same moments serve each baseline. On an included
regular group with $A_g=\max_i|u_i-a_g|>0$, the scaled coordinate
$(u_i-a_g)/A_g$ is affine in the physical band coordinate, with
slope $\lambda D_\Sigma/A_g$ of magnitude at most two; here both the
segment minimum and the represented weights belong to the same
group. Constant groups have no displacement. Hence the analytic
multiplier and the fixed-degree estimates in the strict scan of
Section~\ref{pw:section} apply on each piece with the same
polynomial losses, for all mesh tilts, intercept trials, variance
sums, and dampings. Intercept and other scalar trials use the
physical moments alone. In a raw exception group of minimum $b_I$,
the real damping argument is $\gamma\lambda\ell(w-b_I)$, with
$\min\Sigma\ge b_I$. Its nonnegative constant part is
$\gamma\lambda\ell(\min\Sigma-b_I)$. If the argument at the lower
edge of the band exceeds a large fixed multiple of $J$, the
whole damped contribution there is negligible. On a retained
band both its constant part and $\gamma\lambda\ell D_\Sigma$ are
$O(J)$, and on the zero band only the constant is needed. Thus
the complex multiplier with $|y|\le O_\beta(J)$ also satisfies
the fixed-order analytic estimates of Section~\ref{pw:section}.
An undamped raw group uses a constant multiplier. The external
phase uses the integer frequency $\ell$ times the external index,
so it uses the same period and windows. These arguments cover the
inner exceptional leaves at arbitrary degrees through $k'$.

\paragraph{Step 7: reconstruction of the site sums.}
We give more detail on how the stored moments are used. For each
needed multiplier, expand in the physical displacement in its bin,
and expand $\mathrm e(h d_i/B_j)$ in the phase displacement of
Section~\ref{sp:strict}. Fold the mixed-moment arrays modulo
$B_j$, evaluate their ordinary FFTs, and sum the pieces belonging
to the required group. This evaluates all the ordinary site sums
on each frequency progression. Any row-dependent offsets used
in the integrals are separated by the polylogarithmic Taylor
expansions of Section~\ref{pw:section}, with the fixed mixed moments
providing their site coefficients. The amplitude bounds there
are fixed powers of $n$ on complex neighborhoods of physical
radius $J^{-O(1)}$. A fixed-order Taylor expansion over width
$n^{-\xi}$ therefore suffices, after increasing its degree to absorb
all fixed polynomial losses. For the phase displacement the same
conclusion follows from the negative-power bound, proved in Step~8,
on the ratio of the largest site frequency to $B_j$.

It remains to control the propagation of these approximations
through the path recursions. At a real integration node, a
conditional path recursion is an average, under normalized
nonnegative transition probabilities, of products of factors in the
closed unit disk $\overline{\mathbb D}$. We use two elementary facts:
for $z,\zeta,z',\zeta'\in\overline{\mathbb D}$ one has
$|z\zeta-z'\zeta'|\le|z-z'|+|\zeta-\zeta'|$, and the nearest-point projection onto
the convex set $\overline{\mathbb D}$ does not increase the distance
to a point of $\overline{\mathbb D}$. So project every approximate
factor onto $\overline{\mathbb D}$ (and keep any further rounding
inside $\overline{\mathbb D}$), and round each transition vector to a
nonnegative vector of total mass at most one, so that every computed
node value again lies in $\overline{\mathbb D}$. Rounding a node's
transition vector with $\ell^1$ error $\tau$ then adds at most $\tau$,
since every child value lies in $\overline{\mathbb D}$. By induction on the tree,
the root error is at most the expected sum of the local errors over
a realized tree, hence at most the largest local error times the
number of vertices, which is polynomial in $k'+J$. So
inverse-polynomial local errors give an inverse-polynomial root
error. The stored integer transition masses can be divided
before rounding to any prescribed $O_{\beta,C,\xi}(J)$-bit precision.

\paragraph{Step 8: choice of the powers, and count of operations.}
We now fix the order in which the constants are chosen, and count
the operations. There is an absolute $D_0$ such that all the
polynomial factors from tree sizes, numbers of raw pieces, inner
roots, transitions and per-root arithmetic, including products of
these factors and the number of baselines, are together bounded by
\[
        n^{D_0(\theta+d)}J^{O_{\beta,C,\xi}(1)} .
\]
This follows from the counts in Steps~3, 4 and~6 of the proof of
Lemma~\ref{lem:basegeneral}, at degree cap $k'$.
In Section~\ref{pw:section} each root and mesh tilt has only
$J^{O_\beta(1)}$ count-angle choices, real integration cells,
and quadrature nodes per external frequency. The count-polytope
algorithm in fixed dimension $m$ receives the physical coordinates,
the short tilt and cut data, and polynomial phases. Their bit
lengths are $O_{\beta,C,\xi}(J^{O(1)})$ even at an inner root,
whose physical lower representative is at most $k'W$.
The long integer \emph{mass} is used only in scalar arithmetic or
in normalized tree transitions; it never enters a count polytope.
Consequently lattice-point summation generates no power of $k'$
with an exponent depending on $m$. The number of pieces can be
multiplied by the number of region, group and root queries by
enlarging the same absolute $D_0$. The number of mesh tilts, typical
count frequencies and quadrature nodes depends on $J,m$, as proved
there, independently of the number of allowed exceptional
selections.

Increase the absolute constants $D_0,C_{11}$ to at least one, and
take for example
\[
 \theta=\frac{\xi}{8(D_0+1)},\qquad d=\theta/4,\qquad
 \beta=\min\{\theta/4,\xi/[8(C_{11}+1)]\}.
\]
These choices satisfy
\[
   D_0(\theta+d)+C_{11}\beta<\xi,\qquad d+\beta<\theta,\qquad \theta<\xi.
\]
Small dyadic perturbations preserve these strict inequalities.
In particular $d$ is fixed before the magnitude exponent $C$ is used.
The degree of a local polynomial, the bit precision, and the
number of independent periods may subsequently depend on $C$.
With these choices the trimming work of Step~3 is $o(n)$ after the
shared initial scan.

Next we bound the frequencies. The greatest external Fourier index
has size $O(\varepsilon'^{-1} J^{O(1)})=O(\varepsilon^{-1} n^\xi e^{C_{11}\beta J}J^{O(1)})$, and
the greatest site frequency is this times a raw slot length and a
power of order $O_\beta(J)$. Here:
\begin{itemize}
\item $\varepsilon^{-1}\le n^{1-\sigma}=n^{1-100\xi}$ by hypothesis;
\item $e^{C_{11}\beta J}\le e^{C_{11}\beta}(2n+2)^{C_{11}\beta}=O(n^{\xi/8})$,
because $J=\lceil\log R\rceil\le\log(2n+2)+1$ and $C_{11}\beta\le\xi/8$;
\item slot lengths are at most $k'=O_\beta(n^\theta+J^2)$, with $\theta<\xi/8$;
\item powers of order $O_\beta(J)$ and all other factors are
$J^{O(1)}=n^{o(1)}$.
\end{itemize}
So the greatest site frequency is at most
$n^{1-100\xi+\xi+\xi/8+\xi/8+o(1)}\le n^{1-24\xi}$ for large $n$.
With the window
parameter $\gamma_0=3\xi$ of Section~\ref{sp:strict}, the phase grid
per window has size at most $n^{1-2\gamma_0}=n^{1-6\xi}$, and every
folded period $B_j$ satisfies $B_j\ge n^{1-4\gamma_0}=n^{1-12\xi}$, as
shown in that section. Hence the ratio of the greatest site frequency
to the smallest period is at most $n^{-12\xi}$, which is the negative
power needed there for fixed-degree phase expansions.

Finally we count the work. The moment arrays have one entry per
physical bin and phase bucket. With the $O(n^\xi)$ physical bins, the
phase grid of size at most $n^{1-6\xi}$ and the full multiplier
$n^{D_0(\theta+d)}J^{O_{\beta,C,\xi}(1)}$, all empty-bin
allocations, moment-array passes, transforms and group
reconstructions cost at most
$n^{1-6\xi+\xi+D_0(\theta+d)}J^{O(1)}\le n^{1-4\xi}J^{O(1)}=o(n)$,
using $D_0(\theta+d)<\xi$. The same bound applies to the
frequency-wise path recursions, count polytope sums, and
quadratures, using the smaller number of frequencies in place of the
phase grid. Tables of trigonometric coefficients may be reused across
endpoints. The nonequispaced endpoint evaluation in
Lemma~\ref{bc:cdf} adds only $O(\varepsilon^{-1}J^{O(1)})$ work per retained case
and bisection round, besides the transforms. Finally, use the
normalized quantile bisection of Lemma~\ref{bc:decode}, including its
estimate of $F(v)$; all $O_C(J)$ rounds fit the same budget.
Choosing the internal constant $C_3$ sufficiently large proves the
stated error, also after the monotone pushforward. Bounded values of
$n$ are absorbed in the constants.
\end{proof}

\subsubsection{Assembly on original input words}
We now prove the main theorem. Small inputs, with
$n\le \varepsilon^{-(1+\delta/2)}$, are handled by Proposition~\ref{bc:main}. For
large inputs, Theorem~\ref{lcg:main} reduces the instance to a
fixed-rank core and an exceptional list of degree cap $O_a(n^a)$. The
exceptional list is prepared with Lemma~\ref{fin:ordered} on its
original weights, the core
is encoded once by Lemma~\ref{bc:decode}, the tables at all residual
capacities are built jointly by Lemma~\ref{fin:baselines}, and the
signed sum over the roots is evaluated with the positive correction
trees of Proposition~\ref{bc:mixed}.

\begin{proof}[Proof of Theorem~\ref{fin:main}]
The proof proceeds in seven steps: preliminary reductions, the
structural reduction, the encoding of the core, the core tables, the
correction trees, the error analysis, and the failure probability.

\paragraph{Step 1: preliminary reductions and the small-input branch.}
It suffices to prove the claim for each positive rational $\delta$.
For a prescribed real exponent, use any smaller fixed positive
rational. Its stronger running-time bound implies the stated one, and
the regime test uses integer powers: if $\delta/2=p/q$ in lowest terms,
the test $n\le\varepsilon^{-(1+\delta/2)}$ is the exact integer
comparison $n^q\le\varepsilon^{-(q+p)}$ on a fixed number of words
(taken after $\varepsilon$ is made dyadic below). Handle the exact case of
Section~\ref{s48:model}. Replace $\varepsilon$ by the largest power of two that is at most
$\varepsilon$; this at most halves it, and now $\varepsilon\le1/2$. If
$n\le \varepsilon^{-(1+\delta/2)}$, Proposition~\ref{bc:main} gives
\eqref{eq:target} with success probability at least $2/3$.

\paragraph{Step 2: parameters for large inputs, and the structural reduction.}
It remains to consider $n>\varepsilon^{-(1+\delta/2)}$. Set
$\sigma=\delta/(4+2\delta)$, so that
$0<\sigma<1/2$ and $\varepsilon^{-1}<n^{1-2\sigma}$.
In particular $C_4/\varepsilon\le n^{1-\sigma}$ for large enough $n$
and every fixed internal precision multiplier $C_4$ chosen below.
Let $n_0=n_0(\delta)$ be a threshold beyond which this and the other
``large $n$'' conditions below hold; it depends only on $\delta$ and
the fixed constants. If $n\le n_0$, then also $\varepsilon^{-1}<n\le n_0$, and the
count is computed exactly by enumerating the $2^{n}\le
2^{n_0}$ subsets, in $O_\delta(1)$ word operations. Hence we may assume $n>n_0$.

Let $d$ be given by Lemma~\ref{fin:baselines}, and put
$\xi=\sigma/100$. Choose $a>0$ so small that
$2d_0 a<\min(d,\xi)$, with $d_0$ from \eqref{fin:outercost}. Apply
Theorem~\ref{lcg:main} at internal accuracy $\varepsilon/C_4$, with a
sufficiently large absolute $C_4$. With probability at least $0.95$,
the fixed-rank core and the exceptional list of cap $K=O_a(n^a)$
give a subcount $Z_K$ within $\varepsilon Z/(10C_4)$ of $Z$. Prepare the
exceptional rank structure of Lemma~\ref{fin:ordered} with
$\tau=\zeta=0.01$ and $Q$ a polynomial bound on the number of draws,
which exists by \eqref{fin:outercost}, on the original weights of the exceptional sites, and
answer every query in the returned units by the translation that
follows the proof of Lemma~\ref{fin:ordered}, at $O(1)$ extra cost.
Then build the outer coarse trees. The number of outer
root entries is at most $n^d$ for large $n$, since it is $O_a(n^{d_0a}J^{d_0})$, where
$J=\lceil\log(n+2)\rceil$ and $d_0a<d$. The number $d$ depends only on
$\sigma$. In particular, the subsequent weight bound $n^{C_a}$ may
depend on $a$ without changing $d$.

\paragraph{Step 3: encoding the core once, and the baselines.}
Write $T'$ for the returned capacity, and use the returned weight
units for all core and exceptional sums below. Encode the core
\emph{once}, using the projection rows and balanced radix of
Lemma~\ref{bc:decode} on its physical generators $b$ alone. The norm
budget covers twice every subset coordinate norm, so all physical
comparisons between core sums, including ties, are preserved. In
fixed dimension with polynomial coordinate bounds, the compressed
weights are nonnegative integers of polynomial total magnitude
$n^{C_a}$, and they take $O_a(1)$ operations per site to form. The
physical decoder $\mathcal D$ agrees with the physical sum on
attained sums and is nondecreasing between them. For a residual
$u=u_\omega=T'-b_\omega\ge0$ at an outer root, use binary search on
the encoded interval from zero to its total to find the greatest
$v$ with $\mathcal D(v)\le u$, or use the total if all values
qualify. The encoded empty sum always qualifies. Thus the physical
and encoded prefix masses agree at these baselines. This step costs
$O_a(J)$ per root. Roots of negative residual are ignored.

\paragraph{Step 4: the core tables.}
Apply Lemma~\ref{fin:baselines} at all these baselines, and push
their quantile tables through $\mathcal D$. On the common success
event each table has additive error at most $\varepsilon F_b(u)/C_4$
at every physical prefix through $u$. The construction is linear
in the original input length, even though the base counts can
differ between roots. Since the guarantee is uniform in the prefix,
it covers every residual capacity queried by paths within a root.

\paragraph{Step 5: the correction trees, and the running time.}
Use the augmented positive correction tree of
Proposition~\ref{bc:mixed} for each outer root and its table,
with the same smaller internal error. Median amplification over the
polynomially many roots gives the stipulated relative errors
for all positive contributions jointly, with probability at least
$0.98$. The degree affects the tree sizes polynomially, as shown
in \eqref{fin:outercost}, and the parity cancellation bound holds
at every integer degree. The work following the rank preparation,
including all rank or selection accesses at their worst-case
cost, is
\[
      O_a\big((1+\varepsilon^{-1})(K+1)^{d_0}J^{O(1)}
                           (n^\xi+J)\big)=O_{\delta}(n).
\]
Indeed, the displayed work is $O_a(n^{1-2\sigma+d_0a+\xi}J^{O(1)})$
for large $n$, and $d_0a+\xi<3\xi/2<\sigma$; this strict inequality
absorbs the logarithms. The initial scan, the linear structural
reduction, and the shared core computation all have linear cost in
this regime.

\paragraph{Step 6: the error.}
Write $\eta=\varepsilon/C_4$. On joint success of the core tables
and root estimates, \eqref{bc:rooterror} bounds the error of the
signed sum by $2e^{1/16}\eta(2+\eta)Z_K$. Accumulating each parity
and subtracting as in \eqref{bc:sumround} adds at most
$\varepsilon Z_K/10$. The number of roots is polynomial in $n$ and
$\varepsilon^{-1}<n$, so these operations use $O(\log n)$ significand
bits. The one-sided structural loss is at most $\eta Z/10$, and
$Z_K\le Z$. Thus the relative error before output rounding is at most
\[
 \frac{\eta}{10}+2e^{1/16}\eta(2+\eta)+\frac{\varepsilon}{10}.
\]
Take $C_4\ge16$, and enlarge it for any primitive that requires
a fixed accuracy multiplier. Since $e^{1/16}\le16/15$ and
$\eta\le1/32$, this bound is at most
$(133/30)\eta+\varepsilon/10<\varepsilon/2$.
Rounding the final positive estimate to relative precision $\varepsilon/4$
gives total relative error at most
$\varepsilon/2+\varepsilon(1+\varepsilon/2)/4<\varepsilon$.

\paragraph{Step 7: the failure probability.}
Conditionally on the success of the earlier stages, the failure
probabilities of the structural reduction, the exceptional-list
preparation, the simultaneous baseline construction, and the jointly
amplified root estimates are bounded by $0.05$, $0.01$, $0.02$ and
$0.02$, respectively. The finite-precision couplings cover the capped
uniform draws in Lemma~\ref{fin:ordered} (budget $\zeta$), the $B$-bit
transition sampling of Lemma~\ref{lem:basegeneral}, and the
short-precision Bernoulli and scalar evaluations of the cited
primitives. Each of these has a prescribed budget obtained with
$O(\log n)$ extra bits; we allot $0.01$ to each of these three
groups, at no change in the asymptotic cost. Thus the total failure
probability is at most $0.13<1/3$. If one of these stages terminates
with failure, the counting algorithm returns $\widehat Z=1$ on that
execution. The corresponding stage budget covers this outcome. The small-input branch uses the
separate guarantee of Proposition~\ref{bc:main}. Small $n$ and the fixed
rounding of the accuracy parameter are absorbed into the constants.
\end{proof}

\section{Discussion}\label{sec:discussion}
This section explains two features of the proof of
Theorem~\ref{fin:main}; nothing in it is used there.
Subsection~\ref{rw:slice-needed} shows that relative counting is as
hard as additive estimation on a uniform cardinality layer, which is
why the proof needs slice estimates. Subsection~\ref{sec:limits}
records two limitations of simpler approaches.

\subsection{Why slice estimates are needed}\label{rw:slice-needed}
The next proposition shows that the general relative target is at
least as hard as an additive estimate for a threshold on a uniform
fixed-cardinality subset, and conversely. The idea is to add a large
common base $B$ to every weight, so that the capacity fixes the
cardinality and the residual weights decide only the critical layer.
\begin{proposition}\label{whole:slicereduction}
Fix $0<\alpha\le\kappa<1/2$ and let
$\alpha n\le k\le\kappa n$. For arbitrary nonnegative integers
$v_i$ and cutoff $u$, put
\[
 p=\binom nk^{-1}\#\{A:|A|=k,\ \textstyle\sum_{i\in A}v_i\le u\}.
\]
An unrestricted algorithm satisfying \eqref{eq:target} for relative
knapsack counting gives an additive $\varepsilon$ estimate of $p$
in the same bound up to constants depending on $\alpha,\kappa,\delta$.
Conversely, an additive algorithm for this slice problem gives the
relative target on its lifted family with the same overhead.
\end{proposition}
\begin{proof}
We lift the slice problem to a knapsack instance, relate the two
error notions, and explain how the known layer counts are computed.

Negative or vacuous cutoffs can be handled directly. Choose
$B=1+u+\sum_i v_i$, assign weights $w_i=B+v_i$, and use capacity
$T=kB+u$. These integers use a constant number of original words.
For the knapsack call discard any lifted weight exceeding $T$,
which cannot occur in a feasible subset. The layer sizes
$D=\binom nk$ and $L=\sum_{j<k}\binom nj$ refer to the original
$n$ sites. All smaller layers fit and all larger ones fail, so
\[
              Z=L+Dp,\qquad
 \frac{k}{n-k+1}\le \frac LD
 \le\frac{k}{n+1-2k}.
\]
The upper bound sums a geometric progression using the successive
downward ratio $k/(n-k+1)$; the first term gives the lower bound.

By the lift above, $Z/D$ is bounded above and below by positive constants
depending only on $\alpha,\kappa$. Estimates of $Z,L,D$ to relative
error $c_{\alpha,\kappa}\varepsilon$ therefore yield additive error
$\varepsilon$ in $(Z-L)/D$, by elementary expansion of the quotient.
Conversely, absolute error $c'_{\alpha,\kappa}\varepsilon$ in $p$
gives relative error $\varepsilon$ in $L+Dp$, including the
arithmetic errors, for suitable constants.

The known layer counts use short significands. Apply
Theorem~\ref{thm:fixed} to $n$ unit weights with capacities $k-1$
and $k$, at accuracy $(1-2\kappa)c_{\alpha,\kappa}\varepsilon$. This
estimates $L$ and $L+D$; their difference estimates $D$ with error at
most $(1-2\kappa)c_{\alpha,\kappa}\varepsilon(D+2L)\le
c_{\alpha,\kappa}\varepsilon D$, since $L/D\le k/(n+1-2k)\le
\kappa/(1-2\kappa)$ by the display above. Both calls cost
$O(n+\log^{O(1)}(n+\varepsilon^{-1}+2))$, which fits in
$O_\delta(n+\varepsilon^{-(1+\delta)})$ for fixed $\delta>0$.
\end{proof}

Proposition~\ref{whole:slicereduction} concerns arbitrary residual
weights. On this family the undecided layer has
$D/Z=\Theta_{\alpha,\kappa}(1)$, even when its residual subset sums
are all distinct. A uniform algorithm must therefore, in particular,
estimate arbitrary threshold probabilities for a uniformly chosen
fixed-cardinality subset at the asserted cost.
Theorem~\ref{sp:additive} establishes the required slice estimate at
polynomial residual weight magnitude and gives a soft bound at
polylogarithmic bit length. Conversely, Theorem~\ref{fin:main} and
Proposition~\ref{whole:slicereduction} together give, for
$\alpha n\le k\le\kappa n$, an additive $\varepsilon$ slice estimate
for arbitrary residual weights, correct with probability at least
$2/3$, in time $O_{\delta,\alpha,\kappa}(n+\varepsilon^{-(1+\delta)})$.

\subsection{Two limitations}\label{sec:limits}
The first remark concerns the inward rounding of
Section~\ref{iv:section}, the second the displaced tilt of
Section~\ref{dt:section}.

\begin{remark}[A limitation of exact weight recovery]\label{prop:barrier}
The inward rounding of Proposition~\ref{iv:inner} certifies the
feasibility of the subsets that it retains, and their count is close
to the original feasible count. The following example shows how many
distinct exact sums can be needed to capture most of the tilted mass.
For $n\ge6$, let $B=2^{n+1}$, $w_i=B+2^i$ for $0\le i<n$,
$k=\lfloor n/3\rfloor$ and $T=kB+\lfloor B/6\rfloor$. Distinct
subsets have distinct sums; sets of size below $k$ fit and sets of
size above $k$ do not. At a tilt with $Q_\lambda(S\le T)\in[0.4,0.6]$
we have $\lambda B<\log15$, since otherwise every $p_i\le1/16$ and
$Q_\lambda(S>T)\le Q_\lambda(|A|\ge k)\le n/(16k)<0.4$.
Hence every $p_i$ lies in $[c_0,1/2]$ with $c_0=(1+15^{5/4})^{-1}$,
and every individual sum has tilted probability at most
$(1-c_0)^n$. Any set of sums carrying at least $0.9$ of the
tilted mass therefore has $\exp(\Omega(n))$ elements. This restricts
only the reconstruction of sums: this family itself is easy to count,
by binomial coefficients below cardinality $k$ and a
most-significant-bit scan at cardinality $k$.
\end{remark}

\begin{remark}[A limitation of a single positive envelope]\label{sp:obstruction}
Just above a boundary layer, the comparison of $m_0$ with $r$ in
\eqref{rw:anti} can lose a factor of order $n/k$, even for polynomial,
pairwise distinct weights. Let $n\ge16$, $2\le k\le n/4$, $B=n^3$,
$w_i=B+i$ and $T=kB-1$. All sets of size below $k$ fit, since
$(k-1)n<B$, and every $k$-set fails, so $Z=\sum_{j<k}\binom nj$. At
every trial $0\le\lambda T\le10(n+1)$ of Lemma~\ref{dt:search}(a), a $k$-set $A$ has
$0\le x_A=\lambda(w(A)-T)\le 10(n+1)(1+kn)/(kB-1)\le 2$, and the exact
product law gives
\[
 \frac{m_0}{r}=\frac1Z\sum_A\frac{e^{-x_A}}{1+x_A^2}
       \ge\frac{e^{-2}}5\,\frac{\binom nk}{Z}
       \ge\frac{e^{-2}}5\,\frac{n+1-2k}{k}.
\]
The last inequality is the geometric bound in the proof of
Proposition~\ref{whole:slicereduction}. Section~\ref{pw:section}
controls this boundary layer by conditioning on group counts.
\end{remark}

\appendix
\section{Deterministic algorithms for structured weights}\label{app:structured}
The proof of Theorem~\ref{fin:main} does not use this appendix. It gives two
deterministic results: Section~\ref{sec:fixed} treats a fixed
number of distinct weights, and Section~\ref{lc:section} treats
weights with a fixed number of offsets plus a multiple of a common
unit with coefficient at most $A$. The first is the case $A=0$ of the
second. The algorithm uses the displaced tilt of
Section~\ref{dt:section}, Fourier inversion in the class counts and
the coefficient sum, and exact sums over integer points in fixed
dimension.

\subsection{An arbitrary fixed number of weight values}\label{sec:fixed}
For any fixed number $m$ of distinct weights, a deterministic
relative approximation to the count takes time linear in $n$ plus a
polylogarithmic term.
We state this case first. It is the case $A=0$ of
Theorem~\ref{lc:main} and is proved after that theorem.

\begin{theorem}\label{thm:fixed}
Let $n$ nonnegative integer weights $w_1,\ldots,w_n$ and a
nonnegative integer capacity $T$ be given on the word RAM, and let
$Z=\#\{A\subseteq[n]:\sum_{i\in A}w_i\le T\}$. Let
$0<\varepsilon<1$. For every fixed $m$ there is
a deterministic algorithm with the following guarantee: if at most
$m$ distinct values occur among the positive weights, it returns
$\widehat Z$ with $|\widehat Z-Z|\le\varepsilon Z$ in
$O_m(n+\log^{O_m(1)}(n+\varepsilon^{-1}+2))$ word operations. The distinct weight
values may be arbitrarily large, provided each fits in an input word.
\end{theorem}

\subsection{Linear dependence on the largest lattice coefficient}\label{lc:section}
This subsection gives a deterministic algorithm whose running time
is linear in the largest coefficient $A$ when every weight is one of
$m$ fixed offsets plus a multiple of a common unit, with multiplier
in $[0,A]$. It works at every capacity, with no assumption on the
feasible density or on how the feasible subsets are distributed
over the class counts.

The method uses Fourier inversion in the joint
variables (class counts, centered coefficient sum). The difficulty
is that the frequency grid of the coefficient sum has length about
$A\sqrt n$, which is too long to scan. After centering each class,
the variance orthogonal to the class counts is at least half of
the total variance (inequality~\eqref{lc:variance}). This makes a
computable trigonometric polynomial $E$ strongly convex near each of
its low points, so the few frequencies that matter can be located by
bisection in $O(A)$ cells. These frequencies are then evaluated in
batches. Finally, the count is summed over the slab by the phase
sums \eqref{mc:phasesum}.
Corollary~\ref{lc:hidden} removes the need to supply the
representation.

\begin{theorem}\label{lc:main}
Let $w_1,\ldots,w_n$, $T$, $Z$ and $\varepsilon$
be as in Theorem~\ref{thm:fixed}, and let $m$ be fixed. Suppose that
every positive weight $w_i\le T$ is supplied in the form
\[
             w_i=b_{g(i)}+q a_i,\qquad
             0\le a_i\le A,\quad a_i\in\mathbb Z ,
\]
with a class $g(i)\in\{1,\ldots,m\}$, nonnegative integer offsets
$b_1,\ldots,b_m$ and a positive integer unit $q$. The offsets, the
unit and $T$ may be arbitrary integers that fit in input words. Then
a deterministic algorithm returns $\widehat Z$ with
$|\widehat Z-Z|\le\varepsilon Z$ in
\begin{equation}\label{lc:time}
 O_m(n)+(A+1)\log^{O_m(1)}(n+A+\varepsilon^{-1}+2)
\end{equation}
word operations. In particular $A\le \varepsilon^{-1}$ gives
$O_{\delta,m}(n+\varepsilon^{-(1+\delta)})$ for every fixed $\delta>0$. For
ordinary positive integer weights one may take $m=1$, $b_1=0$, $q=1$
and $A=\max_i w_i$.
\end{theorem}
For residual weights $0\le v_i\le A$, this theorem also gives a
deterministic additive $\varepsilon$ estimate of the slice probability
$p$ defined in Proposition~\ref{whole:slicereduction}, for every
$0\le k\le n$, within \eqref{lc:time}. The layers $k=0,n$ and cutoffs outside
$[0,\sum_i v_i)$ are settled directly. For $k>n/2$, complementation
expresses the answer as one minus the lower-tail probability on
layer $n-k$ at cutoff $\sum_i v_i-u-1$.

For $1\le k\le n/2$, use the lift $w_i=B+v_i$, $T=kB+u$ from the
proof of Proposition~\ref{whole:slicereduction}, with offset $B$,
unit $q=1$ and coefficients $v_i$. The lifted integers fit in words
of a fixed multiple of the original width, which we simulate at
constant cost as in Subsection~\ref{s48:model}. Write $D=\binom nk$
and $L=\sum_{j<k}\binom nj$. The same geometric bound gives
$L/D\le k/(n+1-2k)\le n$. Estimate $Z=L+Dp$ by this theorem and
estimate $L$ and $L+D$ by Theorem~\ref{thm:fixed} on unit weights.
Use relative accuracy $\eta=\varepsilon/[16(n+1)]$ for all three
counts. Since $Z\le L+D$, the differences estimating $Dp$ and $D$
each have absolute error at most $\eta(2L+D)\le\varepsilon D/8$.
Their quotient therefore differs from $p$ by less than $\varepsilon/2$
before arithmetic rounding. The operands have magnitude
$O((n+1)D)$, so $O(\log((n+1)/\varepsilon))$ significand bits suffice
to keep the total error within $\varepsilon$. The finer relative
accuracy changes the
logarithm in \eqref{lc:time} by at most a constant factor, so
$A\le\varepsilon^{-1}$ again gives
$O_\delta(n+\varepsilon^{-(1+\delta)})$ time.
The proof below includes the case $A=0$.

\paragraph{Step 1: preprocessing and the tilted law.}
Form a table of the multiplicities of each pair $(g,a)$ by direct
indexing in $O_m(n+A)$ operations. Weights zero each double $Z$, and
weights above $T$ are never selected; we set both aside. In the variables $k_g=|U\cap g|$
and $z=\sum_{i\in U}a_i$ for a subset $U$, the decision uses the
single inequality $\sum b_g k_g+qz\le T$. Apply the sign compression
of Section~\ref{sec:fdsign} to $(b_1,\ldots,b_m,q,-T)$ with
norm bound $n(A+1)+2$, enlarged by a fixed dimension-dependent constant.
It preserves the decision throughout the attainable box, as well as
the coefficient signs, using $O_m(\log(n+A+2))$ output bits per entry.
Write $w'_i=b'_{g(i)}+q'a_i$, $T'$ for the new data. A retained
positive weight is still positive and at most $T'$: the coefficient
signs and the test on a singleton give these two facts. Trivial
capacities may be settled beforehand.

For $T'\ge\sum w'_i/2$, use the fair law, whose feasible probability
is at least $1/2$. Otherwise choose a displaced tilt for the
compressed costs deterministically, by the bisection of
Lemma~\ref{dt:search}(a). Its oracle uses only means and variances, and
hence evaluates each trial from the histogram. Put $p_i=(1+e^{\lambda w'_i})^{-1}$ and
$C_\lambda=e^{\lambda T'}\prod_i(1+e^{-\lambda w'_i})$. By Lemma~\ref{dt:price}(a),
$C_\lambda/Z\le(n+2)^{C_{\mathrm{tilt}}}$. Thus any required numerical
accuracy for $Z/C_\lambda$ is inverse polynomial
in $R_A=n+A+\varepsilon^{-1}+2$. Let $J$ be $C_3$ times the binary length
of $\lceil R_A\rceil$, for a sufficiently large positive integer $C_3$
depending on $m$. Round all biases to dyadics so that the
total product coupling error is at most $e^{-4J}$, keeping their
values equal at identical table entries. Keep the original $\lambda$
and $C_\lambda$ for the final payoff. Below, $p_i$ denotes the rounded biases,
$v_i=p_i(1-p_i)$, and $V_g=\sum_{i\in g}v_i$. A class with $V_g=0$
has count zero under this rounded law and is omitted.

\paragraph{Step 2: centering within each class.}
Let $c_g$ be a nearest integer to $V_g^{-1}\sum_{i\in g}v_i a_i$.
Define $d_i=a_i-c_g$, $Y=\sum_i d_i X_i$ and $K_g=\sum_{i\in g}X_i$.
Set $s^2=\sum_i v_i d_i^2$ and
\[
     s_\perp^2=\sum_g\left(\sum_{i\in g}v_i d_i^2
                    -\frac{(\sum_{i\in g}v_i d_i)^2}{V_g}\right).
\]
For the probability weights $v_i/V_g$ in one class, the mean
$\bar d$ obeys $|\bar d|\le1/2$. Integrality gives
$\E d^2\ge\E|d|\ge|\bar d|$. Consequently
$\bar d^2\le\E d^2/2$, and summing over the classes gives
\begin{equation}\label{lc:variance}
                        s_\perp^2\ge s^2/2.
\end{equation}
Informally, this estimate prevents a nearly linear relation between
$Y$ and the counts, which would otherwise force a scan of a
frequency grid of length $A\sqrt n$.
The variable $Y$ may take negative values. Since
$\sum_{i\in U}a_i=\sum_gc_gk_g+Y$, from now on the second argument of
$S'$ is the value $t$ of the centered variable $Y$ (not the
uncentered sum $\sum_{i\in U}a_i$ used in the preprocessing). The
compressed physical weight at an integer vector $(k,t)$ is
\[
          S'(k,t)=\sum_g(b'_g+q'c_g)k_g+q't.
\]

\paragraph{Step 3: a typical box and periods.}
Take integer intervals around the means of $Y$ and of each $K_g$
with half-widths $C_4(s\sqrt J+AJ+1)$ and
$C_4(\sqrt{V_g J}+J+1)$, respectively. Bernstein's inequality
and a union bound make the omitted probability at most $2(m+1)e^{-4J}$.
For each interval choose a power-of-two period strictly larger than
its uncut length and at most twice that length plus a constant.
Write $\Pi,\Pi_g$ for these periods. Then intersect each count interval
with $[0,n_g]$, where $n_g$ is the size of class $g$, and the interval
for $Y$ with $[-nA,nA]$. These intersections retain every attainable
point of the box since $|d_i|\le A$. All periods are polynomial in
$R_A$. For any vector in the resulting box, all other vectors
with the same residues lie outside the box. Thus its joint point
mass equals the joint residue probability up to error $O_m(e^{-4J})$.

\paragraph{Step 4: locating the joint frequencies.}
We now show that only $(A+1)\log^{O_m(1)} R_A$ joint frequencies
have non-negligible characteristic values, and we list a set
containing them without enumerating the full grid.
Write $x=2\pi j/\Pi$ and $y_g=2\pi j_g/\Pi_g$. The joint characteristic
product at $(x,y)$ has modulus at most $e^{-D(x,y)}$, where
\[
 D(x,y)=\sum_g(V_g-\Re[e^{iy_g}B_g(x)]),\qquad
                  B_g(x)=\sum_{i\in g}v_i e^{id_i x}.
\]
Its minimum over the angles $y_g$ is $\sum_g(V_g-|B_g(x)|)$.
We use the computable trigonometric polynomial
\begin{align}
 E(x)&=\sum_g\frac{V_g^2-|B_g(x)|^2}{2V_g}  \notag\\
 &=\sum_g\frac1{2V_g}\sum_{i,j\in g} v_i v_j
                      (1-\cos((d_i-d_j)x)).               \label{lc:pair}
\end{align}
It lies between one half and one times the preceding minimum.
It has nonnegative difference weights at frequencies of magnitude
at most $A$, and $E''(0)=s_\perp^2$. Its coefficients are obtained
by one autocorrelation of the variance histogram per class. All
inputs are short dyadics, and the divisions by positive $V_g$ require
only polylogarithmic precision, since the rounded positive biases
are at least $2^{-O(J)}$.

\emph{The first coordinate.} We list a set containing every grid
point $x$ with $E(x)\le3J$. If $s_\perp^2\le C_5 A^2J$, then
\eqref{lc:variance} gives $\Pi=O(AJ+1)$, so we take the entire grid.
Otherwise, split the circle into $O(A)$ equal cells of length at
most $1/(10A)$. In a cell containing a point $x_*$ with
$E(x_*)\le3J$, \eqref{lc:pair} implies
\[
 E''(x_*)\ge s_\perp^2-A^2 E(x_*),\qquad
       |E'''(x)|\le A s_\perp^2 .
\]
For $C_5$ large, $E''\ge 0.8s_\perp^2$ throughout such a cell.
In each cell, use a bounded sign bisection of $E'$ to find the
constrained minimum, including the endpoints of the cell. In a cell
containing a low point this finds the minimum within $1/(1+s_\perp)$;
a near-zero derivative suffices by strong convexity. In a cell with
no low point its result can be arbitrary. Keep the grid indices
within $C_6\sqrt J/s_\perp$ of the reported point in that cell. The
quadratic lower bound relative to a constrained minimum shows that
every low point was kept. By \eqref{lc:variance} and the
large-variance branch, each cell returns
$O(1+\Pi\sqrt J/s_\perp)=O(J)$ indices. It suffices to know $E'$ to
additive accuracy a small constant times $s_\perp^2/(1+s_\perp)$, and to decide
the test $s_\perp^2\le C_5A^2J$ up to a factor $2$, absorbed by $C_5$.
Note that the grid has not been enumerated in this branch.

\emph{The class coordinates.} For every listed $x$, and
independently in each class, retain all $y_g$ if $V_g\le 10J$.
Otherwise use a unit numerical buffer: discard the row if
$V_g-|B_g(x)|>3J+1$, and else retain the $O(J)$ grid points
at circular distance at most $C_7\sqrt{J/V_g}$ from $-\arg B_g(x)$,
enlarged by two spacings. By the same cosine estimate as in
\eqref{sp:cardext}, this includes every possibility with
$V_g-\Re(e^{iy_g}B_g(x))\le 3J$. Thus all omitted joint frequencies
have modulus at most $e^{-3J}$. There are only
$(A+1)\log^{O_m(1)} R_A$ retained tuples. The normalization
$1/(\Pi\prod_g \Pi_g)$ in Fourier inversion bounds the entire omitted
error pointwise by $e^{-3J}$, independently of the full grid volume.

\paragraph{Step 5: evaluating frequencies in batches.}
The queried $x$ need not form a progression. A polynomial
$f(x)=\sum_{|d|\le \Omega}f_d e^{idx}$ with $\Omega=O((A+1)J)$ and
polynomially bounded coefficient one-norm is evaluated at $\mu$
arbitrary angles to inverse-polynomial accuracy in
$\tO(\Omega+\mu+1)$ operations (logarithms here are in $R_A$) by the
transpose of the grid expansion of Section~\ref{sp:section}: on a
power-of-two grid of size $M_0\ge2\Omega+1$, transform the arrays
$f_d(d/M_0)^\ell$, $0\le\ell\le C_8J$, by FFT, and at a query multiply
the $\ell$th value at a nearest grid point $x_0$ by
$(iM_0(x-x_0))^\ell/\ell!$. Since $|d(x-x_0)|<\pi/2$, the same factorial
remainder applies. This also supplies the values
of $E'$ in each bisection round and the class values $B_g(x)$.

For the products at $(x,y_g)$, apply the entire-function and power-sum
calculation from Section~\ref{sp:section} with damping cutoff $O_m(J)$.
Each power sum of degree
$h=O_m(J)$ uses, for $0\le\nu\le h$, the histogram polynomial with
weights $p_i^h$ at turns $\nu d_i x$, multiplied after evaluation by
$e^{i\nu y_g}$. Repeated input types supply their multiplicities.
The damping sums are evaluated in the same way. All histogram
polynomials have degree $O_m(AJ)$, or can instead be queried at $\nu x$,
with inverse-polynomial guard precision for the $2^h$ coefficient
growth. Batching all query points costs
$(A+1)\log^{O_m(1)} R_A$. Adaptively chosen bisection points incur
the same bound over logarithmically many rounds. No further input
scan is needed.

\paragraph{Step 6: recovering the count.}
In the clipped box of Step~3, impose $S'(k,t)\le T'$ and
$\lambda(S'(k,t)-T')\ge-J$. Dropping the remainder of the feasible
halfspace costs at most $e^{-J}$ in the normalized count $Z/C_\lambda$.
On this slab, replace $e^{\lambda(S'-T')}$ by its Taylor polynomial
of degree $O(J)$ with uniform error $e^{-2J}$. The slab inequalities
and the polynomial have rational coefficients since $J$ is an integer
and the tilt is rational ($\lambda=y/T'$ for the dyadic search variable
$y$ of Lemma~\ref{dt:search}(a), or $\lambda=0$ for the fair law).
Substitute the sparse inverse Fourier sum
for the joint point mass, and sum each resulting character times
that polynomial on the slab and box by \eqref{mc:phasesum}, now
in the $m+1$ coordinates $(k,t)$, with $N=n(2A+1)$ since $0\le k_g\le n$
and $|t|\le nA$. The number of integer vectors in the box is at
most $(n+1)^m(2nA+1)$, and the sum of the absolute normalized Fourier
coefficients is at most one before rounding. Thus all inversion,
arithmetic and polynomial errors, including the product-law coupling,
cost at most $\operatorname{poly}_m(R_A)e^{-J}$ in $Z/C_\lambda$.
Taking $C_3$ sufficiently large and multiplying by $C_\lambda$ to short
relative accuracy, with $\log C_\lambda$ summed over the histogram entries,
and by $2$ to the number of zero weights,
proves the relative guarantee and \eqref{lc:time}.
For $A\le \varepsilon^{-1}$, the logarithms on $A+1$ fit the literal bound by
splitting at $n=\varepsilon^{-(1+\delta/2)}$. This proves Theorem~\ref{lc:main}.

\begin{proof}[Proof of Theorem~\ref{thm:fixed}]
Group the positive weights not exceeding $T$ by value in $O_m(n)$
operations, comparing each with the at most $m$ values seen so far.
Apply Theorem~\ref{lc:main} with $q=1$, the distinct values as
offsets $b_g$, and all $a_i=0$, so that $A=0$. Then \eqref{lc:time}
is the asserted bound.
\end{proof}

\subsubsection{Recovering an unknown common unit and unknown classes}
The following corollary shows that the representation required by
Theorem~\ref{lc:main} need not be supplied. Under the promise that
the weights lie in $m$ unknown progressions of a common unknown
step, the algorithm finds a valid representation itself and checks
it exactly.

\begin{corollary}\label{lc:hidden}
In the setting of Theorem~\ref{lc:main}, suppose that only $m$ and an
integer $A\ge0$ are supplied, and that the positive weights not
exceeding $T$ lie in the union of $m$ unknown translates of
$\{0,q,2q,\ldots,Aq\}$ for an unknown common positive integer $q$.
Then a deterministic algorithm returns $\widehat Z$ with
$|\widehat Z-Z|\le\varepsilon Z$ within the bound \eqref{lc:time}.
It constructs and checks an exact representation as in
Theorem~\ref{lc:main}, with at most $m^3$ classes and coefficient
bound $2A$. If the promise fails, it may report failure instead. In
particular, with $A=\lceil 1/\varepsilon\rceil$ no unit and no classes need to be
supplied.
\end{corollary}
\begin{proof}
The plan is as follows. We first cluster the weights and list their
distinct values. We then generate a constant number of candidates
for a multiple of the unit, and for each candidate detect the
correct divisor with a Gram determinant test. Finally we verify the
resulting representation exactly.

Check in $O_m(n)$ operations whether there are at most $m^2$
distinct retained values. If so, use each value as its own offset,
with unit one and coefficients zero, and apply Theorem~\ref{thm:fixed}.
If this check fails and $A=0$, the promise fails and we report failure. For $A\ge1$, use
farthest-point clustering with at most $m$ centers, in $O_m(n)$
operations: start with any retained weight, repeatedly select the
weight of largest distance to the closest chosen center, and after at
most $m$ choices assign each weight to a nearest center. Under the
promise, the covering radius after $m$ choices is at most $Aq$. Otherwise those centers and a remaining farthest
point would be $m+1$ points pairwise farther apart than $Aq$, which is
impossible in the $m$ original progressions. Every resulting cluster
therefore has diameter $D\le2Aq$, without any gap assumption.

For a nonconstant cluster of minimum $a$ and diameter $D$, place
$w$ at index $\lfloor(2A+1)(w-a)/D\rfloor$, with the maximum in
a separate endpoint bin. A bin has width $D/(2A+1)<q$, so it
contains at most one distinct value from each original progression.
Maintain its at most $m$ distinct values and their multiplicities
by direct comparison; a larger number permits failure.
Constant clusters use one bin. This builds a list of all distinct
values in $O_m(n+A)$ operations. Denote its length by $s$;
under the promise $s\le m(A+1)$. Since the initial check found more
than $m^2$ distinct values, one cluster contains at least $m+1$.
Choose any $m+1$ distinct values in such a cluster, and test
the constantly many positive pair differences $q_0$ among them. At
least one such difference equals $kq$ for some $1\le k\le2A$ (in
fact $k\le A$: two of the $m+1$ values lie in the same progression,
whose coefficients lie in $[0,A]$; the weaker range suffices below).

For a trial $q_0$, write $x_j=(w^{(j)}\bmod q_0)/q_0$ for the
$s$ distinct weights. For each $d\le2A$ dividing $q_0$, consider
the Hermitian Gram matrix of order $m+1$
\[
       G_d(u,l)=\sum_{j=1}^s \mathrm e\bigl((u-l)d x_j\bigr),
                         \qquad 0\le u,l\le m .
\]
All entries for all trials $d$ at this $q_0$ are read from the
ordinary Fourier sums at integer frequencies $|h|\le 2mA$.
The grid expansion of Section~\ref{sp:section} computes them in
$(A+1)\log^{O_m(1)}(A+2)$ operations on the short list.
Taking the original-word remainder first allows each $x_j$ to
be rounded to any required $O_m(\log(A+2))$ fractional bits by
one short scaling and division.

Set $\beta=1/[32(A+1)]$ and $\Delta=(4\beta)^{m(m+1)}$.
Compute all determinants to absolute error below $\Delta/8$.
This needs only $O_m(\log(A+2))$ bits after the binary point,
since the dimension is fixed and the entries have magnitude at
most $s$. A divisor passes when the real part of its computed
determinant is below $\Delta/2$. Choose the smallest passing divisor
$d$, or continue to the next trial $q_0$ if none passes.
By Cauchy--Binet, each true determinant is the sum, over choices
of $m+1$ columns, of the squared
absolute Vandermonde determinants of the corresponding $\mathrm e(d x_j)$.
If $m+1$ of these turns have pairwise circular distance at least
$\beta$, each chord has length at least $4\beta$, and hence one summand
is at least $\Delta$. Thus for any passing $d$, all turns are within
$\beta$ of at most $m$ centers: take any maximal
$\beta$-separated subcollection. Conversely, at most $m$ distinct
turns give determinant exactly zero, and such a $d$ necessarily passes.

For a correct trial $q_0=kq$, $d=k$ is a divisor, and the turns
are the original residues modulo $q$, expressed as fractions of $q$.
There are at most $m$ such residues. Therefore the selected $d$
satisfies $d\le k$ and $q'=q_0/d\ge q$. Among items of any one
true residue, the differences of $d x_j$ modulo one are multiples
of $1/k$. Since $2\beta<1/k$, two such turns in an arc of radius
$\beta$ must coincide. Over all true residues and covering arcs there
are therefore at most $m^2$ distinct exact turns, equivalently at
most $m^2$ distinct residues modulo $q'$. The detector need not
distinguish arbitrarily close unequal residues.

Finally, verify the selected $q'$ exactly by listing residues by original-word
remainders, and reject the trial if there are more than $m^2$.
Intersect each residue class with each of the at most $m$ physical
clusters, take its minimum as offset, and check that its span is at
most $2Aq'$. These checks always succeed for the correct trial because
$D\le 2Aq\le 2Aq'$. Any other trial that passes the same exact checks
also gives a valid representation. Apply Theorem~\ref{lc:main},
with $2A$ in place of $A$ and $m^3$ in place of $m$, to the first verified trial, or report failure if there is none.
There are only $O_m(1)$ trials, and all work after constructing
the list takes $(A+1)\log^{O_m(1)}(n+A+\varepsilon^{-1}+2)$. This proves the claim.
\end{proof}

\end{document}